\documentclass[11pt,a4paper]{article}
\pdfoutput=1
\usepackage{jheppub}
\usepackage{rotating}
\usepackage{tablefootnote}
\usepackage{wrapfig}
\usepackage{comment}
\usepackage{graphicx,graphbox}
\usepackage{amsfonts}
\usepackage{amsmath}
\usepackage{slashed}
\usepackage{amssymb}
\usepackage{latexsym}
\usepackage{multirow}
\usepackage{hyperref}
\usepackage{enumitem}
\hypersetup{colorlinks=true,citecolor=red,linkcolor=magenta,urlcolor=blue}
\usepackage{relsize}
\usepackage{booktabs}
\usepackage{subfigure}
\usepackage{cleveref}
\usepackage{fancyhdr}
\usepackage{float}
\usepackage{graphicx}
\usepackage{microtype}
\usepackage{tabularx}
\usepackage{xcolor}
\usepackage{siunitx,adjustbox,bm}
\usepackage[bottom]{footmisc}
\usepackage{makecell}
\usepackage[normalem]{ulem}

\usepackage{mathtools}
\usepackage[thinc]{esdiff}
\usepackage[euler]{textgreek}
\usepackage{braket}
\usepackage{physics}
\usepackage{xfrac}
\usepackage{soul}
\usepackage{stmaryrd}

\newcommand{\ce}[1]{{\color{blue}{#1}}}

\newcommand{\nn}{\nonumber}

\renewcommand{\tr}{\text{tr}}
\renewcommand{\Tr}{\text{Tr}}
\newcommand{\eff}{\text{eff}}
\newcommand{\cpv}{\text{(CPV)}}
\newcommand{\cpc}{\text{(CPC)}}
\renewcommand{\L}{\mathcal{L}}
\renewcommand{\P}{\mathcal{P}}

\usepackage{lineno}

\begin{document}
\allowdisplaybreaks
\title{One-loop Effective Action up to Dimension Eight: Integrating out Heavy Fermion(s)}
\abstract{We present the universal one-loop effective action up to dimension eight after integrating out heavy fermion(s) using the Heat-Kernel method. We have discussed how the Dirac operator being a weak elliptic operator, the fermionic operator still can be written in the form of a strong elliptic one such that the Heat-Kernel coefficients can be used to compute the fermionic effective action. This action captures the footprint of both the CP conserving as well as violating UV interactions. As it does not rely on the specific forms of either UV or low energy theories, can be applicable for a very generic action. Our result encapsulates the effects of heavy fermion loops only.}	
\author[a]{Joydeep Chakrabortty}
\author[a,b]{Shakeel Ur Rahaman}
\author[a]{ Kaanapuli Ramkumar}

\emailAdd{joydeep, shakel@iitk.ac.in, kaanapuliramkumar08@gmail.com}
\affiliation[a]{Indian Institute of Technology Kanpur, Kalyanpur, Kanpur 208016, Uttar Pradesh, India}
\affiliation[b]{Institute of Physics, Sachivalaya Marg, Bhubaneswar, Odisha 751005, India}

	
\preprint

\maketitle
	
\section{Introduction}
\label{sec:intro}
Physics of different lengths, i.e., energy scales are expected to be unfolded gradually. In that case, the most influential technique is the Effective Field Theory (EFT) \cite{Weinberg:1980wa, Georgi:1994qn,Manohar:2018aog,Cohen:2019wxr} as it has an inherent property of capturing the new physics effects even without knowing it exactly. This is the bottom-up approach of EFT. On the other side, employing the Wilsonian method, we can express a full theory into a {\it effective} one after integrating out heavy modes whose mass decides the cut-off scale. In the process of integrating out one needs to be careful about the decoupling limit and validate the truncation of the effective Lagrangian \cite{Banerjee:2023bzl,Brivio:2022pyi}. \\

\noindent
The Standard Model (SM) of particle physics appears to have prevailed after the discovery of the Higgs boson, a decade of collider searches, and many experimental advancements. However, the neutrino mass, the Baryon asymmetry of the universe (BAU), and some other observations, e.g., flavor puzzle may point to the existence of beyond Standard model (BSM) physics around an energy scale that may not be directly accessible by the current experiments. In summary, at this present juncture of particle physics, we have passive tantalizing hints of new physics whose exact nature is yet to be unveiled. This is when EFT can be one of our best tools to take on board in search of BSM physics. In the absence of any direct evidence \footnote{No experiment has not confirmed yet the on-shell existence of any new particle beyond the SM ones.}, we can still capture the indirect effects of heavy new particles, if any,  by extending the renormalisable SM Lagrangian by including the higher mass dimensional effective operators a.k.a. considering the Standard Model Effective Theory (SMEFT) \cite{Brivio:2017vri,Isidori:2023pyp}.\\

\noindent
The method of integrating out the heavy particles and then performing matching between two theories at an energy scale are the sole of Wilsonian EFT \cite{Georgi:1994qn,Wilson:1971,Wilson:1973jj}. The emerged low energy effective Lagrangian captures the footprint of integrated out UV interactions through the higher mass dimensional effective operators that are accompanied by definite coefficients, known as Wilson coefficients (WCs). In the bottom-up approach \cite{BUCHMULLER1986621,Grzadkowski:2010es,Lehman:2014jma,Murphy:2020rsh,Li:2020gnx,Li:2022tec,Banerjee:2020jun,Anisha:2019nzx,Banerjee:2019twi}, we have no clue about the origin of these WCs and thus they are independent of each other. But, in the top-down approach \cite{Gaillard:1985uh,Chan1985}, the WCs are functions of the UV parameters. 
In the Wilsonian EFT, the challenging task is to compute the effective action after integrating out the heavy fields, especially beyond the tree level. Till now many attempts have been made to compute the one-loop effective action consisting of effective operators up to mass dimension six employing different methods \cite{Gaillard:1985uh,Chan1985,Cheyette:1987qz,Henning:2014wua, Drozd:2015rsp,Ellis:2016enq,delAguila:2016zcb,Ellis:2017jns,Kramer:2019fwz,Angelescu:2020yzf,Ellis:2020ivx}. A gauge covariant approach to perform the matching has resulted in a comprehensive formula known as the Universal One-Loop Effective Action (UOLEA) \cite{Henning:2014wua, Drozd:2015rsp,Ellis:2016enq,delAguila:2016zcb,Ellis:2017jns,Kramer:2019fwz,Angelescu:2020yzf,Ellis:2020ivx}. Compared to the traditional approach of matching via the Feynman diagrams, it is straightforward and algorithmic in nature. Based on the UOLEA some automated tools have been developed in aid of the matching procedure \cite{Bakshi:2018ics,Fuentes-Martin:2022jrf,Carmona:2021xtq,Dawson:2022ewj, Aebischer:2023irs}. To date, this formula was restricted up to mass dimension six, until recently in our previous paper \cite{Banerjee:2023iiv} we extend UOLEA up to dimension eight $(D8)$ \footnote{We will use the convention $Di$ to signify operators of mass dimension $i$.}. It is worth mentioning that in recent times, dimension eight effective action has drawn much interest and related phenomenology has been explored \cite{Dawson:2021xei,Hays:2018zze,Corbett:2021eux,DasBakshi:2023htx,DasBakshi:2022mwk,Chala:2021pll,Alioli:2020kez,Degrande:2013kka,Ellis:2020ljj,Hays:2020scx,Dawson:2023ebe,Dawson:2022cmu,Ellis:2019zex,Corbett:2023qtg,Ellis:2023zim,Degrande:2023iob,Banerjee:2023bzl,Jahedi:2023myu,Jahedi:2022duc}. In this previous paper \cite{Banerjee:2023iiv} we introduced the Heat-Kernel (HK) method \cite{Minakshisundaram:1949xg,Minakshisundaram:1953xh,Hadamard2003,DeWitt:1964mxt,Seeley:1969,Schwinger:1951nm,Vassilevich:2003xt,Avramidi:2001ns,Avramidi:2015ch5,Kirsten:2001wz,Fulling:1989nb} to compute the one-loop effective action and provide the UOLEA. We also employed the covariant diagram \cite{Zhang:2016pja} method to testify part of our UOLEA result. \\

\noindent
The one-loop effective action computed in our previous paper using the HK method is really universal in the sense that it does not require to assume of any specific UV or low-energy theory. On top of that, though we have emphasized that our UOLEA is an aftermath of integrating heavy scalars, this result is equally applicable for any action written in the form of a strong elliptic operator ($D^2+U+M^2$) \cite{DeWitt:1964mxt, Fulling:1989nb, Kirsten:2001wz}. In the case of the scalar field, the action contains the Klein-Gordon operator which is already in that desired form. 
In this paper, we want to compute the effective action when heavy fermions are integrated out, and that too using the HK method.  Unfortunately, the Dirac operator is a weak elliptic operator \cite{Esposito:1997mw,Avramidi:1997sh,Avramidi:1997hy}. Thus, we can not directly compute the HK coefficients until we manage to construct a strong elliptic operator in terms of this weak one.  In the absence of non-trivial background when we can use the mass-evenness property of the Dirac operator we can perform bosonization and rewrite the Dirac operator in terms of a Klein-Gordon-like operator \cite{Nieh:1981xk, Obukhov:1982da, Obukhov:1983mm, Cognola:1987wd, Cognola:1988qw,  Yajima:1995jk, Geyer:1999xp, Cognola:1999xv, DeBerredo-Peixoto:2001wkv, Nakonieczny:2018djb,Elizalde:1999zy}. Once we express the fermionic action in the form of a strong elliptic operator, i.e., $(D^2+U+M^2)$, we can use the previously computed UOLEA up to dimension eight.  In the process, the covariant derivative and the potential term ($U$) are modified and contain Clifford generators. Thus, in addition to the method adopted in the case of heavy scalar, we have to further perform the traces over Clifford matrices ($\gamma_{\mu}, \gamma_5$) to achieve the desired and usable form of fermionic UOLEA. We will discuss this in detail in the following sections. \\

Charge(C)-Parity(P) violation (CPV), is one of the most intriguing possible signatures of new physics that goes beyond the SM. ATLAS has recently carried out a comprehensive study of $Zjj$ production and also interpreted the data in terms of CPV SMEFT operators of dimension six \cite{ATLAS:2020nzk, Bakshi:2021prd, Bakshi:2021}. The CP-violation also plays a pivotal role in explaining the BAU as it is one of the necessary conditions \cite{Sakharov:1967dj}. Although SM has a source of CPV, it is not sufficient to explain the observed asymmetry in matter-antimatter. A BSM scenario is necessary to enhance the CPV. Along this thought, the SM extended by Vector-like fermions (VLF) scenarios draw much attention due to its easy incorporation of CPV in the Yukawa sector \cite{ Bakshi:2021prd, Bakshi:2021,Huo:2015exa}.  In VLF models one can directly invoke the heavy fermion mass term without employing any spontaneous symmetry breaking.  For example, even within the SM gauge symmetry framework, we can write down the Dirac mass term for the VLFs and as they are not the outcome of electroweak symmetry breaking vacuum expectation value ($v_{ew}$), can be sufficiently heavier than $v_{ew}$ such that EFT can be validated once we integrate them out. It is important to stress at this point that our results are not restricted to the SM and(or) its VLF extension. Even a similar fermionic UOLEA is applicable for heavy Majorana fermions with suitable modification up to a numerical factor. We must aware the readers about the specialty of interactions involving $\gamma_5$. In the process of fermion integrating out, the presence of $\gamma_5$ requires special attention which we discuss later. The one-loop effective action after fermion integrating out up to dimension six has been discussed in \cite{Angelescu:2020yzf,Ellis:2020ivx}. \\

\noindent
This paper is structured as follows. In Sec.~\ref{sec:effective action}, we briefly introduce the HK method and its connection with the one-loop effective action. We emphasize the subtleties associated with fermion fields. Especially, when bosonization is performed to rewrite the Dirac operator in terms of a strong second-order elliptic operator such that the HK coefficients can be used to compute the effective operators in this case as well. We present the fermionic universal one-loop effective action in Sec.~\ref{sec:fermionic-uolea}. We catalogue our results for different mass dimensions, up to dimension eight, separately. In the following Sec.~\ref{sec:discussion}, we discuss some features of the fermionic effective action - how the flavor dependence emerges, unlike the scalar case. We discuss how the CP violation in the UV theory leaves its footprint in the effective operators at low energy. We also highlight the interplay between CP-conserving and violating operators based on the nature of UV interaction, and also the hierarchical nature of CP violation among equivalent effective operators of different mass dimensions. Then we conclude with a note on possible future directions that need to be explored.

\section{Heat-Kernel and Universal One-Loop Effective Action}
\label{sec:effective action}
Starting from a UV complete action $S$, functional of heavy ($\Phi$) and light ($\phi$) fields, the one-loop effective action obtained by integrating out the heavy field is given by,
{\small\begin{align}\label{eq:effective-action}
        e^{i\,S_{\text{eff}}[\phi]} = \int [\mathcal{D}\Phi]e^{i\,S[\Phi,\phi]}
        &= \int [\mathcal{D}\eta]\, \exp\left[i\,\left(S[\Phi_c,\phi]\,+\,\frac{1}{2}\frac{\delta^2 S}{\delta \Phi^2}\bigg\vert_{\Phi=\Phi_c}\,\eta^2\,+\,\mathcal{O}(\eta^3)\right)\right],\nn\\
        S_{\text{eff}} &\approx S[\Phi_c]\,+\,\frac{i}{2}  \ln\left(\text{Det}\,\frac{\delta^2 S}{\delta \Phi^2}\bigg\vert_{\Phi=\Phi_c}\right).
\end{align}}
Here, the heavy field is expanded  around its classical value ($\Phi_c$), and ($\eta$) is the fluctuation. 
For a generic Lagrangian of the form $\mathcal{L}= \Phi^\dagger\Delta\Phi$, the one-loop contribution to the effective action $(S^{(1)}_{\eff})$ is given by the spectral function,
\begin{equation}\label{eq:generic_action_scalar}
   S^{(1)}_{\eff}= \frac{i}{2} \Tr \ln \Delta = -\frac{i}{2} \Tr\int_0^\infty \frac{dt}{t} e^{-t\Delta}.
\end{equation}
The `$\ln$'-function, as in the above equation, is written in the integral form using the following identity
\begin{equation}\label{eq:identity}
    \ln \lambda = -\int_0^\infty \frac{dt}{t} e^{-t\lambda},
\end{equation}
with $\lambda > 0$.
In the case of scalar fields, the Lagrangian operator $\Delta$ takes the generic form as
\begin{equation}\label{eq:laplacian}
    \Delta = -P^2+U_s+M_s^2,
\end{equation}
where $P$ is the covariant derivative given by $P_\mu = i\,D_\mu=i\,(\partial_\mu - i A_\mu)$ \footnote{Here, the coupling constant is absorbed in the definition of the gauge field $A_\mu$.} and $M_s$ is the mass of the heavy field. Here, $U_s=\delta^2S/\delta \Phi^{\dagger}\delta \Phi$ is a functional of light-fields of mass dimension +2, and contains the information about the UV interactions. We can convert this operator $\Delta$ into a second-order strong elliptic operator invoking Wick rotation and rewriting the same in Euclidean space as $\Delta_E$ \footnote{We will drop the subscript $E$ from $\Delta_E$ from now on wards.}. This allows us to identify the exponent in Eq.~\eqref{eq:generic_action_scalar}, i.e., $e^{-t\Delta}$ with the Heat-Kernel $K(t,x,x,\Delta)$ \cite{Kirsten:2001wz,Vassilevich:2003xt}. Hence, the one-loop contribution to the effective Lagrangian after integrating out heavy scalars can be written in terms of the Heat-Kernel in the Euclidean space as
\begin{eqnarray}
    \mathcal{L}^{\Phi}_{\eff}= \frac{1}{2}\, \tr \int_0^\infty \frac{dt}{t} K(t,x,x,\Delta).
\end{eqnarray}
For a general second-order elliptic operator of the form, $\Delta = D^2 + M^2 + U$, the Heat-Kernel can be written as a power law expansion in the parameter $t$ as \cite{DeWitt:1964mxt,Seeley:1969}
\begin{equation}\label{eq:interaction}
    K(t,x,y,\Delta)=(4\pi t)^{-d/2}\ \mathrm{Exp}\left[\frac{z^2}{4t}-t\,M^2\right] \sum_k \frac{(-t)^k}{k\,!}b_k(x,y),
\end{equation}
where $z_\mu=(x-y)_\mu$, $d=4$ is flat-Euclidean space-time dimension, and $b_k$ are the Heat-Kernel coefficients (HKCs). In the coincidence limit $x\rightarrow y$, the  HKCs are denoted by square brackets ($[b_k]=b_k(x,x)$).

In Ref.~\cite{Banerjee:2023iiv}, we have used the Heat-Kernel expansion to derive the  universal one-loop effective action (UOLEA) expanded up to dimension eight after integrating out scalar fields. In the current paper, we extend the results presented in \cite{Banerjee:2023iiv} to encompass the effect of heavy fermion field integration out at the one-loop leading to {\it Fermionic} UOLEA, i.e., FUOLEA. The massive free fermionic Lagrangian, in terms of the Dirac operator, is given as
\begin{equation}
    \mathcal{L}^{\Psi}= \overline \Psi (\slashed P - M_f) \Psi,
\end{equation}
with $M_f$ being the mass of the fermion. Here, we use the Feynman slash notation to denote contractions with gamma matrices, i.e.,  $\slashed P = \gamma^{\mu}P_\mu$. Following the similar prescription as in Eq.~\eqref{eq:effective-action}, we can compute the effective action after integrating out the fluctuations $(\overline{\eta}, \eta)$ of the heavy fermions over the classical backgrounds $(\overline{\Psi_c}, \Psi_c)$ as 
{\small\begin{align}\label{eq:effective-action_f}
        e^{i\,S_{\text{eff}}[\psi]} = \int [\mathcal{D}\overline\Psi][\mathcal{D}\Psi]e^{i\,S[\Psi,\psi]}
        &= \int [\mathcal{D}\overline\eta][\mathcal{D}\eta]\, \exp \left[i\, \left( S[\Psi_c,\psi]\, + \, \overline\eta\, \frac{\delta^2 S}{\delta \overline\Psi\delta \Psi} \bigg\vert_{\Psi=\Psi_c} \,\eta\, + \,\mathcal{O}(\eta^3)\right)\right],\nn\\
     \text{where}\,\,\,   S_{\text{eff}} &\approx S[\Psi_c]\,-\,i\, \ln\left(\text{Det}\,\frac{\delta^2 S}{\delta \overline\Psi\delta \Psi}\bigg\vert_{\Psi=\Psi_c}\right).
\end{align}}
It is worth noting that apart from different constant pre-factor and different signs, the functional form of the one-loop fermionic effective action possesses a similar mathematical structure of the same computed for the scalar case, see Eq.~\eqref{eq:generic_action_scalar}. But in the case of fermions, the one-loop effective action  comprising the Dirac operator $(\slashed P - M_f)$ is given as
\begin{equation}\label{eq:generic_action_fermion}
   S^{(1)}_{\eff}\big\vert_{\text{fermion}}= - i\, \Tr \ln [\slashed P - M_f].
\end{equation}
The difficulty that arises in working with the Dirac operator is that it is a weakly elliptic first-order differential operator whose spectrum is unbounded, unlike the Klein-Gordon operator in the scalar Lagrangian, which is a strong elliptic operator \cite{Esposito:1997mw,Avramidi:1997sh,Avramidi:1997hy}. The identity in Eq.~\eqref{eq:identity} that allows rewriting effective action in terms of the Heat-Kernel is true if $\lambda > 0$. If we identify $\lambda$ as an eigenvalue of the Dirac operator, then so is $-\lambda$ \cite{Elizalde:1999zy}. Then, these negative eigenvalues of the Dirac operator refrain one to employ the HK method to compute the fermionic effective action.\\

To proceed further we will employ the mass evenness property of $(\slashed P - M_f)$ operator \cite{Nieh:1981xk, Obukhov:1982da, Obukhov:1983mm, Cognola:1987wd, Cognola:1988qw,  Yajima:1995jk, Geyer:1999xp, Cognola:1999xv, DeBerredo-Peixoto:2001wkv, Nakonieczny:2018djb, Elizalde:1999zy} that reads as $\text{Det} [\slashed P - M_f] = \text{Det} [\slashed P + M_f]$ up to a diverging physically irrelevant constant, which is a direct consequence of the symmetry of the spectrum of Dirac operator. This allows one to recast the one-loop effective action in terms of a second-order elliptic operator ($\slashed P^2$) as,
\begin{align}\label{eq:dirac_lag_A}
       S^{(1)}_{\eff}\big\vert_{\text{fermion}} &= - i\, \ln \text{Det} [\slashed P - M_f ] = -\frac{i}{2} \big\{\ln \text{Det}[\slashed P - M_f ] + \ln \text{Det}[-\slashed P - M_f] \big\}\nn\\
       &= -\frac{i}{2}\big\{ \ln \text{Det} \big[ -\slashed P^2 + M_f^2 \big] - \mathcal A\big\} ,
\end{align}
where $\mathcal{A}$ is the multiplicative anomaly defined as
\begin{equation}
    \mathcal A = \ln \text{Det}\big[ -\slashed P^2 + M_f^2 \big] - \ln \text{Det}[\slashed P - M_f ] - \ln \text{Det}[-\slashed P - M_f].
\end{equation}
Note that, though $\ln[ \lambda_i \lambda_j]=\ln \lambda_i + \ln \lambda_j$ works for the individual eigenvalues of the spectrum, it does not hold for the spectral determinant as in general, $\text{Det} [AB]\neq \text{Det} [A] \text{Det} [B]$. Hence, the multiplicative anomaly is in general non-zero. In Ref.~\cite{Cognola:1999xv} the multiplicative anomaly for the massive Dirac operator has been discussed for even dimensions ($d$) in terms of the HKC as
\begin{equation}\label{eq:multiplicative_A}
    \mathcal A = 2\sum_{j=1}^{d/2} \frac{(-1)^j\, M_f^{2j}\, Q_j}{j!}\,[b_{d/2-j}],
\end{equation}
where $[b_k]$ are the HKCs of $(\slashed P^2)$ operator, and \[Q_j = \sum_{l=1}^j \frac{1}{2l-1}.\]
It is worthwhile to note that for the massless case, the multiplicative anomaly is identically zero. For the massive case at $d=4$, the multiplicative anomaly is given by
\begin{eqnarray}
    \mathcal{A} = - 2M_f^2\ [b_1] + \frac{4}{3} M_f^4\ [b_0],
\end{eqnarray}
where the Heat-Kernel initial condition sets the zeroth HKC to identity, i.e.,  $[b_0]= I$ and $[b_1]$ is the first non-trivial HKC. Thus, the multiplicative anomaly only gives correction to the renormalised part of the one-loop effective action operators of mass dimension $\leq 2$, and for the higher dimensional operator computation, we can focus only on $(-\slashed P^2+M_f^2)$ operator. This whole game of expressing the Dirac operator  in terms of a strongly elliptic second-order operator is called the bosonization and is equally valid for an interacting theory as well. In this present work, we consider the interaction of the following form $\overline \Psi \Sigma \Psi$ in the fermionic Lagrangian \footnote{$\Sigma$ contains scalar as well as pseudo-scalar interactions.}. 

For the rest of the paper, we will not pay attention to this multiplicative anomaly as it is not of much relevance here.  Employing the properties of  the strongly elliptic second-order operator,  the fermionic one-loop effective action can be derived from  Eq.~\eqref{eq:generic_action_scalar}. This allows us to write a unified one-loop effective Lagrangian in the Euclidean space as
\begin{eqnarray}\label{eq:unified_lag}
    \mathcal{L}_{\eff}= c_s \tr \ln[ -\P^2 + M^2 + U] = c_s \, \tr \int_0^\infty \frac{dt}{t} K(t,x,x,\Delta),
\end{eqnarray}
where $c_s=+1/2$, $+1$, and $-\frac{1}{2}$ for real scalar and complex scalar and fermionic backgrounds respectively. Here, the second-order elliptic operator ($\Delta$) has a generic form $\Delta = -\P^2 + M^2 + U$ where $\P$  and $U$ are the generalised covariant derivative and the field-dependent functional whose structure depends on the background fields being integrated, respectively. This leads to the universal one-loop effective action (UOLEA) in terms of the HKCs \cite{Banerjee:2023iiv}
\begin{equation}\label{eq:heat_ker_eff}
     \mathcal{L}_{\eff}=\frac{c_s}{(4\pi)^{d/2}}\sum_{k=0}^\infty M^{d-2k}\frac{(-1)^k}{k!}\ \Gamma[k-d/2]\; \tr [b_k]. 
\end{equation}
In appendix \ref{App:UOLEA},  we provide the explicit form of the UOLEA in terms of the generalised operators $\P$ and $U$, mimicking the same given in Ref.~\cite{Banerjee:2023iiv}. Substituting the field-specific form of the generalised operators, $\P$ and $U$, one can obtain one-loop effective action operators up to dimension eight for both scalar and fermionic background fields. For the fermion case, we must perform additional ``traces (tr)" over Clifford matrices to bring the effective action in usable form. 

It is important to note that for $k\leq 2$ in Eq.~\eqref{eq:heat_ker_eff}, the effective Lagrangian is divergent due to the occurrence of simple poles in the gamma function at zero and negative arguments. Thus, one requires to regularise and renormalise the Lagrangian simultaneously. In the case of integrating out scalar fields, dimensional regularisation, and $\overline{MS}$ renormalisation scheme have been used, see Ref.~\cite{Banerjee:2023iiv}. In the case of fermionic fields, as mentioned earlier, one has to be careful about the additional contributions from the multiplicative anomaly, see Eq.~\eqref{eq:multiplicative_A}. 
While employing the HK method, we used dimensional regularisation which is equivalent to the zeta function regularisation \cite{Hawking:1976ja}. As we are aiming for higher mass dimensional effective operator computation, the loop contributions are finite and consistent with the renormalisation theorem. Hence, we do not require to invoke any analytical continuation into $d\neq4$ dimensions.  This ensures that the renormalised results will equally be valid for the fermionic case even in the presence of $\gamma^5$. 
\noindent
In the following subsections, we show how the scalar fermion cases can be brought on the same footing at the operator level such that we can propose the UOLEA for both of them. We further mention why the fermion case requires special attention.

\subsection{Generic Lagrangian for Heavy Scalar}
While integrating out heavy scalar we start with generic Lagrangian in the following form
\begin{equation}
    \L^\Phi = \Phi^\dagger (-P^2 + M_s^2 + U_s) \Phi,
\end{equation}
the generalised covariant derivative is $\P_\mu \equiv P_\mu=i D_\mu$ and the stress tensor is given as the commutator of a generalised covariant derivative is given by,
\[G_{\mu\nu}=[\P_\mu,\P_\nu]\equiv [P_\mu,P_\nu]=F_{\mu\nu}.\]
We do not require the specific form of the field-dependent  functional $U$ that contains information about the interaction between heavy scalar and other light fields, ie., infrared  (IR) ones. 
\subsection{Generic Lagrangian for Heavy Fermion}
The heavy fermion field Lagrangian of our interest is written as
\begin{equation}\label{eq:fermion_lag}
    \mathcal{L}^{\Psi}= \overline \Psi (\slashed P - M_f - \Sigma) \Psi,
\end{equation}
with $M_f$ being the mass of the heavy fermion $(\Psi)$ and $\Sigma$ capturing all its interaction with the light (IR) fields. Note that, our method of fermionic UOLEA (FUOLEA) computation is independent of the inner structure of $\Sigma$. In this paper, we emphasize the interaction between heavy fermion and light (pseudo)-scalars. From that  perspective, we assume the following form of interaction: \[\Sigma=S+iR\gamma^5,\] where, $S$ and $R$ are scalar and pseudo-scalar respectively. 

As we have discussed earlier, this form of action will not allow us to use the HK method. Thus, our first aim is to perform a consistent bosonization procedure to recast the first-order operator in terms of the strong elliptic second-order operator. The bosonized fermionic Lagrangian reads as
\begin{equation}\label{eq:dirac_lag}
\begin{split}
        \mathcal{L}^{\Psi}_{\eff} &= - i\, \tr \ln [\slashed P - M_f - S -i\,\gamma^5 R]\\
        &= \frac{-i}{2}\tr \Big\{\ln [\slashed P - M_f - S -i\,\gamma^5 R] + \ln [-\slashed P - M_f - S -i\,\gamma^5 R]\Big\}\\
        &= \frac{-i}{2}\tr\, \ln\Big[-P^2 - \frac{1}{2}\sigma_{\mu\nu}F_{\mu\nu} +2i \gamma^5 R \slashed P +M_f^2 +S^2 -R^2 -(\slashed P S) \\
        &\hspace{35mm} +2M(S+i\gamma^5 R)+i\gamma^5(RS+SR) +i \gamma^5 (\slashed P R)\Big],
\end{split}
\end{equation}
where,
\[F_{\mu\nu}=[P_\mu,P_\nu],\ \sigma_{\mu\nu} = \frac{1}{2}[\gamma_\mu,\gamma_\nu],\ (P_\mu S) = [P_\mu,S],\]
and we have ignored the contribution from the multiplicative anomaly. Note that in the above equation, Eq.~\eqref{eq:dirac_lag},  the presence of a first-order derivative operator $(\slashed P)$ refrains us from directly identifying it with the unified one-loop effective action, see Eq.~\eqref{eq:unified_lag}. This issue can be resolved by a redefinition of the covariant derivative 
\begin{equation}\label{eq:derv_redef}
    \tilde P_\mu = P_\mu -i\gamma^5\gamma_\mu R.
\end{equation}
Now we can write the fermionic one-loop effective Lagrangian in terms of a Laplace-type operator as
\begin{equation}\label{eq:fermion_EL}
    \mathcal{L}^{\Psi}_{\eff} = \frac{-i}{2}\tr\, \ln[-\tilde P^2 + M_f^2 + U_f],
\end{equation}
where,
\begin{gather}
    U_f = Y + 2M_f \Sigma,\notag\\
    Y= -\frac{1}{2}\sigma_{\mu\nu}F_{\mu\nu} + S^2+3R^2-(\slashed P S) +i\gamma^5(RS+SR). \label{eq:U_f_def}
\end{gather}
At this point, we can identify the operator in  Eq.~\eqref{eq:fermion_EL} with same depicted in Eq.~\eqref{eq:unified_lag}, with the generalised covariant derivative $\P \equiv  \tilde P_\mu = P_\mu -i\gamma^5\gamma_\mu R$, and the $U$ equivalent  term is given by $U_f$ in Eq.~\eqref{eq:U_f_def}. The commutator of the generalised covariant derivative $G_{\mu\nu}$ leads to
\begin{eqnarray}
    G_{\mu\nu} = [\P_\mu,\P_\nu] = [P_\mu -i\gamma^5\gamma_\mu R,P_\nu -i\gamma^5\gamma_\nu R] = F_{\mu\nu} + \Gamma_{\mu\nu},
\end{eqnarray}
where,
\begin{eqnarray}
    \Gamma_{\mu\nu}= i \gamma^5\gamma_\mu (P_\nu R) - i \gamma^5\gamma_\nu (P_\mu R) + 2\sigma_{\mu\nu}R^2.
\end{eqnarray}

We summarise how the generic Laplacian operator can be mapped for scalar and fermion cases in Table~\ref{tab:1-loop_operators}. 
\begingroup
\setlength{\tabcolsep}{10pt} 
\renewcommand{\arraystretch}{1.5} 
\begin{table}[h]
    \centering
    \begin{tabular}{|c |c| c|}
        \hline
        & Scalar & Fermion \\
        \hline
        $c_s$ & 1 or 1/2 & -1/2\\
        \hline
        $\P_\mu$ & $P_\mu$ & $P_\mu -i\gamma^5\gamma_\mu R$\\
        \hline
        \multirow{2}{1em}{$U$} & \multirow{2}{1em}{$U_s$} & $U_f=Y + 2M\Sigma$,\\
        & & $Y= -\frac{1}{2}\sigma_{\mu\nu}G_{\mu\nu} + S^2+3R^2-(\slashed P S) +i\gamma^5(RS+SR),\ \Sigma=S+i\gamma^5 R$\\
        \hline
        \multirow{2}{2em}{$G_{\mu\nu}$} & \multirow{2}{1em}{$F_{\mu\nu}$} & $F_{\mu\nu} + \Gamma_{\mu\nu}$,\\
        & & $\Gamma_{\mu\nu}= i \gamma^5\gamma_\mu (P_\nu R) - i \gamma^5\gamma_\nu (P_\mu R) + 2\sigma_{\mu\nu}R^2$\\
        \hline
    \end{tabular}
    \caption{Generalisation of one-loop effective Lagrangian.}
    \label{tab:1-loop_operators}
\end{table}
\endgroup

Note that,  in the case of fermion, $U_f$ can be decomposed into two parts, see Eq.~\eqref{eq:U_f_def}. Among them,  $Y$  is a mass dimension two operators while $\Sigma$ is an operator with mass dimension one. Hence, when one  expands $\frac{1}{M^n}U_f^m$, it contributes to operators of different dimensions starting  from $\mathcal{O}(1/M^n)$.  Here, contributions to different mass dimensional operators from $Y$ and $\Sigma$ are non-universal due to the involvement of $M$ with $\Sigma$. For example, let us consider the operator $\frac{1}{M^4}U^4$ from the FUOLEA \footnote{In the following sections, we will use $U$ instead of $U_f$ equivalently.}. The contributions from $Y$ and $\Sigma$ can be collected in the following form
\begin{align}
   \frac{1}{M^4} \tr[U^4]=\tr\bigg[16 \,\Sigma ^4 + \frac{32}{M} \, Y\,\Sigma ^3 + \frac{8}{M^2}\, \{2 Y^2\,\Sigma ^2+  (Y\,\Sigma)^2\}+ \frac{8}{M^3} \, Y^3\,\Sigma + \frac{1}{M^4} \,Y^4\bigg].\label{eq:U_to_muti_op}
\end{align}
It is evident that, $\frac{1}{M^4} U^4$ operator in the FUOLEA contributes to dimension four ($\Sigma^4$), dimension five ($Y\,\Sigma ^3$), dimension six ($Y^2\,\Sigma ^2,\, (Y\,\Sigma)^2$), dimension seven ($Y^3\,\Sigma$), and dimension eight ($Y^4$) operators. 
\section{FUOLEA: UOLEA after Integrating Out Heavy Fermions} \label{sec:fermionic-uolea}
This section presents the fermionic one-loop effective Lagrangian operators in terms of the fermion Lagrangian operators. We present the explicit calculation for obtaining dimension five operators from the general one-loop effective Lagrangian given in Appendix \ref{App:UOLEA}. Then We provide the results for higher dimension operators up to dimension eight.

The initial Lagrangian Eq.~\eqref{eq:fermion_lag} is written in the Minkowski space with $(+---)$ metric signature. As the HK method is defined for the Euclidean metric signature, we use the following convention for the gamma matrices in Euclidean space ($\gamma^\mu_E$). 
\begin{align}
    \gamma^i_E&=\gamma^i, &\gamma^4_E&= i \gamma^0, &\gamma^5_E =-(\gamma^1 \gamma^2 &\gamma^3 \gamma^4)_E = \gamma^5,\notag\\
    \{\gamma_E^\mu,\gamma_E^\nu\}&=2 g^{\mu\nu}, &{\gamma^5_E}^\dagger&=\gamma_E^5, &(\gamma^5_E)^2&=1.
\end{align}
Here the gamma matrices without subscripts are in the Minkowski metric. After Wick rotation, the metric tensor is given by $g^{\mu\nu}=-\delta^{\mu\nu}$. We drop the subscript $E$ from the Euclidean gamma matrices in future sections.
\subsection{Dimension Five operators in FUOLEA}\label{sec:D5}

Dimension five operators are $\mathcal{O}(1/M)$ in the inverse mass expansion of the one-loop effective Lagrangian. Following the above-mentioned power counting procedure, $1/M^2$ operators of UOLEA containing at least one $U$ operator, $1/M^4$ operators containing at least three $U$ operators, and $1/M^6$ operators containing at least five $U$ operators contribute to the D5 fermionic one-loop effective Lagrangian operator. These include the following operators from the UOLEA mentioned in Appendix \ref{App:UOLEA},
\begin{align}\label{eq:D5_UOLEA}
    \L_\eff =\cfrac{c_s}{(4\pi)^{2}}\tr\,\bigg\{& \frac{1}{M^2} \frac{1}{6}  \,\Big[ -U^3 - \frac{1}{2} (\P_\mu U)^2-\frac{1}{2}U\,(G_{\mu\nu})^2 \Big] + \frac{1}{M^4} \frac{1}{24} \Big[U^4 - U^2 (\P^2 U)\Big]\nn\\ & + \frac{1}{M^6} \frac{1}{60}  \,\Big[ -U^5\Big]\bigg\}.
\end{align}
Here the trace is over all internal indices including the spinor indices. Below, we systematically show how the fermionic effective Lagrangian operators can be obtained from Eq.~\eqref{eq:D5_UOLEA}.
\subsubsection*{\underline{Contributions from $1/M^2$ terms in Eq.~\eqref{eq:D5_UOLEA}}}

Considering definitions of interaction function $U$ and covariant derivative $\P$ from Table \ref{tab:1-loop_operators}, terms proportional to $1/M^2$ in Eq.~\eqref{eq:D5_UOLEA} can be expanded as,
\begin{align}\label{eq:M^2_D5}
    \L^{\Psi}_\eff \llbracket 1/M^2 \rrbracket = \cfrac{c_s}{(4\pi)^{2}}\tr\,\frac{1}{M^2} \frac{1}{6}  \,\Big[& -(Y+2M\Sigma)^3 - \frac{1}{2} [P_\mu-i\gamma^5\gamma_\mu R,(Y+2M\Sigma)]^2\nn\\
    &-\frac{1}{2}(Y+2M\Sigma)(F_{\mu\nu}+\Gamma_{\mu\nu})^2 \Big].
\end{align}
Here we have used the fact that the closed derivatives in Eq.~\eqref{eq:D5_UOLEA} are commutators, i.e. $(\P U) = [\P,U]$. Since we are considering $1/M^2$ terms of the UOLEA, only the terms of $\mathcal O(M)$ contribute to $D5$ and hence we would neglect terms of other orders in mass. When expanding the operators in Eq.~\eqref{eq:D5_UOLEA}, we encounter terms such as $Y^2$ and $(\Gamma_{\mu\nu})^2$. Hence we give their expansion below.
\begin{align}
        Y^2 =& \frac{1}{4} (\sigma_{\mu\nu}F_{\mu\nu})^2 - \frac{1}{2} \sigma_{\mu\nu}F_{\mu\nu}S^2 - \frac{3}{2} \sigma_{\mu\nu}F_{\mu\nu}R^2 -\frac{i}{2}\sigma_{\mu\nu}\gamma^5 F_{\mu\nu} (RS+SR) - \frac{1}{2} \sigma_{\mu\nu}S^2 F_{\mu\nu}\nn\\
        & +\frac{1}{2} \sigma_{\mu\nu} F_{\mu\nu}(\slashed P S) +\frac{1}{2} (\slashed P S)\sigma_{\mu\nu}F_{\mu\nu} - \frac{3}{2} \sigma_{\mu\nu}R^2 F_{\mu\nu} - \frac{i}{2}\gamma^5 \sigma_{\mu\nu}(RS+SR) F_{\mu\nu} +S^4\nn\\
        & + 9R^4  +3S^2R^2 - RSRS - SRSR - RS^2R - SR^2S +(\slashed P S)^2 + 3R^2S^2 + i\gamma^5 RS^3\nn\\
        & + i\gamma^5 S^2RS + i\gamma^5 S^3R + 3i\gamma^5 R^3S + 3i\gamma^5 R^2SR  + 3i\gamma^5 RSR^2 + i\gamma^5 SRS^2 + 3i\gamma^5 SR^3 \nn\\
        &  -i \gamma^5 (RS+SR) (\slashed P S) -i (\slashed P S)\gamma^5 (RS+RS) - (\slashed P S)S^2 -3 (\slashed P S)R^2 - S^2 (\slashed P S) \nn\\
        & - 3R^2(\slashed P S).
\end{align}
\begin{equation}
    \begin{split}
        (\Gamma_{\mu\nu})^2 = 8(P_\nu R)^2 - 2 (\slashed P R)^2 + 4i\gamma^5\gamma_\mu\sigma_{\mu\nu} (P_\nu R) R^2 + 4i\gamma^5\sigma_{\mu\nu}\gamma_\mu R^2 (P_\nu R) - 48 R^4.
    \end{split}
\end{equation}

The trace over the spinor indices has been performed when providing the final expression. Therefore, we first provide results of some frequently occurring spinor traces of a few terms that would be helpful in the calculation of the $D5$ operators.
\begin{align}
    &\text{tr}^s\, Y =4S^2+12R^3 , \quad \text{tr}^s\, \gamma^5 Y = 4i(RS+SR), \quad\text{tr}^s\, \gamma_\mu Y = -4(P_\mu S), \quad\text{tr}^s\, \gamma^5\gamma_\mu Y = 0,\nn\\
    &\text{tr}^s\, \Gamma_{\mu\nu} = 0, \quad \text{tr}^s\, \gamma^5\Gamma_{\mu\nu} = 0, \quad\text{tr}^s\, \gamma^5(\Gamma_{\mu\nu})^2 = 0, \quad \text{tr}^s\, (\Gamma_{\mu\nu})^2 = 24(P_\mu R)^2 - 192 P^4,\nn\\
    &\text{tr}^s\, Y^2 = 4\Big\{3S^2R^2 +S^4+3R^2S^2+9R^4-RSRS-SRSR-RS^2R-SR^2S+(P_\mu S)^2-\frac{1}{2} (F_{\mu\nu})^2 \Big\},\nn\\
    &\text{tr}^s\, \gamma^5 Y^2 = 4i\Big\{\frac{1}{2} \tilde F_{\mu\nu} F_{\mu\nu} +  S^2RS + S^3R + 3R^3S + 3R^2SR +  RS^3 + SRS^2 +3 RSR^2 +3SR^3\Big\}.\nn
\end{align}
Here we have used,
\begin{equation}
    \tilde F_{\mu\nu}= \frac{1}{2}\epsilon _{\mu\nu\rho\sigma} F_{\rho\sigma}.
\end{equation}
The trace over all internal gauge and field indices is represented by $\tr^i$, trace over spinor indices is given by $\tr^s$ and $\tr\equiv\tr^i \cdot \tr^s$ represents trace over all internal indices. Since $\tr^i$ and $\tr^s$ act on different space, internal index, and Clifford space respectively, they can be performed in any order. Now getting back to Eq.~\eqref{eq:M^2_D5}, we expand and simplify the individual terms.
\begin{align}
       \tr\ (Y+2M\Sigma)^3 \llbracket M \rrbracket = &\, \tr \big[Y^2(2M\Sigma) + Y(2M\Sigma)Y + (2M\Sigma)Y^2\big] \nn \\
       =&\, \tr \big[6M Y^2\Sigma\big] = \tr \big[6M\, Y^2(S+i\gamma^5 R)\big] \nn\\
       =&\, \tr^i \Big[24M\Big\{-\frac{1}{2}S(F_{\mu\nu})^2+\frac{1}{2}R\tilde F_{\mu\nu} F_{\mu\nu} +S^5 - 3SR^4 \nn \\ &\,
        + 3S^3R^2 -5 S^2RSR + S(P_\mu S)^2\Big\}\Big].
\end{align}
\begin{align}
   \tr\ [P_\mu-i\gamma^5\gamma_\mu R,(Y+2M\Sigma)]^2 \llbracket M \rrbracket &=\tr\big[2 [P_\mu-i\gamma^5\gamma_\mu R,Y][P_\mu-i\gamma^5\gamma_\mu R,2M\Sigma]\big]\nn\\
    &=  \tr \big[4M\big\{(P_\mu Y)(P_\mu \Sigma)-i\gamma^5\gamma_\mu RY (P_\mu \Sigma)+iYR\gamma^5\gamma_\mu(P_\mu \Sigma)\nn\\
    &\quad-i(P_\mu Y)\gamma^5\gamma_\mu R\Sigma + i(P_\mu Y) \Sigma R \gamma^5\gamma_\mu - \gamma^5\gamma_\mu RY\gamma^5\gamma_\mu R\Sigma\nn\\
    &\quad -4 R^2Y \Sigma  -4 YR^2\Sigma -YR \gamma^5\gamma_\mu \Sigma R\gamma^5\gamma_\mu\big\}\big] \nn\\
    &= \tr^i \big[16M\big\{-2S(P_\mu R)^2+2S(P_\mu S)^2 +8S^2RSR + 32SR^4\nn\\
    &\quad -8S^3R^2 - R(P_\mu R)(P_\mu S) - R(P_\mu S)(P_\mu R) \big\}\big].
\end{align}
\begin{align}
   \tr\ (Y+2M\Sigma)(F_{\mu\nu}+\Gamma_{\mu\nu})^2 \llbracket M \rrbracket &= \tr \big[2M\Sigma(F_{\mu\nu}+\Gamma_{\mu\nu})^2\big]\nn\\
    &= \tr \big[2M\Sigma\big\{(F_{\mu\nu})^2+(\Gamma_{\mu\nu})^2+ F_{\mu\nu}\Gamma_{\mu\nu}+\Gamma_{\mu\nu}F_{\mu\nu}\big\}\big]\nn\\
    &=\tr^i \big[8M\big\{S(F_{\mu\nu})^2+6S(P_{\mu}R)^2-48SR^4\big\}\big].
\end{align}

Here we use double brackets $\llbracket M \rrbracket$ to represent terms of $\mathcal O (M)$. In the above calculation, we have performed trace over spinor indices and have used trace properties over other internal indices to shuffle the structures in order to simplify.
\subsubsection*{\underline{Contributions from $1/M^4$ terms in Eq.~\eqref{eq:D5_UOLEA}}}
Terms proportional to $1/M^4$ in Eq.~\eqref{eq:D5_UOLEA} are expanded as,
\begin{align}\label{eq:M^4_D5}
    \L^{\Psi}_\eff \llbracket 1/M^4 \rrbracket = & \cfrac{c_s}{(4\pi)^{2}}\tr\,\frac{1}{M^4} \frac{1}{24}  \,\Big[ (Y+2M\Sigma)^4 - (Y+2M\Sigma)^2 \nn \\ & [P_\mu-i\gamma^5\gamma_\mu R,[P_\mu-i\gamma^5\gamma_\mu R,(Y+2M\Sigma)]]\Big].
\end{align}
Since we are considering $1/M^4$ terms of the UOLEA, only the terms of $\mathcal O(M^3)$ contribute to $D5$ and hence we would neglect terms of other orders in mass. Expanding the individual terms in Eq.~\eqref{eq:M^4_D5} and performing trace over the spinor indices ($\tr^s$),
\begin{align}
   \tr\ (Y+2M\Sigma)^4 \llbracket M^3 \rrbracket &= \tr \big[4 (2M)^3 Y \Sigma^3\big]\nn\\
    &= \tr \big[32M^3 Y \{S^3-SR^2-RSR-R^S+i\gamma^5(S^2R+SRS+RS^2-R^3)\}\big]\nn\\
    & =\tr^i \big[ 128 M^3\{S^5-S^3R^2-5S^2RSR-7SR^4\}\big],\nn
\end{align}
\begin{align}
   \tr\ (Y+2M\Sigma)^2 &[P_\mu-i\gamma^5\gamma_\mu R,[P_\mu-i\gamma^5\gamma_\mu R,(Y+2M\Sigma)]] \llbracket M^3 \rrbracket\nn\\
    &= \tr \big\{(2M)^3\Sigma^2 [P_\mu-i\gamma^5\gamma_\mu R,[P_\mu-i\gamma^5\gamma_\mu R,\Sigma]]\big\}\nn\\
    &=\tr \big\{(2M)^3\Sigma^2[P_\mu-i\gamma^5\gamma_\mu R,(P_\mu\Sigma)-i\gamma^5\gamma_\mu R\Sigma+i\Sigma R\gamma^5\gamma_\mu]\big\} \nn\\
    &=\tr \big[ (2M)^3\Sigma^2\{(P^2\Sigma)-i\gamma^5\gamma_\mu (P_\mu R)\Sigma-i\gamma^5\gamma_\mu R(P_\mu\Sigma)\nn\\
    &\quad +i(P_\mu\Sigma) R\gamma^5\gamma_\mu+i\Sigma(P_\mu R)\gamma^5\gamma_\mu + 4R^2\Sigma+ 2\gamma^5\gamma_\mu R \Sigma R\gamma^5\gamma_\mu \nn\\
    &\quad-i\gamma^5\gamma_\mu R(P_\mu\Sigma)+i(P_\mu\Sigma)R\gamma^5\gamma_\mu +4\Sigma R^2\}\big]\nn\\
    &= \tr^i \big[32M^3\big\{-32SR^4+8S^3R^2-8S^2RSR-2S(P_\mu S)^2+2(P_\mu R)^2\nn\\
    &\hspace{20mm}+2R(P_\mu S)(P_\mu R)+2R(P_\mu R)(P_\mu S)\big\}\big].
\end{align}
\subsubsection*{\underline{Contributions from $1/M^6$ terms in Eq.~\eqref{eq:D5_UOLEA}}}
The term proportional to $1/M^6$ in Eq.~\eqref{eq:D5_UOLEA} is expanded as,
\begin{align}\label{eq:M^5_D5}
    \L^{\Psi}_\eff \llbracket 1/M^6 \rrbracket = -\cfrac{c_s}{(4\pi)^{2}}\tr\,\frac{1}{M^6} \frac{1}{60}  \,\Big[ U^5\Big].
\end{align}
Here, only terms of $\mathcal O(M^5)$ contribute to $D5$ operators. Hence we consider only the $2M\Sigma$ term in $U$. Expanding and simplifying spinor traces, we get,
\begin{align}
    U^5 \llbracket M^5 \rrbracket= \tr \big[(2M)^5\Sigma^5\big]=\tr^i \big[4(2M)^5\big\{S^5+5SR^4-5S^2RSR-5S^3R^2\big\}\big].\nn
\end{align}

Combining all the contributions, fermionic one-loop effective Lagrangian can be expressed in terms of operators of dimension five as
\begin{align}
    \L_\eff^{\Psi{(D5)}}= \frac{c_s}{(4\pi)^2}\tr^i\, \frac{1}{M} \Big\{&-2RF_{\mu\nu}\tilde F_{\mu\nu} + \frac{4}{3}S (F_{\mu\nu})^2-\frac{4}{5}S^5-4S(P_\mu S)^2-4S(P_\mu R)^2-4SR^4\nn\\
    &+4S^2RSR-\frac{20}{3}S^3R^2-\frac{4}{3}R(P_\mu S)(P_\mu R)-\frac{4}{3}R(P_\mu R)(P_\mu S)\Big\}.
\end{align}

\subsection{Dimension Six operators in FUOLEA}\label{sec:D6}

Following the power counting procedure mentioned in Sec.~\ref{sec:effective action}, terms that contribute to $D6$ operators from the UOLEA given in Appendix \ref{App:UOLEA} are,
\begin{align}
    \L_\eff=\frac{c_s}{(4\pi)^2}& \tr\bigg\{ \frac{1}{M^2} \frac{1}{6}  \,\bigg[ -U^3 - \frac{1}{2} (\P_\mu U)^2-\frac{1}{2}U\,(G_{\mu\nu})^2  - \frac{1}{10}(J_\nu)^2   + \frac{1}{15}\,G_{\mu\nu}\,G_{\nu\rho}\,G_{\rho\mu} \bigg]\nn\\
    &+ \frac{1}{M^4} \frac{1}{24} \bigg[U^4 - U^2 (\P^2 U) + \frac{4}{5}U^2 (G_{\mu\nu})^2 + \frac{1}{5} (U\,G_{\mu\nu})^2  -\frac{2}{5} U\, (\P_\mu U)\,J_{\mu} \nn\\ 
    & + \frac{1}{5} (\P^2 U)^2\bigg] + \frac{1}{M^6} \frac{1}{60}  \,\bigg[ -U^5 + 2\,U^3 (\P^2 U) + U^2(\P_\mu U)^2\bigg]+ \frac{1}{M^8} \frac{1}{120} \,\bigg[U^6\bigg]\bigg\}.
\end{align}
Expanding the generalised operators and collecting the $\mathcal O(1/M^2)$ terms, we get,
\begin{align}
    \L_\eff^{\Psi{(D6)}}=\frac{c_s}{(4\pi)^2} \tr\frac{1}{M^2}\bigg\{&-\frac{1}{6}Y^3-\frac{4}{3}Y\,\Sigma^4-\frac{1}{12}Y\, (F_{\mu\nu})^2 -\frac{1}{12}Y\, (\Gamma_{\mu\nu})^2+\frac{2}{3}Y^2\,\Sigma^2\nn\\
    & +\frac{2}{15}\Sigma^2\, (\Gamma_{\mu\nu})^2+\frac{4}{15}\Sigma^2\,[P_\mu-i\gamma^5\gamma_\mu R,\Sigma]^2+\frac{2}{15}\Sigma^2\, (F_{\mu\nu})^2\nn\\
    & -\frac{1}{6}\Sigma^2\,[P_\mu-i\gamma^5\gamma_\mu R,[P_\mu-i\gamma^5\gamma_\mu R,Y]]+\frac{1}{3} (Y\,\Sigma)^2  \nn\\
    & +\frac{8}{15}\Sigma^3\,[P_\mu-i\gamma^5\gamma_\mu R,[P_\mu-i\gamma^5\gamma_\mu R,\Sigma]]+\frac{1}{30} (\Sigma\, F_{\mu\nu})^2\nn\\
    & -\frac{1}{60}[P_\mu-i\gamma^5\gamma_\mu R,F_{\mu\nu}][P_\rho-i\gamma^5\gamma_\rho R,F_{\rho\nu}]+ \frac{1}{15} \Sigma\, F_{\mu\nu}\Sigma\, \Gamma_{\mu\nu}\nn\\
    & -\frac{1}{60}[P_\mu-i\gamma^5\gamma_\mu R,F_{\mu\nu}][P_\rho-i\gamma^5\gamma_\rho R,\Gamma_{\rho\nu}] +\frac{1}{30} (\Sigma\, \Gamma_{\mu\nu})^2\nn\\
    & -\frac{1}{30}[P_\mu-i\gamma^5\gamma_\mu R,\Gamma_{\mu\nu}][P_\rho-i\gamma^5\gamma_\rho R,\Gamma_{\rho\nu}]+\frac{1}{90}F_{\mu\nu}F_{\nu\rho}F_{\rho\mu}\nn\\
    & -\frac{1}{6}Y\Sigma\,[P_\mu-i\gamma^5\gamma_\mu R,[P_\mu-i\gamma^5\gamma_\mu R,\Sigma]]-\frac{1}{12}Y\,F_{\mu\nu}\,\Gamma_{\mu\nu}\nn\\
    & -\frac{1}{6}\Sigma \,Y\,[P_\mu-i\gamma^5\gamma_\mu R,[P_\mu-i\gamma^5\gamma_\mu R,\Sigma]]-\frac{1}{12}Y\,\Gamma_{\mu\nu}\,F_{\mu\nu}\nn\\
    & -\frac{1}{15}\Sigma\, [P_\mu-i\gamma^5\gamma_\mu R,\Sigma][P_\nu-i\gamma^5\gamma_\nu R,\Gamma_{\nu\mu}]+\frac{2}{15}\Sigma^2 F_{\mu\nu}\Gamma_{\mu\nu}\nn\\
    & -\frac{1}{15}\Sigma\, [P_\mu-i\gamma^5\gamma_\mu R,\Sigma][P_\nu-i\gamma^5\gamma_\nu R,F_{\nu\mu}]+\frac{2}{15}\Sigma^2 \Gamma_{\mu\nu}F_{\mu\nu}\nn\\
    & -\frac{1}{12} [P_\mu-i\gamma^5\gamma_\mu R,Y]^2 + \frac{1}{30} [P_\mu-i\gamma^5\gamma_\mu R,[P_\mu-i\gamma^5\gamma_\mu R,\Sigma]^2\nn\\
    & +\frac{2}{15}\Sigma^2\, F_{\mu\nu}\,\Gamma_{\mu\nu} +\frac{2}{15}\Sigma^2\, \Gamma_{\mu\nu}\,F_{\mu\nu}+\frac{1}{90}F_{\mu\nu}\,F_{\nu\rho}\,F_{\rho\mu}\nn\\
    & +\frac{1}{30} F_{\mu\nu}\,F_{\nu\rho}\,\Gamma_{\rho\mu}+\frac{1}{30} F_{\mu\nu}\,\Gamma_{\nu\rho}\,\Gamma_{\rho\mu}+\frac{1}{90} \Gamma_{\mu\nu}\,\Gamma_{\nu\rho}\,\Gamma_{\rho\mu}\bigg\}.
\end{align}
Following the steps explained in the previous section, the dimension six fermionic one-loop effective operators are given by,
\begin{align}
    \L_\eff^{\Psi{(D6)}}= \frac{c_s}{(4\pi)^2}&\tr^i\,\frac{1}{M^2}\Big\{\frac{2}{15}S^6 + \frac{4}{3}S^4R^2 + \frac{4}{3}S^3RSR - 2(S^2R)^2-\frac{4}{3} S^2R^4 + 4 R^3SRS \nn\\
    & -\frac{2}{3}(R^2S)^2-\frac{2}{3}R^6 + \frac{4}{5} S^2(P_\mu S)^2 - \frac{4}{3} SR (P_\mu R)(P_\mu S) - \frac{4}{3} R^2(P_\mu S)^2\nn \\
    & - \frac{4}{3} RS(P_\mu S)(P_\mu R)+\frac{8}{3} SR(P_\mu S)(P_\mu R) + \frac{8}{3} RS (P_\mu R)(P_\mu S) \nn\\
    & - \frac{8}{3} R^2(P_\mu R)^2 + \frac{6}{5}(S(P_\mu S))^2 + 2(S(P_\mu R))^2 +\frac{2}{3}(R(P_\mu S))^2 \nn\\
    & -\frac{2}{3}(R(P_\mu R))^2 - \frac{1}{5} (P^2S)^2 -\frac{1}{3}(P^2R)^2 - \frac{7}{15}S^2 (F_{\mu\nu})^2 - \frac{1}{3}R^2 (F_{\mu\nu})^2 \nn\\
    & - \frac{1}{5} (S F_{\mu\nu})^2 + \frac{1}{3} (R F_{\mu\nu})^2 + \frac{16}{15} S (P_\mu S) (P_\nu F_{\nu\mu}) + \frac{4}{3} R (P_\mu R) (P_\nu F_{\nu\mu}) \nn\\
    & + \frac{2}{3}(SR+RS)\tilde F_{\mu\nu} F_{\mu\nu} + \frac{2}{3} S\tilde F_{\mu\nu} R F_{\mu\nu} + \frac{4}{15} (P_\nu F_{\nu\mu})^2 + \frac{2}{45} F_{\mu\nu}F_{\nu\rho}F_{\rho\mu}
    \Big\}.
\end{align}

The above-given dimension six operators can be classified into different classes in the Green's basis \cite{Chala:2021cgt}
\begin{eqnarray}
    \Phi^6,\; \Phi^4 D^2,\; \Phi^2 D^4,\; F \Phi^2 D^2,\; F^2 \Phi^2,\; F^2 D^2,\; F^3,
\end{eqnarray}
where, $[S, R] \in \Phi$, $D$ represents the covariant derivative $P_\mu$, and $F$ represents the field tensor $F_{\mu \nu}$.

\subsection{Dimension Seven operators in FUOLEA}\label{sec:D7}
 
Following the power counting procedure mentioned in Sec.~\ref{sec:effective action}, the dimension seven fermionic one-loop effective Lagrangian operators get contribution from the following terms of the UOLEA in Appendix \ref{App:UOLEA}.
\begin{align}
    \L_\eff = & \frac{c_s}{(4\pi)^{2}}\,\tr\bigg\{\frac{1}{M^4} \frac{1}{24} \bigg[U^4 - U^2 (\P^2 U) + \frac{4}{5}U^2 (G_{\mu\nu})^2 + \frac{1}{5} (U\,G_{\mu\nu})^2 - \frac{2}{5} U\, (\P_\mu U)\,J_{\mu}  \nn\\ 
    & \hspace{2.5cm} + \frac{1}{5} (\P^2 U)^2 + \frac{2}{5} U(J_\mu)^2 - \frac{2}{15} (\P^2 U) (G_{\rho\sigma})^2 - \frac{4}{15} U\,G_{\mu\nu}G_{\nu\rho} G_{\rho\mu} \nn\\ 
    & \hspace{2.5cm}  - \frac{8}{15} (\P_\mu \P_\nu U)\, G_{\rho\mu} G_{\rho\nu} \bigg]  \nn\\
    & + \frac{1}{M^6} \frac{1}{60}  \,\bigg[ -U^5 + 2\,U^3 (\P^2 U) + U^2(\P_\mu U)^2 - \frac{2}{3} U^2 G_{\mu\nu} U\,G_{\mu\nu}  - U^3 (G_{\mu\nu})^2  \nn \\
    & \hspace{2cm} + \frac{1}{3} U^2 (\P_\mu U)J_\mu - \frac{1}{3} U\,(\P_\mu U)(\P_\nu U)\,G_{\mu\nu}   - \frac{1}{3} U^2 J_\mu (\P_\mu U)  \nn\\
    & \hspace{2cm} -  \frac{1}{3} U\,G_{\mu\nu}(\P_\mu U)(\P_\nu U) - U\,(\P^2 U)^2  -  \frac{2}{3} (\P^2 U) (\P_\nu U)^2 \bigg]  \nn\\
    & + \frac{1}{M^8} \frac{1}{120}  \,\bigg[U^6 - 3\,U^4 (\P^2 U) - 2\,U^3(\P_\nu U)^2 \bigg] + \frac{1}{M^{10}} \frac{1}{210}  \,\bigg[-U^7 \bigg]\bigg\}.
 \end{align}%
\noindent
Expanding the functional $U$ and collecting the terms of $\mathcal O(1/M^3)$, the one-loop effective Lagrangian operators in terms of $\Sigma$, $Y$ and the generalised covariant derivative ($\P$) are given in Appendix \ref{App:D7}. These are more generalised forms of the effective Lagrangian operators where $\P$ contains any gauge fields and additional operators that arise due to the bosonization of the Dirac operator, $\Sigma$ contains any mass dimension one interaction and $Y$ is the corresponding mass dimension two operators produced during the bosonization. Due to a large number of operators in dimension seven, we present the effective Lagrangian in terms of Green's operator classes \cite{Chala:2021cgt}: 
\begin{eqnarray}
    \Phi^7,\; \Phi^5 D^2,\; \Phi^3 D^4,\; F \Phi^3 D^2,\; F^2 \Phi^3,\; F^2 \Phi D^2,\; F^3 \Phi.
\end{eqnarray}
Operators of a particular class are represented by the double brackets $\L \llbracket class \rrbracket$. The individual classes are further classified based on the CP-conserving nature of the operators and the presence of Levi-Civita tensor $(\varepsilon_{\alpha\beta\mu\nu})$. CP-violating operators are represented by a superscript $CPV$ and the CP-conserving operators of the same class are represented by the superscript $CPC$. Other classes with operators having Levi-Civita tensor are placed in two categories $I$ and $II$ where $II$ contains operators with the Levi-Civita tensor. Hence the dimension seven one-loop effective Lagrangian operators are given by,
\begin{align}
    \L_\eff^{\Phi(D7)} =& \frac{c_s}{(4\pi)^{2}}\,\tr\bigg[\L\llbracket \Phi^7 \rrbracket+\L \llbracket \Phi^5 D^2 \rrbracket+\L \llbracket \Phi^3 D^4 \rrbracket+\L_{I}\llbracket F \Phi^3 D^2 \rrbracket +\L_{II}\llbracket F \Phi^3 D^2 \rrbracket \nn \\ &
    + \L^{CPC}\llbracket F^2 \Phi^3 \rrbracket  
     +\L^{CPV}\llbracket F^2 \Phi^3 \rrbracket + \L_{I}\llbracket  F^2 \Phi D^2 \rrbracket + \L_{II}\llbracket  F^2 \Phi D^2\rrbracket \nn \\ & + \L^{CPC}\llbracket F^3\Phi\rrbracket+ \L^{CPV}\llbracket F^3\Phi\rrbracket \bigg].
\end{align}
For the sake of clarity in the text, we use the following notation hereafter.
\begin{align*}
    S_{\mu_1\mu_2...\mu_n} &\equiv [P_{\mu_n}...,[P_{\mu_2},[P_{\mu_1},S]]],\\
    R_{\mu_1\mu_2...\mu_n} &\equiv [P_{\mu_n}...,[P_{\mu_2},[P_{\mu_1},R]]],\\
    F_{\alpha\beta\mu_1...\mu_n} &\equiv [P_{\mu_n}...,[P_{\mu_1},F_{\alpha\beta}]].
\end{align*}
Dimension seven operators of different classes are given below.
\begin{align}
    \L \llbracket \Phi^7 \rrbracket = &\,\tr^i \bigg[  \frac{1}{M^3}\bigg\{ \frac{4}{3} \,SR^6-\frac{4}{3} \,SRSRSR^2-\frac{4}{3} \,S^2R^3SR+\frac{4}{3} \,S^2R^2SR^2-\frac{4}{3} \,S^2RSR^3\nn \\ 
    & +\frac{4}{3} \,S^3R^4 +\frac{4}{3} \,S^3RS^2R-\frac{4}{3} \,S^4RSR-\frac{4}{15} \,S^5R^2-\frac{4}{105} \,S^7\bigg\}\bigg].
\end{align}
\begin{align}
    \L \llbracket \Phi^5 D^2 \rrbracket = &\,\tr^i \bigg[\frac{1}{M^3}\bigg\{-\frac{4}{3} \,S^2R_{\mu }S_{\mu }R + \frac{16}{15} \,S^2S_{\mu }R_{\mu }R + \frac{2}{3} \,SR_{\mu }RR_{\mu }R - 2 \,SR_{\mu }SS_{\mu }R \nn \\
    & + \frac{4}{3} \,SR_{\mu }R_{\mu }R^2 - \frac{2}{3} \,SR_{\mu }S_{\mu }SR - \frac{2}{3} \,SS_{\mu }RS_{\mu }R + \frac{6}{5} \,SS_{\mu }SR_{\mu }R - \frac{2}{3} \,SS_{\mu }R_{\mu }SR  \nn \\
    & + \frac{16}{15} \,SS_{\mu }S_{\mu }R^2 + \frac{2}{3} \,R_{\mu }RS_{\mu }R^2 + \frac{2}{3} \,R_{\mu }SRR_{\mu }R + \frac{2}{5} \,R_{\mu }S^2S_{\mu }R + 2 \,R_{\mu }SR_{\mu }R^2  \nn \\
    & - 2 \,R_{\mu }SR_{\mu }S^2 + \frac{6}{5} \,R_{\mu }SS_{\mu }SR + \frac{2}{3} \,R_{\mu }R_{\mu }RSR + \frac{4}{3} \,R_{\mu }R_{\mu }SR^2 + \frac{2}{3} \,R_{\mu }R_{\mu }S^3 \nn \\
    & + \frac{2}{3} \,R_{\mu }S_{\mu }R^3 + \frac{16}{15} \,R_{\mu }S_{\mu }S^2R + \frac{2}{3} \,S_{\mu }RR_{\mu }R^2 - \frac{2}{3} \,S_{\mu }SRS_{\mu }R + \frac{2}{5} \,S_{\mu }S^2R_{\mu }R  \nn \\
    & - 2 \,S_{\mu }SR_{\mu }SR + \frac{6}{5} \,S_{\mu }SS_{\mu }R^2 + \frac{2}{3} \,S_{\mu }R_{\mu }R^3 - \frac{4}{3} \,S_{\mu }R_{\mu }S^2R - \frac{2}{3} \,S_{\mu }S_{\mu }RSR  \nn \\
    & + \frac{16}{15} \,S_{\mu }S_{\mu }SR^2 - \frac{6}{5} \,S_{\mu }SS_{\mu }S^2 - \frac{2}{15} \,S_{\mu }S_{\mu }S^3\bigg\}\bigg].
\end{align}
\begin{align}
    \L \llbracket \Phi^3 D^4 \rrbracket = &\,\tr^i \bigg[ \frac{1}{M^3}\bigg\{\frac{2}{5} \,R_{\mu }S_{\mu \nu \nu }R - \frac{2}{45} \,S_{\mu }R_{\nu \nu }R_{\mu } + \frac{8}{15} \,R_{\mu \mu }R_{\nu \nu }S + \frac{8}{15} \,R_{\mu \mu }S_{\nu \nu }R \nn \\
    & - \frac{2}{15} \,R_{\mu \nu }S_{\nu }R_{\mu } + \frac{4}{45} \,R_{\mu \mu }S_{\nu }R_{\nu } + \frac{2}{15} \,R_{\mu \nu }R_{\mu \nu }S + \frac{2}{5} \,S_{\mu\nu \nu }R_{\mu }R - \frac{1}{3} \,S_{\mu }S_{\mu\nu \nu }S \nn \\
    & + \frac{32}{45} \,S_{\mu \mu }R_{\nu }R_{\nu } - \frac{16}{45} \,S_{\mu \nu }R_{\nu }R_{\mu } - \frac{19}{45} \,S_{\mu \mu }S_{\nu }S_{\nu } + \frac{2}{5} \,S_{\mu \mu }R_{\nu \nu }R + \frac{2}{15} \,S_{\mu\nu }R_{\nu \mu }R  \nn \\
    & - \frac{4}{15} \,S_{\mu \mu}S_{\nu \nu }S - \frac{1}{3} \,S_{\mu \nu \nu }S_{\mu }S\bigg\} \bigg].
\end{align} 
\begin{align}
    \L_{I} \llbracket F \Phi^3 D^2 \rrbracket = &\, \tr^i \bigg[ \frac{1}{M^3}\bigg\{ \frac{56}{45} \,R_{\mu }R_{\nu }F_{\mu \nu }S - \frac{32}{45} \,R_{\mu }S_{\nu }F_{\mu \nu }R - \frac{4}{5} \,R_{\mu }F_{\mu \nu }R_{\nu }S + \frac{4}{15} \,R_{\mu }F_{\mu \nu }S_{\nu }R \nn \\
    & + \frac{26}{45} \,S_{\mu }R_{\nu }F_{\mu \nu }R + \frac{2}{15} \,S_{\mu }F_{\mu \nu }R_{\nu }R - \frac{56}{45} \,S_{\mu }F_{\mu \nu }S_{\nu }S - \frac{32}{45} \,F_{\mu \nu }S_{\mu }R_{\nu }R  \nn \\
    & + \frac{8}{9} \,F_{\mu \nu }R_{\mu }R_{\nu }S + \frac{4}{9} \,F_{\mu \nu }R_{\mu }S_{\nu }R + \frac{10}{9} \,S_{\mu }S_{\nu }F_{\mu \nu }S + \frac{10}{9} \,F_{\mu \nu }S_{\mu }S_{\nu }S\bigg\} \bigg].
\end{align}

\begin{align}
    \L_{II} \llbracket F \Phi^3 D^2 \rrbracket = &\, \tr^i \bigg[ \frac{1}{M^3}\bigg\{\frac{2}{3} \,R_{\mu }\tilde F_{\beta \mu }S_{\beta }S + \frac{2}{3} \,S_{\mu }R_{\nu } \tilde F_{\mu \nu }S - \frac{2}{3} \,S_{\mu }S_{\nu }\tilde F_{\mu \nu }R + \frac{2}{3}\,S_{\mu }\tilde F_{\beta \mu }R_{\beta }S \nn \\
    & + \frac{2}{3} \,\tilde F_{\alpha \beta }R_{\alpha }S_{\beta }S - \frac{2}{3} \,\tilde F_{\alpha \beta }S_{\alpha }S_{\beta }R - \frac{4}{3} \,R_{\mu }R_{\nu }\tilde F_{\mu \nu }R - \frac{4}{3} \,R_{\mu }F_{\beta \mu }R_{\beta }R \nn \\
    & - \frac{4}{3} \,\tilde F_{\alpha \beta }R_{\alpha }R_{\beta }R\bigg\} \bigg].
\end{align}
\begin{align}
    \L^{CPC}\llbracket F^2 \Phi^3 \rrbracket = &\, \tr^i \bigg[ \frac{1}{M^3}\bigg\{ \frac{4}{15} \,SF_{\mu \nu }RF_{\mu \nu }R - \frac{2}{3} \,SF_{\mu \nu }F_{\mu \nu }R^2 + \frac{4}{15} \,F_{\mu \nu }SRF_{\mu \nu }R - \frac{4}{15} \,F_{\mu \nu }SF_{\mu \nu }R^2 \nn \\
    & - \frac{4}{15} \,F_{\mu \nu }F_{\mu \nu }RSR - \frac{2}{3} \,F_{\mu \nu }F_{\mu \nu }SR^2 + \frac{2}{5} \,F_{\mu \nu }SF_{\mu \nu }S^2 + \frac{2}{45} \,F_{\mu \nu }F_{\mu \nu }S^3\bigg\} \bigg].
\end{align}

\begin{align}
    \L^{CPV}\llbracket F^2 \Phi^3 \rrbracket = &\,\tr^i \bigg[\frac{1}{M^3}\bigg\{-\frac{1}{3} \,S^2\,\tilde F_{\mu \nu }F_{\mu \nu }R - \frac{2}{3} \,S \tilde F_{\mu \nu }SF_{\mu \nu }R - \frac{2}{3}\,S\tilde F_{\mu \nu }F_{\mu \nu }SR + \frac{2}{3} \,\tilde F_{\mu \nu }S^2F_{\mu \nu }R \nn \\
    & - \frac{2}{3}\,\tilde F_{\mu \nu }SF_{\mu \nu }SR - \frac{1}{3} \,\tilde F_{\mu \nu }F_{\mu \nu }S^2R - \frac{2}{3} \,\tilde F_{\mu \nu }RF_{\mu \nu }R^2 + \frac{4}{3} \,\tilde F_{\mu \nu }F_{\mu \nu }R^3\bigg\} \bigg].
\end{align}
\begin{align}
    \L_{I}\llbracket F^2 \Phi D^2 \rrbracket = &\,\tr^i \bigg[ \frac{1}{M^3}\bigg\{\frac{1}{6} \,F_{\mu \nu \alpha \alpha }F_{\mu \nu }S + \frac{2}{15} \,F_{\mu \nu \mu }F_{\alpha \nu \alpha }S + \frac{1}{6} \,F_{\mu \nu }F_{\mu \nu \alpha \alpha }S + \frac{8}{45} \,F_{\mu \nu }F_{\alpha \mu }S_{\alpha \nu } \nn \\
    & + \frac{11}{90} \,F_{\mu \nu }F_{\mu \nu }S_{\alpha \alpha } \bigg\}\bigg].
\end{align}

\begin{align}
    \L_{II}\llbracket F^2 \Phi D^2 \rrbracket = &\,\tr^i \bigg[ \frac{1}{M^3}\bigg\{ -\frac{1}{6}\,\tilde F_{\mu \nu \alpha \alpha }F_{\mu \nu }R - \frac{1}{6} \,\tilde F_{\mu \nu }F_{\mu \nu \kappa \kappa }R - \frac{1}{6} \,\tilde F_{\mu \nu }F_{\mu \nu }R_{\kappa \kappa }\bigg\}\bigg].
\end{align}
\begin{align}
    \L^{CPC} \llbracket F^3 \Phi \rrbracket = &\,\tr^i \bigg[ \frac{1}{M^3}\bigg\{ - \frac{64}{45} \,F_{\mu \nu }F_{\nu\alpha}F_{\alpha \mu }S\bigg\}\bigg].
\end{align}

\begin{align}
    \L^{CPV}\llbracket F^3 \Phi \rrbracket = &\,\tr^i \bigg[ \frac{1}{M^3}\bigg\{ \frac{2}{3} \,\tilde F_{\mu \nu }F_{\nu \kappa }F_{\kappa \mu }R + \frac{1}{6} \,F_{\mu \nu }\tilde F_{\nu \kappa }F_{\kappa \mu }R + \frac{1}{2} \,F_{\mu \nu }F_{\nu \kappa }\tilde F_{\kappa \mu }R \bigg\}\bigg].
\end{align}
\subsection{Dimension Eight operators in FUOLEA}\label{sec:D8}
Following the power counting procedure mentioned in Sec.~\ref{sec:effective action}, the dimension eight fermionic one-loop effective Lagrangian operators get contribution from the following terms of the UOLEA in Appendix \ref{App:UOLEA}.
\begin{align}
    \L_\eff = & \frac{c_s}{(4\pi)^{2}}\,\tr\bigg\{\frac{1}{M^4} \frac{1}{24} \bigg[U^4 - U^2 (\P^2 U) + \frac{4}{5}U^2 (G_{\mu\nu})^2 + \frac{1}{5} (U\,G_{\mu\nu})^2 +  \frac{1}{5} (\P^2 U)^2  \nn\\ 
    & \hspace{2cm} -\frac{2}{5} U\, (\P_\mu U)\,J_{\mu} + \frac{2}{5} U(J_\mu)^2 - \frac{2}{15} (\P^2 U) (G_{\rho\sigma})^2 +\frac{1}{35}(\P_\nu J_{\mu})^2  \nn\\ 
    & \hspace{2cm} - \frac{4}{15} U\,G_{\mu\nu}G_{\nu\rho} G_{\rho\mu} - \frac{8}{15} (\P_\mu \P_\nu U)\, G_{\rho\mu} G_{\rho\nu} + \frac{16}{105}G_{\mu\nu}J_{\mu}J_{\nu}   \nn\\
    & \hspace{2cm} + \frac{1}{420} (G_{\mu\nu}G_{\rho\sigma})^2 +\frac{17}{210}(G_{\mu\nu})^2(G_{\rho\sigma})^2 +\frac{2}{35}(G_{\mu\nu}G_{\nu\rho})^2 \nn\\
    & \hspace{2cm} + \frac{1}{105} G_{\mu\nu}G_{\nu\rho}G_{\rho\sigma}G_{\sigma\mu} +\frac{16}{105} (\P_\mu J_{\nu}) G_{\nu\sigma}G_{\sigma\mu} \bigg]  \nn\\
    & + \frac{1}{M^6} \frac{1}{60}  \,\bigg[ -U^5 + 2\,U^3 (\P^2 U) + U^2(\P_\mu U)^2 - \frac{2}{3} U^2 G_{\mu\nu} U\,G_{\mu\nu}  - U^3 (G_{\mu\nu})^2  \nn \\
    & \hspace{2cm} + \frac{1}{3} U^2 (\P_\mu U)J_\mu - \frac{1}{3} U\,(\P_\mu U)(\P_\nu U)\,G_{\mu\nu}   - \frac{1}{3} U^2 J_\mu (\P_\mu U)  \nn\\
    & \hspace{2cm} -  \frac{1}{3} U\,G_{\mu\nu}(\P_\mu U)(\P_\nu U) - U\,(\P^2 U)^2  -  \frac{2}{3} (\P^2 U) (\P_\nu U)^2  - \frac{1}{7} ((\P_\mu U)G_{\mu\alpha})^2 \nn\\
    & \hspace{2cm} +\frac{2}{7} U^2 G_{\mu\nu}G_{\nu\alpha}G_{\alpha\mu}+\frac{8}{21}U\,G_{\mu\nu}U\,G_{\nu\alpha}G_{\alpha\mu}-\frac{4}{7}U^2(J_\mu)^2 -\frac{3}{7} (U\,J_\mu)^2 \nn \\
    & \hspace{2cm} +\frac{4}{7}U\,(\P^2U)(G_{\mu\nu})^2 +\frac{4}{7}(\P^2U)U(G_{\mu\nu})^2 -\frac{2}{7}U\,(\P_\mu U)J_\nu G_{\mu\nu} \nn \\
    & \hspace{2cm} -\frac{2}{7}(\P_\mu U)U\,G_{\mu\nu} J_\nu -\frac{4}{7}U\,(\P_\mu U)G_{\mu\nu} J_\nu -\frac{4}{7}(\P_\mu U)U\, J_\nu G_{\mu\nu} \nn \\
    & \hspace{2cm} +\frac{4}{21}U\,G_{\mu\nu}(\P^2U)G_{\mu\nu}  +\frac{11}{21}(\P_\alpha U)^2(G_{\mu\nu})^2 - \frac{10}{21}(\P_\mu U)J_\nu U\, G_{\mu\nu}  \nn \\
    & \hspace{2cm} - \frac{10}{21}(\P_\mu U) G_{\mu\nu} U \,J_\nu - \frac{2}{21} (\P_\mu U)(\P_\nu U)G_{\mu\alpha}G_{\alpha\nu} + \frac{10}{21} (\P_\nu U)(\P_\mu U)G_{\mu\alpha}G_{\alpha\nu}  \nn \\
    & \hspace{2cm}-\frac{1}{7} (G_{\alpha\mu}(\P_\mu U))^2 - \frac{1}{42} ((\P_\alpha U)G_{\mu\nu})^2 -\frac{1}{14} (\P_\mu \P^2 U)^2 -\frac{4}{21} (\P^2U) (\P_\mu U)J_\mu \nn \\
    & \hspace{2cm} +\frac{4}{21} (\P_\mu U)(\P^2U)J_\mu +\frac{2}{21} (\P_\mu U) (\P_\nu U)(\P_\mu J_{\nu}) - \frac{2}{21} (\P_\nu U) (\P_\mu U)(\P_\mu J_{\nu}) \bigg]  \nn\\
    & + \frac{1}{M^8} \frac{1}{120}  \,\bigg[U^6 - 3\,U^4 (\P^2 U) - 2\,U^3(\P_\nu U)^2 + \frac{12}{7}U^2 (\P_\mu \P_\nu U)(\P_\nu \P_\mu U)  \nn\\
    & \hspace{.5cm}  +\frac{26}{7} (\P_\mu \P_\nu U) U\,(\P_\mu U)(\P_\nu U) +\frac{26}{7} (\P_\mu \P_\nu U) (\P_\mu U)(\P_\nu U)U  + \frac{9}{7} (\P_\mu U)^2(\P_\nu U)^2  \nn\\
    & \hspace{2cm} + \frac{9}{7} U\,(\P_\mu \P_\nu U)U\,(\P_\nu \P_\mu U)  + \frac{17}{14} ((\P_\mu U)(\P_\nu U))^2 + \frac{8}{7} U^3G_{\mu\nu}U\,G_{\mu\nu}  \nn\\
    & \hspace{2cm} + \frac{5}{7} U^4(G_{\mu\nu})^2 + \frac{18}{7} G_{\mu\nu}(\P_\mu U)U^2(\P_\nu U) + \frac{9}{14} (U^2G_{\mu\nu})^2   \nn\\
    & \hspace{2cm} + \frac{18}{7} G_{\mu\nu}U\,(\P_\mu U)(\P_\nu U)U + \frac{18}{7} (\P_\mu \P_\nu U) (\P_\mu U)U\,(\P_\nu U)   \nn\\
    & \hspace{2cm} +  \bigg( \frac{8}{7} G_{\mu\nu}U\,(\P_\mu U)U\,(\P_\nu U) +  \frac{26}{7} G_{\mu\nu}(\P_\mu U)U\,(\P_\nu U)U \bigg) \nn\\
    & \hspace{2cm} +  \bigg( \frac{24}{7} G_{\mu\nu}(\P_\mu U)(\P_\nu U)U^2 - \frac{2}{7} G_{\mu\nu}U^2(\P_\mu U)(\P_\nu U)\bigg)\bigg] \nn\\
    & + \frac{1}{M^{10}} \frac{1}{210}  \,\bigg[-U^7 - 5\, U^4 (\P_\nu U)^2 - 8\,U^3(\P_\mu U)U(\P_\mu U)  -\frac{9}{2} (U^2 (\P_\mu U))^2 \bigg] \nn\\
    & + \frac{1}{M^{12}} \frac{1}{336} \,\bigg[U^8\bigg] \bigg\}.
\end{align}

Expanding the functional $U$ and collecting the terms of $\mathcal O(1/M^4)$, the one-loop effective Lagrangian operators in terms of $\Sigma$, $Y$ and the generalised covariant derivative ($\P$) are given in Appendix \ref{App:D8}. 

We  will present the effective Lagrangian in terms of the dimension eight Green's basis operator classes \cite{Chala:2021cgt}
\begin{eqnarray}
\Phi^8,\; \Phi^6 D^2,\; \Phi^4 D^4,\;\Phi^2D^6,\;  F\Phi^4D^2,\; F\Phi^2 D^4,\; F^2\Phi^4,\; F^2\Phi^2D^2,\; F^2 D^4\; F^3\Phi^2,\;  F^3D^2,\;F^4 .  \nn
\end{eqnarray}
Hence the dimension eight one-loop effective Lagrangian operators are given by,
\begin{align}
    L_\eff^{\Psi{(D8)}} =& \frac{c_s}{(4\pi)^{2}}\bigg[\L \llbracket \Phi^8 \rrbracket+\L \llbracket \Phi^6 D^2 \rrbracket+\L_I \llbracket \Phi^4 D^4 \rrbracket +\L_{II} \llbracket \Phi^4 D^4 \rrbracket +\L \llbracket \Phi^2 D^6 \rrbracket+\L_I \llbracket F\Phi^4D^2 \rrbracket \nn\\
    & +\L_{II} \llbracket F\Phi^4D^2 \rrbracket + \L_I \llbracket F\Phi^2 D^4 \rrbracket +\L_{II} \llbracket F\Phi^2 D^4 \rrbracket +\L^{CPC} \llbracket F^2 \Phi^4 \rrbracket+\L^{CPV} \llbracket F^2 \Phi^4 \rrbracket \nn \\
    & +\L_I \llbracket F^2 \Phi^2 D^2 \rrbracket +\L_{II} \llbracket F^2 \Phi^2 D^2 \rrbracket +\L \llbracket F^2D^4 \rrbracket+\L^{CPC} \llbracket F^3\phi^2 \rrbracket +\L^{CPV} \llbracket F^3\phi^2 \rrbracket \nn \\
    & + \L \llbracket F^3D^2 \rrbracket +\L \llbracket F^4 \rrbracket \bigg].
\end{align}
Here we have followed the notations of $CPC$, $CPV$, $I$, and $II$ as described in the previous section. The dimension eight operators of different classes are given below.
\begin{align}
    \L \llbracket \Phi^8 \rrbracket  =& \tr^i \bigg[ \frac{1}{M^4} \bigg\{- \frac{2754}{35} \,(SR^3)^2 - \frac{2048}{105} \,SR^2SR^4 - \frac{6284}{105} \,SRSR^5 - \frac{339}{7} \,(SR)^4 \nn\\
    & - \frac{10240}{21} \,S^2R^6 - \frac{3072}{35} \,S^2R^2SRSR + \frac{6614}{105} \,S^2R^2S^2R^2 - \frac{8564}{105} \,S^2RSR^2SR  \nn\\
    & - \frac{3072}{35} \,S^2RSRSR^2 + \frac{3100}{21} \,S^2RS^2R^3 + \frac{512}{21} \,S^3R^3SR - \frac{5204}{105} \,S^3R^2SR^2  \nn\\
    & + \frac{512}{21} \,S^3RSR^3 - \frac{2}{3} \,S^3RS^3R + \frac{20396}{105} \,S^4R^4 - \frac{512}{21} \,S^4RS^2R + \frac{5204}{105} \,S^5RSR \nn \\
    & - \frac{512}{21} \,S^6R^2 + \frac{24611}{210} \,R^8 + \frac{1}{70} \,S^8  \bigg\} \bigg].
\end{align}
\begin{align}
     \L \llbracket \Phi^6 D^2 \rrbracket = & \tr^i \bigg[ \frac{1}{M^4} \bigg\{-\frac{121}{21} \,S^3R_{\mu }S_{\mu }R - \frac{703}{105} \,S^3S_{\mu }R_{\mu }R + \frac{2}{15} \,S^2S_{\mu }RS_{\mu }R - \frac{6}{5} \,S^2S_{\mu }SR_{\mu }R \nn \\
     & - \frac{206}{105} \,S^2R_{\mu }RR_{\mu }R + \frac{134}{35} \,S^2R_{\mu }SS_{\mu }R + \frac{121}{21} \,S^2R_{\mu }R_{\mu }R^2 - \frac{107}{21} \,S^2R_{\mu }S_{\mu }SR   \nn \\
     & - \frac{661}{105} \,S^2S_{\mu }R_{\mu }SR - \frac{703}{105} \,S^2S_{\mu }S_{\mu }R^2 - \frac{4}{3} \,SR_{\mu }R^2S_{\mu }R + \frac{2}{15} \,SR_{\mu }RSR_{\mu }R  \nn \\
     & - \frac{2}{15} \,SR_{\mu }RS_{\mu }R^2 - \frac{4}{3} \,SR_{\mu }SRR_{\mu }R + \frac{4}{5} \,SR_{\mu }S^2S_{\mu }R - \frac{82}{105} \,SR_{\mu }SR_{\mu }R^2  \nn \\
     & - \frac{10}{7} \,SR_{\mu }SS_{\mu }SR + \frac{107}{21} \,SR_{\mu }R_{\mu }RSR + \frac{159}{35} \,SR_{\mu }R_{\mu }SR^2 + \frac{221}{35} \,SR_{\mu }S_{\mu }R^3 \nn \\
     & - \frac{661}{105} \,SR_{\mu }S_{\mu }S^2R + \frac{4}{15} \,SS_{\mu }R^2R_{\mu }R + \frac{2}{3} \,SS_{\mu }RSS_{\mu }R - \frac{206}{105} \,SS_{\mu }RR_{\mu }R^2  \nn \\
     & - \frac{4}{15} \,SS_{\mu }SRS_{\mu }R - \frac{4}{5} \,SS_{\mu }S^2R_{\mu }R - \frac{10}{7} \,SS_{\mu }SR_{\mu }SR - \frac{6}{5} \,SS_{\mu }SS_{\mu }R^2  \nn \\
     & + \frac{661}{105} \,SS_{\mu }R_{\mu }R^3 - \frac{107}{21} \,SS_{\mu }R_{\mu }S^2R - \frac{661}{105} \,SS_{\mu }S_{\mu }RSR - \frac{703}{105} \,SS_{\mu }S_{\mu }SR^2  \nn \\
     & + \frac{326}{105} \,R_{\mu }R^2R_{\mu }R^2 + \frac{1222}{105} \,R_{\mu }RR_{\mu }R^3 + \frac{122}{105} \,R_{\mu }RS_{\mu }RSR - \frac{2}{15} \,R_{\mu }SR^2S_{\mu }R  \nn \\
     & + \frac{314}{105} \,R_{\mu }SRSR_{\mu }R + \frac{2}{15} \,R_{\mu }SRR_{\mu }SR - \frac{4}{3} \,R_{\mu }SRS_{\mu }R^2 - \frac{206}{105} \,R_{\mu }S^2RR_{\mu }R \nn \\
     & + \frac{598}{105} \,R_{\mu }SR_{\mu }RSR - \frac{82}{105} \,R_{\mu }SR_{\mu }SR^2 + \frac{2}{5} \,R_{\mu }SR_{\mu }S^3 - \frac{142}{35} \,R_{\mu }SS_{\mu }R^3  \nn \\
     & - \frac{6}{5} \,R_{\mu }SS_{\mu }S^2R + \frac{361}{105} \,R_{\mu }R_{\mu }R^4 + \frac{661}{105} \,R_{\mu }R_{\mu }RS^2R + \frac{107}{21} \,R_{\mu }R_{\mu }SRSR  \nn \\
     & + \frac{121}{21} \,R_{\mu }R_{\mu }S^2R^2 - \frac{703}{105} \,R_{\mu }R_{\mu }S^4 + \frac{27}{7} \,R_{\mu }S_{\mu }R^2SR + \frac{107}{21} \,R_{\mu }S_{\mu }RSR^2  \nn \\
     & + \frac{661}{105} \,R_{\mu }S_{\mu }SR^3 - \frac{703}{105} \,R_{\mu }S_{\mu }S^3R - \frac{242}{105} \,S_{\mu }R^2S_{\mu }R^2 + \frac{122}{105} \,S_{\mu }RR_{\mu }RSR \nn \\
     & - \frac{2}{3} \,S_{\mu }RS_{\mu }R^3 - \frac{206}{105} \,S_{\mu }SR^2R_{\mu }R + \frac{2}{3} \,S_{\mu }SRSS_{\mu }R + \frac{4}{15} \,S_{\mu }SRR_{\mu }R^2  \nn \\
     & + \frac{2}{3} \,S_{\mu }SRS_{\mu }SR + \frac{2}{15} \,S_{\mu }S^2RS_{\mu }R - \frac{2}{5} \,S_{\mu }S^3R_{\mu }R + \frac{4}{5} \,S_{\mu }S^2R_{\mu }SR  \nn \\
     & - \frac{4}{5} \,S_{\mu }S^2S_{\mu }R^2 + \frac{18}{35} \,S_{\mu }S^2S_{\mu }S^2 - \frac{142}{35} \,S_{\mu }SR_{\mu }R^3 + \frac{134}{35} \,S_{\mu }SR_{\mu }S^2R  \nn \\
     & + \frac{2}{5} \,S_{\mu }SS_{\mu }RSR - \frac{6}{5} \,S_{\mu }SS_{\mu }SR^2 + \frac{18}{35} \,S_{\mu }SS_{\mu }S^3 + \frac{107}{21} \,S_{\mu }R_{\mu }R^2SR  \nn \\
     & + \frac{27}{7} \,S_{\mu }R_{\mu }RSR^2 + \frac{221}{35} \,S_{\mu }R_{\mu }SR^3 - \frac{121}{21} \,S_{\mu }R_{\mu }S^3R + \frac{661}{105} \,S_{\mu }S_{\mu }R^4 \nn \\
     &- \frac{107}{21} \,S_{\mu }S_{\mu }RS^2R - \frac{661}{105} \,S_{\mu }S_{\mu }SRSR - \frac{703}{105} \,S_{\mu }S_{\mu }S^2R^2 + \frac{91}{15} \,S_{\mu }S_{\mu }S^4\nn \\
     & - \frac{2}{5} \,R_{\mu }S^3S_{\mu }R - \frac{4}{5} \,R_{\mu }S^2R_{\mu }R^2 + \frac{6}{5} \,R_{\mu }S^2R_{\mu }S^2 - \frac{4}{5} \,R_{\mu }S^2S_{\mu }SR \bigg\} \bigg].
\end{align}
\begin{align}
    \L_{I} \llbracket \Phi^4 D^4 \rrbracket = & \tr^i \bigg[ \frac{1}{M^4}\bigg\{\frac{29}{63} \,SR_{\mu \mu }S_{\nu \nu }R - \frac{271}{315} \,SR_{\mu \nu }S_{\mu \nu }R + \frac{22}{45} \,SS_{\mu \mu }R_{\nu \nu }R - \frac{16}{45} \,SS_{\mu \nu }R_{\mu \nu }R \nn \\
    & + \frac{83}{315} \,R_{\mu }R_{\mu }R_{\nu }R_{\nu } + \frac{89}{315} \,R_{\mu }R_{\mu }R_{\nu \nu }R + \frac{32}{315} \,R_{\mu }R_{\mu }S_{\nu \nu }S - \frac{313}{630} \,R_{\mu }R_{\nu }R_{\mu }R_{\nu }  \nn \\
    & - \frac{22}{63} \,R_{\mu }R_{\nu }R_{\mu \nu }R - \frac{59}{63} \,R_{\mu }R_{\nu }S_{\mu \nu }S + \frac{137}{315} \,R_{\mu }S_{\mu }R_{\nu \nu }S + \frac{22}{315} \,R_{\mu }S_{\mu }S_{\nu \nu }R  \nn \\
    & - \frac{111}{70} \,R_{\mu }S_{\nu }S_{\mu }R_{\nu } - \frac{724}{315} \,R_{\mu }S_{\nu }R_{\mu \nu }S - \frac{319}{210} \,R_{\mu }S_{\nu }S_{\mu \nu }R + \frac{52}{105} \,R_{\mu }R_{\mu \nu }R_{\nu }R \nn \\
    & - \frac{85}{63} \,R_{\mu }R_{\mu \nu }S_{\nu }S - \frac{2}{3} \,R_{\mu }R_{\nu \nu }R_{\mu }R + \frac{47}{105} \,R_{\mu }R_{\nu \nu }S_{\mu }S - \frac{334}{315} \,R_{\mu }S_{\mu \nu }R_{\nu }S  \nn \\
    & - \frac{451}{630} \,R_{\mu }S_{\mu \nu }S_{\nu }R - \frac{32}{63} \,R_{\mu }S_{\nu \nu }R_{\mu }S - \frac{8}{15} \,R_{\mu }S_{\nu \nu }S_{\mu }R - \frac{208}{315} \,S_{\mu }R_{\mu }S_{\nu }R_{\nu } \nn \\
    & + \frac{59}{105} \,S_{\mu }R_{\mu }R_{\nu \nu }S  + \frac{11}{63} \,S_{\mu }R_{\mu }S_{\nu \nu }R - \frac{58}{45} \,S_{\mu }R_{\nu }S_{\mu }R_{\nu } - \frac{208}{315} \,S_{\mu }R_{\nu }S_{\nu }R_{\mu } \nn \\
    & - \frac{608}{315} \,S_{\mu }R_{\nu }R_{\mu \nu }S - \frac{479}{315} \,S_{\mu }R_{\nu }S_{\mu \nu }R  - \frac{68}{45} \,S_{\mu }S_{\mu }R_{\nu }R_{\nu } + \frac{46}{63} \,S_{\mu }S_{\mu }S_{\nu }S_{\nu }  \nn \\
    & + \frac{37}{126} \,S_{\mu }S_{\mu }R_{\nu \nu }R - \frac{7}{18} \,S_{\mu }S_{\mu }S_{\nu \nu }S - \frac{727}{630} \,S_{\mu }S_{\nu }R_{\mu }R_{\nu } - \frac{289}{315} \,S_{\mu }S_{\nu }R_{\nu }R_{\mu } \nn \\
    & + \frac{101}{210} \,S_{\mu }S_{\nu }S_{\mu }S_{\nu } - \frac{68}{35} \,S_{\mu }S_{\nu }R_{\mu \nu }R + \frac{283}{210} \,S_{\mu }S_{\nu }S_{\mu \nu }S - \frac{404}{315} \,S_{\mu }R_{\mu \nu }R_{\nu }S  \nn \\
    & - \frac{173}{105} \,S_{\mu }R_{\mu \nu }S_{\nu }R + \frac{8}{21} \,S_{\mu }R_{\nu \nu }R_{\mu }S + \frac{32}{315} \,S_{\mu }R_{\nu \nu }S_{\mu }R - \frac{703}{630} \,S_{\mu }S_{\mu \nu }R_{\nu }R  \nn \\
    & + \frac{41}{35} \,S_{\mu }S_{\mu \nu }S_{\nu }S - \frac{1}{3} \,S_{\mu }S_{\nu \nu }R_{\mu }R + \frac{8}{45} \,S_{\mu }S_{\nu \nu }S_{\mu }S + \frac{1}{35} \,R_{\mu \mu }RR_{\nu \nu }R  \nn \\
    & + \frac{5}{21} \,R_{\mu \mu }SR_{\nu \nu }S + \frac{16}{21} \,R_{\mu \mu }SS_{\nu \nu }R - \frac{103}{315} \,R_{\mu \mu }R_{\nu }R_{\nu }R + \frac{59}{105} \,R_{\mu \mu }R_{\nu }S_{\nu }S  \nn \\
    & + \frac{158}{315} \,R_{\mu \mu }S_{\nu }R_{\nu }S + \frac{353}{630} \,R_{\mu \mu }S_{\nu }S_{\nu }R + \frac{11}{315} \,R_{\mu \mu }R_{\nu \nu }R^2 + \frac{59}{105} \,R_{\mu \mu }R_{\nu \nu }S^2 \nn \\
    & + \frac{5}{9} \,R_{\mu \mu }S_{\nu \nu }SR + \frac{4}{105} \,R_{\mu \nu }RR_{\mu \nu }R - \frac{88}{105} \,R_{\mu \nu }SR_{\mu \nu }S - \frac{94}{105} \,R_{\mu \nu }SS_{\mu \nu }R  \nn \\
    & + \frac{82}{315} \,R_{\mu \nu }R_{\mu }R_{\nu }R - \frac{608}{315} \,R_{\mu \nu }R_{\mu }S_{\nu }S - \frac{149}{63} \,R_{\mu \nu }S_{\mu }R_{\nu }S - \frac{211}{105} \,R_{\mu \nu }S_{\mu }S_{\nu }R  \nn \\
    & + \frac{23}{63} \,R_{\mu \nu }R_{\mu \nu }R^2 - \frac{101}{105} \,R_{\mu \nu }R_{\mu \nu }S^2 - \frac{19}{45} \,R_{\mu \nu }S_{\mu \nu }SR + \frac{5}{21} \,S_{\mu \mu }RS_{\nu \nu }R  \nn \\
    & + \frac{16}{21} \,S_{\mu \mu }SR_{\nu \nu }R - \frac{7}{15} \,S_{\mu \mu }SS_{\nu \nu }S + \frac{32}{315} \,S_{\mu \mu }R_{\nu }R_{\nu }S + \frac{11}{63} \,S_{\mu \mu }R_{\nu }S_{\nu }R  \nn \\
    & - \frac{41}{315} \,S_{\mu \mu }S_{\nu }R_{\nu }R - \frac{7}{18} \,S_{\mu \mu }S_{\nu }S_{\nu }S + \frac{124}{315} \,S_{\mu \mu }R_{\nu \nu }SR + \frac{79}{315} \,S_{\mu \mu }S_{\nu \nu }R^2 \nn \\
    & - \frac{11}{15} \,S_{\mu \mu }S_{\nu \nu }S^2 - \frac{32}{105} \,S_{\mu \nu }RS_{\mu \nu }R - \frac{94}{105} \,S_{\mu \nu }SR_{\mu \nu }R + \frac{16}{105} \,S_{\mu \nu }SS_{\mu \nu }S  \nn \\
    & - \frac{59}{63} \,S_{\mu \nu }R_{\mu }R_{\nu }S - \frac{479}{315} \,S_{\mu \nu }R_{\mu }S_{\nu }R - \frac{319}{210} \,S_{\mu \nu }S_{\mu }R_{\nu }R + \frac{283}{210} \,S_{\mu \nu }S_{\mu }S_{\nu }S  \nn \\
    & - \frac{50}{63} \,S_{\mu \nu }R_{\mu \nu }SR - \frac{37}{315} \,S_{\mu \nu }S_{\mu \nu }R^2 + \frac{47}{105} \,S_{\mu \nu }S_{\mu \nu }S^2\bigg\}\bigg].
\end{align}

\begin{align} \label{eq:phi4d4-2}
    \L_{II} \llbracket \Phi^4 D^4 \rrbracket = &  \tr^i \bigg[ \frac{1}{M^4} \bigg\{ \frac{7}{45} \varepsilon _{\alpha \beta \mu \nu } \,R_{\mu }R_{\nu }S_{\alpha }R_{\beta }  + \frac{4}{45} \varepsilon _{\alpha \beta \mu \nu } \,R_{\mu }R_{\nu }S_{\alpha \beta }R  - \frac{1}{5} \varepsilon _{\alpha \beta \mu \nu } \,R_{\mu }S_{\nu }R_{\alpha \beta }R  \nn\\
    & - \frac{1}{5} \varepsilon _{\alpha \beta \mu \nu } \,R_{\mu }R_{\nu \alpha }S_{\beta }R  + \frac{8}{45} \varepsilon _{\alpha \beta \mu \nu } \,R_{\mu }S_{\nu \alpha \beta }R^2  - \frac{7}{45} \varepsilon _{\alpha \beta \mu \nu } \,S_{\mu }R_{\nu }R_{\alpha }R_{\beta }  \nn\\
    & - \frac{1}{3} \varepsilon _{\alpha \beta \mu \nu } \,S_{\mu }S_{\nu }S_{\alpha \beta }R  + \frac{1}{5} \varepsilon _{\alpha \beta \mu \nu } \,S_{\mu }R_{\nu \alpha }R_{\beta }R
    + \frac{1}{5} \varepsilon _{\alpha \beta \mu \nu } \,R_{\mu \nu }S_{\alpha }R_{\beta }R  \nn\\
    & - \frac{4}{45} \varepsilon _{\alpha \beta \mu \nu } \,S_{\mu \nu }R_{\alpha }R_{\beta }R + \frac{1}{3} \varepsilon _{\alpha \beta \mu \nu } \,S_{\mu \nu }S_{\alpha }S_{\beta }R + \frac{8}{45} \varepsilon _{\alpha \beta \mu \nu } \,S_{\mu \nu \alpha }R_{\beta }R^2 \bigg\} \bigg].
\end{align}
\begin{align}
    \L \llbracket \Phi^2D^6 \rrbracket =&\, \tr^i \bigg[ \cfrac{1}{M^4} \bigg\{\frac{1}{35} \,R_{\mu \mu \nu }R_{\alpha \alpha \nu } + \frac{1}{210} \,R_{\mu \nu \alpha }R_{\mu \nu \alpha } -\frac{2}{105} \,S_{\mu \mu \nu }S_{\alpha \alpha \nu } + \frac{1}{30} \,S_{\mu \nu \nu }S_{\mu \alpha \alpha } \bigg\} \bigg].
\end{align}
\begin{align}
    \L_{I} \llbracket F\Phi^4D^2 \rrbracket = & \,  \tr^i \bigg[ \frac{1}{M^4} \bigg\{ \frac{59}{63} \,S^2F_{\mu \nu }R_{\mu \nu }R - \frac{473}{315} \,SR_{\mu }S_{\nu }F_{\mu \nu }R + \frac{4}{7} \,R_{\mu \nu }RF_{\mu \nu }R^2 + \frac{1}{5} \,R_{\mu }F_{\mu \nu }S_{\nu }SR \nn \\
    & + \frac{2}{15} \,SS_{\mu }F_{\mu \nu }R_{\nu }R - \frac{24}{35} \,SR_{\mu \nu }SF_{\mu \nu }R - \frac{48}{35} \,SS_{\mu \nu }RF_{\mu \nu }R - \frac{16}{315} \,SF_{\mu \nu }RS_{\mu \nu }R  \nn \\
    & + \frac{137}{315} \,SF_{\mu \nu }SR_{\mu \nu }R - \frac{97}{315} \,SF_{\mu \nu }R_{\mu }S_{\nu }R + \frac{23}{21} \,SF_{\mu \nu }S_{\mu }R_{\nu }R + \frac{59}{63} \,SF_{\mu \nu }R_{\mu \nu }SR  \nn \\
    & + \frac{68}{105} \,SF_{\mu \nu }S_{\mu \nu }R^2 + \frac{5}{9} \,R_{\mu }SS_{\nu }F_{\mu \nu }R + \frac{4}{315} \,R_{\mu }SS_{\nu }F_{\nu \mu }R + \frac{647}{630} \,R_{\mu }SF_{\mu \nu }S_{\nu }R  \nn \\
    & + \frac{317}{315} \,R_{\mu }R_{\nu }RF_{\mu \nu }R - \frac{142}{63} \,R_{\mu }R_{\nu }SF_{\mu \nu }S + \frac{68}{105} \,R_{\mu }R_{\nu }F_{\mu \nu }R^2 - \frac{136}{105} \,R_{\mu }R_{\nu }F_{\mu \nu }S^2  \nn \\
    & - \frac{73}{90} \,R_{\mu }S_{\nu }SF_{\mu \nu }R - \frac{47}{210} \,R_{\mu }S_{\nu }F_{\mu \nu }SR - \frac{7}{15} \,R_{\mu }F_{\mu \nu }RR_{\nu }R + \frac{338}{315} \,R_{\mu }F_{\mu \nu }SR_{\nu }S  \nn \\
    & + \frac{37}{126} \,R_{\mu }F_{\mu \nu }SS_{\nu }R - \frac{241}{315} \,R_{\mu }F_{\mu \nu }R_{\nu }R^2 + \frac{23}{21} \,R_{\mu }F_{\mu \nu }R_{\nu }S^2 - \frac{362}{315} \,SS_{\mu }R_{\nu }F_{\mu \nu }R \nn \\
    & - \frac{13}{105} \,S_{\mu }SR_{\nu }F_{\mu \nu }R + \frac{8}{105} \,S_{\mu }SR_{\nu }F_{\nu \mu }R + \frac{127}{105} \,S_{\mu }SF_{\mu \nu }R_{\nu }R - \frac{431}{315} \,S_{\mu }R_{\nu }SF_{\mu \nu }R  \nn \\
    & - \frac{37}{21} \,S_{\mu }R_{\nu }F_{\mu \nu }SR - \frac{317}{315} \,S_{\mu }S_{\nu }RF_{\mu \nu }R - \frac{19}{126} \,S_{\mu }S_{\nu }SF_{\mu \nu }S - \frac{64}{105} \,S_{\mu }S_{\nu }F_{\mu \nu }R^2  \nn \\
    & + \frac{277}{630} \,S_{\mu }S_{\nu }F_{\mu \nu }S^2 - \frac{44}{105} \,S_{\mu }F_{\mu \nu }RS_{\nu }R + \frac{1}{9} \,S_{\mu }F_{\mu \nu }SR_{\nu }R + \frac{229}{210} \,S_{\mu }F_{\mu \nu }SS_{\nu }S  \nn \\
    & + \frac{376}{315} \,S_{\mu }F_{\mu \nu }R_{\nu }SR - \frac{11}{105} \,S_{\mu }F_{\mu \nu }S_{\nu }R^2 - \frac{5}{21} \,S_{\mu }F_{\mu \nu }S_{\nu }S^2 + \frac{397}{315} \,SR_{\mu }F_{\mu \nu }S_{\nu }R \nn \\
    & - \frac{32}{35} \,R_{\mu \nu }S^2F_{\mu \nu }R - \frac{316}{315} \,S_{\mu \nu }SRF_{\mu \nu }R + \frac{24}{35} \,S_{\mu \nu }SF_{\mu \nu }S^2 + \frac{88}{315} \,F_{\mu \nu }RR_{\mu \nu }R^2  \nn \\
    & - \frac{20}{21} \,F_{\mu \nu }SRS_{\mu \nu }R - \frac{48}{35} \,F_{\mu \nu }S^2R_{\mu \nu }R - \frac{11}{21} \,F_{\mu \nu }SR_{\mu }S_{\nu }R - \frac{1}{5} \,F_{\mu \nu }SS_{\mu }R_{\nu }R  \nn \\
    & - \frac{79}{315} \,F_{\mu \nu }SR_{\mu \nu }SR + \frac{76}{105} \,F_{\mu \nu }SS_{\mu \nu }R^2 + \frac{8}{35} \,F_{\mu \nu }SS_{\mu \nu }S^2 + \frac{83}{105} \,F_{\mu \nu }R_{\mu }RR_{\nu }R  \nn \\
    & - \frac{103}{45} \,F_{\mu \nu }R_{\mu }SR_{\nu }S - \frac{157}{105} \,F_{\mu \nu }R_{\mu }SS_{\nu }R + \frac{314}{315} \,F_{\mu \nu }R_{\mu }R_{\nu }R^2 - \frac{113}{315} \,F_{\mu \nu }R_{\mu }R_{\nu }S^2  \nn \\
    & + \frac{5}{7} \,F_{\mu \nu }R_{\mu }S_{\nu }SR + \frac{46}{105} \,F_{\mu \nu }R_{\nu }RR_{\mu }R - \frac{7}{45} \,F_{\mu \nu }R_{\nu }SR_{\mu }S - \frac{4}{315} \,F_{\mu \nu }R_{\nu }SS_{\mu }R  \nn \\
    & - \frac{37}{35} \,F_{\mu \nu }S_{\mu }RS_{\nu }R - \frac{257}{315} \,F_{\mu \nu }S_{\mu }SR_{\nu }R + \frac{19}{105} \,F_{\mu \nu }S_{\mu }SS_{\nu }S + \frac{272}{315} \,F_{\mu \nu }S_{\mu }R_{\nu }SR  \nn \\
    & + \frac{8}{5} \,F_{\mu \nu }S_{\mu }S_{\nu }R^2 - \frac{286}{315} \,F_{\mu \nu }S_{\mu }S_{\nu }S^2 + \frac{11}{105} \,F_{\mu \nu }S_{\nu }RS_{\mu }R - \frac{8}{105} \,F_{\mu \nu }S_{\nu }SR_{\mu }R  \nn \\
    & + \frac{1}{10} \,F_{\mu \nu }S_{\nu }SS_{\mu }S - \frac{268}{315} \,F_{\mu \nu }R_{\mu \nu }R^3 + \frac{32}{35} \,F_{\mu \nu }R_{\mu \nu }S^2R + \frac{316}{315} \,F_{\mu \nu }S_{\mu \nu }RSR  \nn \\
    & + \frac{316}{315} \,F_{\mu \nu }S_{\mu \nu }SR^2 - \frac{32}{35} \,F_{\mu \nu }S_{\mu \nu }S^3 \bigg\} \bigg].
\end{align}

\begin{align}
    \L_{II} \llbracket F\Phi^4D^2 \rrbracket = & \, \tr^i \bigg[ \frac{1}{M^4}\bigg\{ SR_{\mu }R_{\nu }\tilde F_{\mu \nu }R + SR_{\mu }F_{\beta \mu }R_{\beta }R + S\tilde F_{\mu \nu }R_{\mu }R_{\nu }R + R_{\mu }SR_{\nu }\tilde F_{\mu \nu }R \nn \\
    &  + \frac{4}{5} \,SS_{\mu }S_{\nu }\tilde F_{\mu \nu }R + \frac{1}{15} \,SS_{\mu }\tilde F_{\beta \mu }S_{\beta }R + \frac{2}{15} \,S\tilde F_{\mu \nu }S_{\mu }S_{\nu }R + R_{\mu }S\tilde F_{\beta \mu }R_{\beta }R \nn \\
    & + R_{\mu }R_{\nu }S\tilde F_{\mu \nu }R  + R_{\mu }R_{\nu }\tilde F_{\mu \nu }SR \nn - \frac{2}{5} \,R_{\mu }S_{\nu }R\tilde F_{\mu \nu }R - \frac{4}{15} \,R_{\mu }S_{\nu }S\tilde F_{\mu \nu }S  \nn \\
    & - \frac{8}{15} \,R_{\mu }S_{\nu }\tilde F_{\mu \nu }R^2 + \frac{4}{15} \,R_{\mu }S_{\nu }\tilde F_{\mu \nu }S^2 + \frac{4}{15} \,R_{\mu }\tilde F_{\beta \mu }RS_{\beta }R + \,R_{\mu }\tilde F_{\beta \mu }SR_{\beta }R \nn \\
    & - \frac{4}{5} \,R_{\mu }\tilde F_{\beta \mu }SS_{\beta }S + R_{\mu }\tilde F_{\beta \mu }R_{\beta }SR - \frac{7}{15} \,R_{\mu }\tilde F_{\beta \mu }S_{\beta }S^2 + \frac{4}{15} \,\tilde F_{\mu \nu }S_{\mu }R_{\nu }S^2 \nn \\
    & + \frac{3}{5} \,S_{\mu }SS_{\nu }\tilde F_{\mu \nu }R - \frac{1}{15} \,S_{\mu }SF_{\beta \mu }S_{\beta }R - \frac{2}{5}  \,S_{\mu }R_{\nu }R\tilde F_{\mu \nu }R - \frac{4}{15} \,S_{\mu }R_{\nu }S\tilde F_{\mu \nu }S \nn \\
    & + \frac{19}{15} \,S_{\mu }R_{\nu }\tilde F_{\mu \nu }R^2 - \frac{7}{15} \,S_{\mu }R_{\nu }\tilde F_{\mu \nu }S^2 + \frac{7}{15} \,S_{\mu }S_{\nu }S\tilde F_{\mu \nu }R + \frac{2}{15} \,S_{\mu }S_{\nu }\tilde F_{\mu \nu }SR  \nn \\
    & + \frac{2}{5} \,S_{\mu }\tilde F_{\beta \mu }RR_{\beta }R - \frac{4}{15}\,S_{\mu }\tilde F_{\beta \mu }SR_{\beta }S - \frac{1}{15} \,S_{\mu }\tilde F_{\beta \mu }SS_{\beta }R + \frac{61}{45} \,S_{\mu }\tilde F_{\beta \mu }R_{\beta }R^2 \nn \\
    & - \frac{7}{15} \,S_{\mu }\tilde F_{\beta \mu }R_{\beta }S^2 + \frac{1}{15} \,S_{\mu }\tilde F_{\beta \mu }S_{\beta }SR + \tilde F_{\mu \nu }SR_{\mu }R_{\nu }R + \frac{7}{15} \,\tilde F_{\mu \nu }SS_{\mu }S_{\nu }R \nn \\
    & + \frac{4}{15} \,\tilde F_{\mu \nu }R_{\mu }RS_{\nu }R - \frac{4}{5} \,\tilde F_{\mu \nu }R_{\mu }SS_{\nu }S + \frac{3}{5} \,\tilde F_{\mu \nu }S_{\mu }SS_{\nu }R - \frac{8}{45} \,\tilde F_{\mu \nu }S_{\mu }R_{\nu }R^2  \nn \\
    & + \frac{19}{15}  \,\tilde F_{\mu \nu }R_{\mu }S_{\nu }R^2 - \frac{7}{15} \,\tilde F_{\mu \nu }R_{\mu }S_{\nu }S^2 + \frac{2}{5} \,\tilde F_{\mu \nu }S_{\mu }RR_{\nu }R - \frac{4}{15} \,\tilde F_{\mu \nu }S_{\mu }SR_{\nu }S \nn \\
    & + \tilde F_{\mu \nu }R_{\mu}R_{\mu }SR + \tilde F_{\mu \nu }R_{\mu }SR_{\nu }R + R_{\mu }\tilde F_{\beta \mu }S_{\beta }R^2 + \frac{4}{5} \,\tilde F_{\mu \nu }S_{\mu }S_{\nu }SR\bigg\}\bigg].
\end{align}
\begin{align}
    \L_{I} \llbracket F\Phi^2D^4 \rrbracket = & \, \tr^i \bigg[ \cfrac{1}{M^4} \bigg\{ \frac{1}{21} \,R_{\mu }F_{\alpha \nu \nu \mu }R_{\alpha } + \frac{1}{45} \,R_{\mu }F_{\mu \nu \alpha \alpha }R_{\nu } + \frac{4}{35} \,R_{\mu }F_{\mu \nu \nu \alpha }R_{\alpha } + \frac{1}{6} \,S_{\mu }F_{\mu \nu \alpha \alpha }S_{\nu }  \nn \\
    &  - \frac{8}{315} \,S_{\mu }F_{\mu \nu \nu \alpha }S_{\alpha } + \frac{32}{315} \,R_{\mu \mu }F_{\alpha \nu \nu }R_{\alpha }  \nn + \frac{8}{315} \,R_{\mu \mu }F_{\alpha \nu }R_{\alpha \nu } - \frac{4}{105} \,R_{\mu \nu }F_{\alpha \mu \alpha \nu }R  \nn \\
    & - \frac{8}{315} \,R_{\mu \nu }F_{\alpha \nu \alpha }R_{\mu } + \frac{8}{315} \,R_{\mu \nu }F_{\alpha \mu }R_{\alpha \nu } + \frac{1}{15} \,R_{\mu \nu }F_{\alpha \nu }R_{\mu \alpha } + \frac{8}{315} \,R_{\mu \nu }F_{\mu \nu }R_{\alpha \alpha }  \nn \\
    & - \frac{16}{315} \,S_{\mu \mu }F_{\alpha \nu \nu }S_{\alpha } - \frac{1}{15} \,S_{\mu \nu }F_{\alpha \nu \alpha }S_{\mu } - \frac{1}{105} \,R_{\mu \mu \nu }F_{\alpha \nu \alpha }R- \frac{8}{315} \,R_{\mu \mu \nu }F_{\alpha \nu }R_{\alpha } \nn \\
    & - \frac{1}{105} \,R_{\mu \nu \alpha }F_{\mu \nu \alpha }R + \frac{1}{63} \,R_{\mu \nu \alpha }F_{\mu \nu }R_{\alpha } + \frac{1}{105} \,R_{\mu \nu \nu }F_{\alpha \mu \alpha }R + \frac{8}{315} \,R_{\mu \nu \nu }F_{\alpha \mu }R_{\alpha } \nn \\
    & - \frac{1}{6} \,S_{\mu \nu \nu }F_{\alpha \mu }S_{\alpha }  + \frac{1}{45} \,F_{\mu \nu \alpha \alpha }R_{\nu }R_{\mu } + \frac{4}{45} \,F_{\mu \nu \alpha \mu }R_{\nu }R_{\alpha } - \frac{1}{35} \,F_{\mu \nu \mu \alpha }R_{\alpha }R_{\nu }  \nn \\
    & + \frac{16}{315} \,F_{\mu \nu \mu \alpha }R_{\nu }R_{\alpha }- \frac{8}{315} \,F_{\mu \nu \mu \alpha }S_{\nu }S_{\alpha }  + \frac{4}{105} \,F_{\mu \nu \mu \alpha }R_{\nu \alpha }R - \frac{1}{105} \,F_{\mu \nu \alpha }R_{\nu \mu \alpha }R   \nn \\
    & + \frac{32}{315} \,F_{\mu \nu \mu }R_{\alpha \alpha }R_{\nu } + \frac{8}{315} \,F_{\mu \nu \mu }R_{\alpha \nu }R_{\alpha } - \frac{16}{315} \,F_{\mu \nu \mu }S_{\alpha \alpha }S_{\nu } + \frac{1}{105} \,F_{\mu \nu \mu }R_{\alpha \alpha \nu }R  \nn \\
    & - \frac{1}{105} \,F_{\mu \nu \mu }R_{\nu \alpha \alpha }R + \frac{1}{15} \,F_{\mu \nu }R_{\alpha \alpha }R_{\nu \mu } - \frac{16}{315} \,F_{\mu \nu }R_{\alpha \alpha \nu }R_{\mu } - \frac{8}{315} \,F_{\mu \nu }R_{\alpha \mu \nu }R_{\alpha } \nn \\
    & - \frac{1}{105} \,F_{\mu \nu }R_{\nu \mu \alpha }R_{\alpha } + \frac{1}{6} \,F_{\mu \nu }S_{\nu \alpha \alpha }S_{\mu } \bigg\} \bigg].
\end{align}

\begin{align}
    \L_{II} \llbracket F\Phi^2D^4 \rrbracket = & \, \tr^i \bigg[ \cfrac{1}{M^4} \bigg\{ - \frac{1}{15} \,S_{\mu }\tilde F_{\kappa \mu \beta \beta }R_{\kappa } + \frac{2}{15} \,S_{\mu \nu }\tilde F_{\mu \nu \kappa \kappa }R  + \frac{1}{15} \,S_{\mu \nu \nu }\tilde F_{\kappa \mu }R_{\kappa } - \frac{1}{15} \,\tilde F_{\mu \nu \alpha \alpha }S_{\mu }R_{\nu } \nn \\
    & - \frac{2}{15} \,\tilde F_{\mu \nu \alpha \alpha }S_{\mu \nu }R - \frac{1}{15} \,\tilde F_{\mu \nu \alpha }R_{\alpha \mu }S_{\nu } + \frac{1}{15} \,\tilde F_{\mu \nu }S_{\mu \beta }R_{\beta \nu } - \frac{1}{15} \,\tilde F_{\mu \nu }S_{\mu \nu }R_{\kappa \kappa } \nn \\
    & + \frac{1}{15} \,\tilde F_{\mu \nu }S_{\mu \beta \beta }R_{\nu}
     \bigg\} \bigg].
\end{align}
\begin{align}
    \L^{CPC} \llbracket F^2\phi^4 \rrbracket = & \,\tr^i \bigg[ \frac{1}{M^4}\bigg\{ -\frac{2}{21} \,S^2F_{\mu \nu }RF_{\mu \nu }R + \frac{239}{630} \,S^2F_{\mu \nu }F_{\mu \nu }R^2 - \frac{37}{210} \,SF_{\mu \nu }RSF_{\mu \nu }R \nn \\
    & - \frac{3}{7} \,SF_{\mu \nu }SRF_{\mu \nu }R + \frac{67}{315} \,SF_{\mu \nu }SF_{\mu \nu }R^2 + \frac{33}{70} \,SF_{\mu \nu }F_{\mu \nu }RSR + \frac{223}{630} \,SF_{\mu \nu }F_{\mu \nu }SR^2  \nn \\
    & + \frac{6}{35} \,F_{\mu \nu }SRSF_{\mu \nu }R - \frac{37}{210} \,F_{\mu \nu }SRF_{\mu \nu }SR - \frac{2}{21} \,F_{\mu \nu }S^2RF_{\mu \nu }R + \frac{2}{63} \,F_{\mu \nu }S^2F_{\mu \nu }R^2 \nn \\
    & + \frac{14}{45} \,F_{\mu \nu }SF_{\mu \nu }RSR  + \frac{67}{315} \,F_{\mu \nu }SF_{\mu \nu }SR^2 - \frac{1}{42} \,F_{\mu \nu }F_{\mu \nu }RS^2R + \frac{33}{70} \,F_{\mu \nu }F_{\mu \nu }SRSR \nn \\
    & + \frac{239}{630} \,F_{\mu \nu }F_{\mu \nu }S^2R^2 - \frac{9}{14} \,F_{\mu \nu }F_{\mu \nu }R^4 + \frac{1}{42} \,F_{\mu \nu }R^2F_{\mu \nu }R^2 + \frac{346}{315} \,F_{\mu \nu }RF_{\mu \nu }R^3  \nn \\
    & - \frac{11}{70} \,F_{\mu \nu }S^2F_{\mu \nu }S^2 - \frac{4}{21} \,F_{\mu \nu }SF_{\mu \nu }S^3  + \frac{1}{70} \,F_{\mu \nu }F_{\mu \nu }S^4\bigg\}\bigg].
\end{align}

\begin{align}
    \L^{CPV} \llbracket F^2\phi^4 \rrbracket = & \tr^i \bigg[\frac{1}{M^4} \bigg\{ \frac{1}{30} \,S^3\tilde F_{\mu \nu }F_{\mu \nu }R + \frac{3}{5} \,S^2 \tilde F_{\mu \nu }SF_{\mu \nu }R + \frac{3}{10} \,S^2 \tilde F_{\mu \nu }F_{\mu \nu }SR- \frac{4}{45} \,S\tilde F_{\mu \nu }R^2F_{\mu \nu }R  \nn \\
    & + \frac{34}{45} \,S\tilde F_{\mu \nu }RF_{\mu \nu }R^2 + \frac{4}{5} \,S\tilde F_{\mu \nu }SF_{\mu \nu }SR  - \frac{83}{90} \,S\tilde F_{\mu \nu }F_{\mu \nu }R^3 + \frac{3}{10} \,S\tilde F_{\mu \nu }F_{\mu \nu }S^2R \nn \\
    & + \frac{2}{3} \,\tilde F_{\mu \nu }RF_{\mu \nu }RSR + \frac{34}{45} \,\tilde F_{\mu \nu }SR^2F_{\mu \nu }R - \frac{4}{45} \,\tilde F_{\mu \nu }SRF_{\mu \nu }R^2 - \frac{2}{3} \,\tilde F_{\mu \nu }S^3F_{\mu \nu }R  \nn \\
    & - \frac{22}{45} \,\tilde F_{\mu \nu }SF_{\mu \nu }R^3 + \frac{3}{5} \,\tilde F_{\mu \nu }SF_{\mu \nu }S^2R - \frac{5}{6} \,\tilde F_{\mu \nu }F_{\mu \nu }R^2SR - \frac{5}{6} \,\tilde F_{\mu \nu }F_{\mu \nu }RSR^2 \nn \\
    & - \frac{83}{90} \,\tilde F_{\mu \nu }F_{\mu \nu }SR^3 + \frac{1}{30} \,\tilde F_{\mu \nu }F_{\mu \nu }S^3R \bigg\} \bigg].
\end{align}
\begin{align}
    \L_{I} \llbracket F^2\phi^2D^2 \rrbracket = & \, \tr^i \bigg[ \cfrac{1}{M^4} \bigg\{ \frac{4}{15} \,F_{\mu \nu }F_{\alpha \mu }R_{\alpha \nu }R  - \frac{2}{15} \,F_{\mu \nu }F_{\mu \nu \alpha }S_{\alpha }S - \frac{1}{9} \,F_{\mu \nu }F_{\alpha \mu }R_{\nu }R_{\alpha } - \frac{1}{12} \,F_{\mu \nu }F_{\alpha \mu }S_{\nu }S_{\alpha }  \nn \\
    &+ \frac{17}{45} \,R_{\mu }F_{\mu \nu }F_{\alpha \nu \alpha }R + \frac{23}{126} \,R_{\mu }F_{\mu \nu }F_{\alpha \nu }R_{\alpha } + \frac{61}{630} \,R_{\mu }F_{\nu \alpha }F_{\mu \alpha \nu }R - \frac{11}{126} \,R_{\mu }F_{\nu \alpha }F_{\nu \alpha \mu }R \nn \\
    & - \frac{89}{630} \,R_{\mu }F_{\nu \alpha }F_{\nu \mu \alpha }R - \frac{107}{630} \,R_{\mu }F_{\nu \alpha }F_{\mu \alpha }R_{\nu } + \frac{17}{70} \,R_{\mu }F_{\nu \mu }F_{\alpha \nu }R_{\alpha } - \frac{8}{105} \,S_{\mu }F_{\nu \alpha \alpha }F_{\nu \mu }S  \nn \\
    & - \frac{2}{15} \,S_{\mu }F_{\nu \alpha \mu }F_{\nu \alpha }S - \frac{16}{105} \,S_{\mu }F_{\mu \nu }F_{\alpha \nu \alpha }S - \frac{179}{252} \,S_{\mu }F_{\mu \nu }F_{\alpha \nu }S_{\alpha } + \frac{29}{180} \,S_{\mu }F_{\nu \alpha }F_{\mu \alpha }S_{\nu }  \nn \\
    & + \frac{31}{180} \,S_{\mu }F_{\nu \mu }F_{\alpha \nu }S_{\alpha } - \frac{1}{420} \,R_{\mu \mu }F_{\nu \alpha }F_{\nu \alpha }R + \frac{1}{105} \,R_{\mu \nu }F_{\alpha \mu }F_{\alpha \nu }R - \frac{97}{315} \,R_{\mu \nu }F_{\alpha \nu }F_{\alpha \mu }R  \nn \\
    & - \frac{4}{45} \,S_{\mu \nu }F_{\alpha \nu }F_{\alpha \mu }S + \frac{61}{180} \,F_{\mu \nu \alpha \alpha }F_{\mu \nu }R^2 - \frac{11}{60} \,F_{\mu \nu \alpha \alpha }F_{\mu \nu }S^2 - \frac{5}{63} \,F_{\mu \nu \mu \alpha }F_{\alpha \nu }R^2  \nn \\
    & + \frac{4}{45} \,F_{\mu \nu \nu \alpha }F_{\alpha \mu }R^2 - \frac{2}{15} \,F_{\mu \nu \alpha }RF_{\alpha \mu \nu }R + \frac{2}{15} \,F_{\mu \nu \alpha }RF_{\alpha \nu \mu }R - \frac{13}{105} \,F_{\mu \nu \alpha }RF_{\mu \nu \alpha }R  \nn \\
    & + \frac{52}{315} \,F_{\mu \nu \alpha }R_{\alpha }F_{\mu \nu }R + \frac{1}{9} \,F_{\mu \nu \alpha }R_{\mu }F_{\alpha \nu }R - \frac{43}{315} \,F_{\mu \nu \alpha }R_{\nu }F_{\alpha \mu }R - \frac{2}{15} \,F_{\mu \nu \alpha }S_{\alpha }F_{\mu \nu }S  \nn \\
    & + \frac{2}{15} \,F_{\mu \nu \alpha }F_{\alpha \mu \nu }R^2 - \frac{2}{15} \,F_{\mu \nu \alpha }F_{\alpha \nu \mu }R^2 + \frac{9}{35} \,F_{\mu \nu \alpha }F_{\mu \nu \alpha }R^2 - \frac{2}{15} \,F_{\mu \nu \alpha }F_{\mu \nu \alpha }S^2  \nn \\
    & + \frac{1}{45} \,F_{\mu \nu \alpha }F_{\alpha \mu }R_{\nu }R - \frac{1}{45} \,F_{\mu \nu \alpha }F_{\alpha \nu }R_{\mu }R + \frac{6}{35} \,F_{\mu \nu \alpha }F_{\mu \nu }R_{\alpha }R + \frac{32}{105} \,F_{\mu \nu \mu }RF_{\alpha \nu \alpha }R \nn \\
    & - \frac{4}{35} \,F_{\mu \nu \mu }SF_{\alpha \nu \alpha }S + \frac{64}{315} \,F_{\mu \nu \mu }R_{\alpha }F_{\alpha \nu }R - \frac{8}{63} \,F_{\mu \nu \mu }S_{\alpha }F_{\alpha \nu }S + \frac{3}{7} \,F_{\mu \nu \mu }F_{\alpha \nu \alpha }R^2  \nn \\
    & - \frac{3}{35} \,F_{\mu \nu \mu }F_{\alpha \nu \alpha }S^2 + \frac{121}{315} \,F_{\mu \nu \mu }F_{\alpha \nu }R_{\alpha }R - \frac{16}{105} \,F_{\mu \nu \mu }F_{\alpha \nu }S_{\alpha }S - \frac{2}{15} \,F_{\mu \nu \nu }RF_{\alpha \mu \alpha }R  \nn \\
    & - \frac{1}{15} \,F_{\mu \nu \nu }R_{\alpha }F_{\alpha \mu }R - \frac{2}{15} \,F_{\mu \nu \nu }F_{\alpha \mu \alpha }R^2 - \frac{1}{15} \,F_{\mu \nu \nu }F_{\alpha \mu }R_{\alpha }R - \frac{2}{105} \,F_{\mu \nu }RF_{\alpha \nu \alpha \mu }R  \nn \\
    & + \frac{4}{15} \,F_{\mu \nu }RF_{\mu \nu \alpha \alpha }R - \frac{4}{15} \,F_{\mu \nu }SF_{\mu \nu \alpha \alpha }S - \frac{7}{45} \,F_{\mu \nu }R_{\alpha }F_{\alpha \mu \nu }R + \frac{1}{15} \,F_{\mu \nu }R_{\alpha }F_{\alpha \nu \mu }R  \nn \\
    & + \frac{44}{315} \,F_{\mu \nu }R_{\alpha }F_{\mu \nu \alpha }R + \frac{23}{1260} \,F_{\mu \nu }R_{\alpha }F_{\alpha \nu }R_{\mu }+\frac{31}{210} \,F_{\mu \nu }R_{\alpha }F_{\mu \nu }R_{\alpha } + \frac{10}{63} \,F_{\mu \nu }R_{\mu }F_{\alpha \nu \alpha }R  \nn \\
    & + \frac{25}{252} \,F_{\mu \nu }R_{\mu }F_{\alpha \nu }R_{\alpha } - \frac{5}{21} \,F_{\mu \nu }R_{\nu }F_{\alpha \mu \alpha }R - \frac{2}{15} \,F_{\mu \nu }S_{\alpha }F_{\mu \nu \alpha }S + \frac{9}{70} \,F_{\mu \nu }S_{\alpha }F_{\alpha \nu }S_{\mu }  \nn \\
    & - \frac{44}{315} \,F_{\mu \nu }S_{\alpha }F_{\mu \nu }S_{\alpha } - \frac{1}{6} \,F_{\mu \nu }S_{\alpha }F_{\nu \alpha }S_{\mu } - \frac{8}{63} \,F_{\mu \nu }S_{\mu }F_{\alpha \nu \alpha }S + \frac{1}{6} \,F_{\mu \nu }S_{\mu }F_{\alpha \nu }S_{\alpha }  \nn \\
    & - \frac{9}{70} \,F_{\mu \nu }S_{\nu }F_{\alpha \mu }S_{\alpha } + \frac{199}{630} \,F_{\mu \nu }R_{\alpha \alpha }F_{\mu \nu }R + \frac{8}{315} \,F_{\mu \nu }R_{\alpha \mu }F_{\alpha \nu }R + \frac{37}{630} \,F_{\mu \nu }R_{\mu \alpha }F_{\alpha \nu }R  \nn \\
    & + \frac{1}{30} \,F_{\mu \nu }R_{\nu \alpha }F_{\alpha \mu }R - \frac{4}{15} \,F_{\mu \nu }S_{\alpha \alpha }F_{\mu \nu }S + \frac{8}{45} \,F_{\mu \nu }F_{\alpha \mu \nu \alpha }R^2 + \frac{1}{105} \,F_{\mu \nu }F_{\alpha \nu \alpha \mu }R^2  \nn \\
    & + \frac{61}{180} \,F_{\mu \nu }F_{\mu \nu \alpha \alpha }R^2 - \frac{11}{60} \,F_{\mu \nu }F_{\mu \nu \alpha \alpha }S^2 + \frac{32}{315} \,F_{\mu \nu }F_{\alpha \mu \alpha }R_{\nu }R - \frac{1}{10} \,F_{\mu \nu }F_{\alpha \mu \nu }R_{\alpha }R  \nn \\
    & + \frac{8}{105} \,F_{\mu \nu }F_{\alpha \nu \alpha }R_{\mu }R - \frac{8}{105} \,F_{\mu \nu }F_{\alpha \nu \alpha }S_{\mu }S + \frac{1}{10} \,F_{\mu \nu }F_{\alpha \nu \mu }R_{\alpha }R + \frac{7}{30} \,F_{\mu \nu }F_{\mu \nu \alpha }R_{\alpha }R \nn \\
    & - \frac{1}{15} \,F_{\mu \nu }F_{\alpha \mu }R_{\nu \alpha }R + \frac{4}{45} \,F_{\mu \nu }F_{\alpha \mu }S_{\alpha \nu }S + \frac{23}{210} \,F_{\mu \nu }F_{\alpha \nu }R_{\mu }R_{\alpha } + \frac{1}{12} \,F_{\mu \nu }F_{\alpha \nu }S_{\mu }S_{\alpha }   \nn \\
    & - \frac{8}{315} \,F_{\mu \nu }F_{\alpha \nu }R_{\alpha \mu }R - \frac{32}{315} \,F_{\mu \nu }F_{\alpha \nu }R_{\mu \alpha }R - \frac{8}{45} \,F_{\mu \nu }F_{\mu \alpha }R_{\nu }R_{\alpha } + \frac{347}{1260} \,F_{\mu \nu }F_{\mu \alpha }S_{\nu }S_{\alpha } \nn \\
    & +\frac{4}{63} \,F_{\mu \nu }F_{\mu \nu }R_{\alpha }R_{\alpha } - \frac{1}{14} \,F_{\mu \nu }F_{\mu \nu }S_{\alpha }S_{\alpha } + \frac{71}{252} \,F_{\mu \nu }F_{\mu \nu }R_{\alpha \alpha }R - \frac{67}{1260} \,S_{\mu \mu }F_{\nu \alpha }F_{\nu \alpha }S \nn \\
    & -\frac{121}{1260} \,F_{\mu \nu }R_{\alpha }F_{\nu \alpha }R_{\mu }-\frac{83}{1260} \,F_{\mu \nu }R_{\nu }F_{\alpha \mu }R_{\alpha }-\frac{67}{1260} \,F_{\mu \nu }F_{\mu \nu }S_{\alpha \alpha }S  \bigg\}\bigg].
\end{align}

\begin{align}
    \L_{II} \llbracket F^2\phi^2D^2 \rrbracket = & \, \tr^i \bigg[ \cfrac{1}{M^4} \bigg\{\frac{11}{60} S\tilde F_{\mu \nu \alpha \alpha }F_{\mu \nu }R + \frac{2}{15} S\tilde F_{\mu \nu \alpha }F_{\mu \nu \alpha }R + \frac{11}{60} S\tilde F_{\mu \nu }F_{\mu \nu \kappa \kappa }R \nn \\
    & + \frac{2}{15} R_{\mu }\tilde F_{\alpha \nu \mu }F_{\alpha \nu }S + \frac{1}{6} S_{\beta }\tilde F_{\kappa \mu }F_{\beta \mu }R_{\kappa } + \frac{1}{60} S_{\mu }F_{\alpha \nu \alpha }\tilde F_{\mu \nu }R - \frac{1}{15} S_{\mu }\tilde F_{\beta \mu \kappa }F_{\beta \kappa }R  \nn \\
    & + \frac{2}{15} S_{\mu }F_{\alpha \nu \mu }\tilde F_{\alpha \nu }R + \frac{1}{12} S_{\mu }F_{\alpha \nu }\tilde F_{\mu \nu \alpha }R + \frac{1}{3} S_{\mu }\tilde F_{\beta \mu }F_{\beta \kappa \kappa }R + \frac{1}{6} S_{\mu }F_{\alpha \nu }\tilde F_{\alpha \mu \nu }R  \nn \\
    & + \frac{3}{20} S_{\mu }F_{\alpha \nu }\tilde F_{\mu \nu }R_{\alpha } + \frac{4}{15} S_{\mu }\tilde F_{\beta \mu }F_{\beta \kappa }R_{\kappa } - \frac{7}{60} S_{\mu }\tilde F_{\alpha \nu }F_{\alpha \nu }R_{\mu } + \frac{1}{15} S_{\mu }F_{\alpha \nu }\tilde F_{\alpha \mu }R_{\nu } \nn \\
    & + \frac{1}{12} S_{\mu }F_{\mu \nu }\tilde F_{\kappa \nu }R_{\kappa } + \frac{11}{60} R_{\mu \mu }\tilde F_{\alpha \nu }F_{\alpha \nu }S - \frac{1}{6} S_{\alpha \nu }F_{\alpha \mu }\tilde F_{\mu \nu }R - \frac{1}{3} S_{\kappa \nu }\tilde F_{\mu \nu }F_{\kappa \mu }R  \nn \\
    & - \frac{1}{6} S_{\mu \alpha }F_{\alpha \nu }\tilde F_{\mu \nu }R - \frac{1}{3} S_{\mu \kappa }\tilde F_{\mu \nu }F_{\kappa \nu }R + \frac{1}{60} S_{\mu \mu }\tilde F_{\alpha \nu }F_{\alpha \nu }R + \frac{4}{15} \tilde F_{\mu \nu \alpha \alpha }SF_{\mu \nu }R  \nn \\
    & + \frac{11}{60} \tilde F_{\mu \nu \alpha \alpha }F_{\mu \nu }SR + \frac{2}{15} \tilde F_{\mu \nu \alpha }R_{\alpha }F_{\mu \nu }S + \frac{2}{15} \tilde F_{\mu \nu \alpha }S_{\alpha }F_{\mu \nu }R  - \frac{7}{12} \tilde F_{\mu \nu \alpha }S_{\mu }F_{\alpha \nu }R  \nn \\
    & + \frac{2}{15} \tilde F_{\mu \nu \alpha }F_{\mu \nu \alpha }SR + \frac{1}{3} \tilde F_{\mu \nu \alpha }F_{\alpha \mu }S_{\nu }R + \frac{1}{6} \tilde F_{\mu \nu \alpha }F_{\mu \nu }S_{\alpha }R + \frac{1}{6} \tilde F_{\mu \nu \mu }F_{\beta \nu }S_{\beta }R  \nn \\
    & + \frac{1}{3} F_{\mu \nu \alpha }\tilde F_{\alpha \mu }S_{\nu }R + \frac{2}{5} F_{\mu \nu \mu }S_{\alpha }\tilde F_{\alpha \nu }R + \frac{2}{5} F_{\mu \nu \mu }\tilde F_{\kappa \nu }S_{\kappa }R + \frac{4}{15} \tilde F_{\mu \nu }SF_{\mu \nu \kappa \kappa }R \nn \\
    & + \frac{2}{15} \tilde F_{\mu \nu }R_{\alpha }F_{\mu \nu \alpha }S - \frac{1}{30} \tilde F_{\mu \nu }S_{\alpha }F_{\mu \nu \alpha }R + \frac{1}{3} \tilde F_{\mu \nu }S_{\mu }F_{\beta \nu \beta }R + \frac{7}{30} F_{\mu \nu }S_{\alpha }\tilde F_{\alpha \nu \mu }R  \nn \\
    & - \frac{1}{12} \tilde F_{\mu \nu }S_{\alpha }F_{\mu \nu }R_{\alpha }  - \frac{4}{15}\tilde F_{\mu \nu }S_{\mu }F_{\nu \kappa }R_{\kappa } + \frac{31}{60} F_{\mu \nu }S_{\alpha }\tilde F_{\alpha \nu }R_{\mu } + \frac{1}{3} \tilde F_{\mu \nu }S_{\beta }F_{\mu \beta \nu }R  \nn \\
    & + \frac{1}{6} \tilde F_{\mu \nu }S_{\beta }F_{\mu \beta }R_{\nu } - \frac{1}{6} F_{\mu \nu }S_{\mu }\tilde F_{\kappa \nu \kappa }R - \frac{1}{12} F_{\mu \nu }S_{\mu }\tilde F_{\kappa \nu }R_{\kappa } + \frac{4}{15} \tilde F_{\mu \nu }R_{\alpha \alpha }F_{\mu \nu }S  \nn \\
    & + \frac{4}{15} \tilde F_{\mu \nu }S_{\alpha \alpha }F_{\mu \nu }R + \frac{1}{15} \tilde F_{\mu \nu }S_{\mu \kappa }F_{\nu \kappa }R + \frac{11}{60} \tilde F_{\mu \nu }F_{\mu \nu \kappa \kappa }SR - \frac{1}{15} \tilde F_{\mu \nu }F_{\mu \beta \beta }S_{\nu }R \nn \\
    & + \frac{2}{15} \tilde F_{\mu \nu }F_{\mu \nu \kappa }R_{\kappa }S + \frac{2}{15} \tilde F_{\alpha \beta }F_{\alpha \beta \kappa }S_{\kappa }R - \frac{1}{12} \tilde F_{\alpha \kappa }F_{\alpha \beta }S_{\beta }R_{\kappa } + \frac{13}{60} \tilde F_{\alpha \kappa }F_{\alpha \beta }S_{\kappa }R_{\beta } \nn \\
    & - \frac{1}{30} F_{\mu \nu }\tilde F_{\mu \nu }S_{\kappa }R_{\kappa } - \frac{31}{60} F_{\mu \nu }\tilde F_{\kappa \mu }S_{\kappa }R_{\nu }  + \frac{1}{4} F_{\mu \nu }\tilde F_{\kappa \mu }S_{\nu }R_{\kappa } + \frac{11}{60} \tilde F_{\alpha \beta }F_{\alpha \beta }R_{\kappa \kappa }S  \nn \\
    & + \frac{1}{3} \tilde F_{\beta \kappa }F_{\alpha \beta }S_{\alpha \kappa }R + \frac{1}{3} \tilde F_{\alpha \kappa }F_{\alpha \beta }S_{\kappa \beta }R + \frac{1}{60} F_{\mu \nu }\tilde F_{\mu \nu }S_{\kappa \kappa }R + \frac{7}{30} F_{\mu \nu }\tilde F_{\kappa \nu }S_{\kappa \mu }R  \nn \\
    & - \frac{1}{6} F_{\mu \nu }\tilde F_{\kappa \nu }S_{\mu \kappa }R + \frac{1}{6} F_{\mu \nu }S_{\alpha }\tilde F_{\alpha \mu \nu }R - \frac{1}{6} S_{\mu }F_{\alpha \nu }\tilde F_{\alpha \mu \nu }R + \frac{1}{6} \tilde F_{\beta \kappa \nu }F_{\beta \nu }S_{\kappa }R \nn \\ 
    & - \frac{1}{3} \tilde F_{\beta \kappa \nu }S_{\beta }F_{\kappa \nu }R - \frac{1}{6} S_{\mu }F_{\alpha \nu }\tilde F_{\mu \nu \alpha }R +\frac{1}{6}F_{\mu \nu }\tilde F_{\beta \nu \mu }S_\beta R + \frac{1}{4}F_{\mu\nu}S_\alpha \tilde F_{\alpha\mu\nu}R \bigg\}\bigg].
\end{align}
\begin{align}
    \L \llbracket F^2 D^4 \rrbracket = & \, \tr^i \bigg[ \cfrac{1}{M^4} \bigg\{ \frac{1}{210} \,F_{\mu \nu \mu \alpha }F_{\beta \nu \beta \alpha } - \frac{1}{60} \,F_{\mu \nu \alpha \alpha }F_{\mu \nu \beta \beta } \bigg\} \bigg].
\end{align}
\begin{align}
    \L^{CPC} \llbracket F^3\Phi^2 \rrbracket = & \, \tr^i \bigg[ \cfrac{1}{M^4} \bigg\{ \frac{44}{63} \,F_{\mu \nu }F_{\nu \alpha }F_{\alpha \mu }S^2 - \frac{397}{315} \,F_{\mu \nu }F_{\nu \alpha }RF_{\alpha \mu }R + \frac{452}{315} \,F_{\mu \nu }F_{\nu \alpha }SF_{\alpha \mu }S \nn \\ & - \frac{527}{315} \,F_{\mu \nu }F_{\nu \alpha }F_{\alpha \mu }R^2 \bigg\} \bigg].
\end{align}

\begin{align}
    \L^{CPV} \llbracket F^3\Phi^2 \rrbracket = & \, \tr^i \bigg[ \cfrac{1}{M^4} \bigg\{-\frac{11}{21}\,S\tilde F_{\mu \nu }F_{\nu \alpha } F_{\alpha \mu }R + \frac{1}{12}\,SF_{\mu \nu }\tilde F_{\nu \alpha }F_{\alpha \mu }R - \frac{17}{60} \,SF_{\mu \nu }F_{\nu \alpha }\tilde F_{\alpha \mu }R \nn \\ &
    - \frac{7}{12} \tilde F_{\mu \nu }SF_{\nu \alpha }F_{\alpha \mu }R 
     - \frac{3}{20} F_{\mu \nu }S\tilde F_{\nu \alpha }F_{\alpha \mu }R - \frac{4}{15} F_{\mu \nu }SF_{\nu \alpha }\tilde F_{\alpha \mu }R \nn \\ & 
     - \frac{2}{5} \tilde  F_{\mu \nu }F_{\nu \alpha }SF_{\alpha \mu }R - \frac{31}{60} F_{\mu \nu }\tilde F_{\nu \alpha }SF_{\alpha \mu }R 
     - \frac{2}{3} F_{\mu \nu }F_{\nu \alpha }S\tilde F_{\alpha \mu }R  \nn \\ & - \frac{7}{15}\tilde  F_{\mu \nu }F_{\nu \alpha } F_{\alpha \mu }SR + \frac{1}{12}  F_{\mu \nu }\tilde F_{\nu \alpha } F_{\alpha \mu }SR - \frac{17}{60} F_{\mu \nu }F_{\nu \alpha }\tilde F_{\alpha \mu }SR \bigg\} \bigg].
\end{align}
\begin{align}
    \L \llbracket F^3D^2 \rrbracket = & \, \tr^i \bigg[ \cfrac{1}{M^4} \bigg\{\frac{1}{30} \,F_{\mu \nu \mu }F_{\alpha \beta }F_{\alpha \beta \nu } + \frac{1}{6} \,F_{\mu \nu }F_{\alpha \mu }F_{\alpha \nu \beta \beta } + \frac{8}{315} \,F_{\mu \nu \alpha }F_{\alpha \nu }F_{\beta \mu \beta } \bigg\} \bigg].
\end{align}
\begin{align}
    \L_{I} \llbracket F^4 \rrbracket =&\, \tr^i \bigg[ \cfrac{1}{M^4} \bigg\{\frac{19}{630} \,(F_{\mu \nu })^2(F_{\alpha \beta })^2 + \frac{8}{315} \,(F_{\mu \nu }F_{\alpha \beta })^2 + \frac{53}{315} \,F_{\mu \nu }F_{\nu \alpha }F_{\alpha \beta }F_{\beta \mu } \nn \\ & - \frac{34}{105} \,(F_{\mu \alpha }F_{\alpha\mu })^2\bigg\} \bigg].
\end{align}
\subsection{Dimension One-Four operators in FUOLEA}\label{sec:renorm}
Using the results obtained by dimensional regularisation and $\overline{MS}$ renormalisation scheme, the finite part of the renormalisable Lagrangian, operators of dimension $\leq4$, are calculated for the fermionic case. Here we have neglected contributions due to the multiplicative anomaly discussed in Sec.~\ref{sec:effective action}. 
\subsection*{\underline{Dimension One operators in FUOLEA}}
Operators from UOLEA given in Appendix \ref{App:UOLEA} contributing to dimension one operators are,
\begin{equation}
    \L_\eff=\frac{c_s}{(4\pi)^2} \tr\bigg\{-M^2\ \left(\ln\left[\frac{M^2}{\mu^2}\right]-1\right)\, U \bigg\}.
\end{equation}
Expanding the generalised operators and retaining only the terms $\mathcal O (M^3)$, we get,
\begin{align}
    \L_\eff^{\Psi(D1)}=\frac{c_s}{(4\pi)^2} \tr\bigg\{-2 M^3 \left(\ln\left[\frac{M^2}{\mu^2}\right]-1\right)\, \Sigma \bigg\}.
\end{align}
Performing the trace over the spinor indices, the dimension one operator in terms of the functionals in the fermionic Lagrangian is,
\begin{align}
    \L_\eff^{\Psi(D1)}=\frac{c_s}{(4\pi)^2} \tr^i\bigg\{8 M^3 \left(1-\ln\left[\frac{M^2}{\mu^2}\right]\right)\, S \bigg\}.
\end{align}
\subsection*{\underline{Dimension Two operators in FUOLEA}}
The dimension two operators get contribution from the following operators of the UOLEA given in Appendix \ref{App:UOLEA}.
\begin{equation}
    \L_\eff=\frac{c_s}{(4\pi)^2} \tr\bigg\{-M^2\,\left(\ln\left[\frac{M^2}{\mu^2}\right]-1\right)\, U -\frac{1}{2}  \ln\left[\frac{M^2}{\mu^2}\right] \, U^2 \bigg\}.
\end{equation}
Expanding the generalised operators and retaining only the terms $\mathcal O (M^3)$, we get,
\begin{equation}
    \L_\eff^{\Psi(D2)}=\frac{c_s}{(4\pi)^2} \tr\bigg\{-M^2\bigg[\left(\ln\left[\frac{M^2}{\mu^2}\right]-1\right)\, Y + 2  \ln\left[\frac{M^2}{\mu^2}\right] \, \Sigma^2\bigg] \bigg\}.
\end{equation}
Performing the spin trace, the dimension two operators are given by,
\begin{equation}
    \L_\eff^{\Psi(D2)}=\frac{c_s}{(4\pi)^2} \tr^i\bigg\{4M^2\bigg[\left(1-3\ln\left[\frac{M^2}{\mu^2}\right]\right)\, S^2 +  \left(3-\ln\left[\frac{M^2}{\mu^2}\right]\right)\,R^2 \bigg] \bigg\}.
\end{equation}
\subsection*{\underline{Dimension Three operators in FUOLEA}}
Following the power counting mentioned in Sec.~\ref{sec:effective action}, the operators from the UOLEA that contribute to the dimension three operators are,
\begin{align}
    \L_\eff=&\frac{c_s}{(4\pi)^2} \tr\bigg\{ - \frac{1}{2}   \ln\left[\frac{M^2}{\mu^2}\right] \, U^2 - \frac{1}{M^2} \frac{U^3}{6} \bigg\}.
\end{align}
Expanding the generalised operators and collecting terms of $\mathcal{O}(M)$, we get,
\begin{align}
    \L_\eff=&\frac{c_s}{(4\pi)^2} \tr\bigg\{ - M   \ln\left[\frac{M^2}{\mu^2}\right] \, \Sigma Y - \frac{4}{3} M \Sigma^3 \bigg\}.
\end{align}
Writing in terms of the functionals in the Lagrangian, the dimension three operators, after taking trace over spinor indices are given by,
\begin{align}
    \L_\eff=&\frac{c_s}{(4\pi)^2} \tr^i\bigg\{ 8M\bigg[ \bigg( 2 - \ln\left[\frac{M^2}{\mu^2}\right]\bigg) \, SR^2 - \bigg(\ln\left[\frac{M^2}{\mu^2}\right] + \frac{2}{3}\bigg) \, S^3  \bigg]\bigg\}.
\end{align}
\subsection*{\underline{Dimension  Four operators in FUOLEA}}
Operators from UOLEA given in Appendix \ref{App:UOLEA} contributing to dimension one operators are,
\begin{align}
    \L_\eff=&\frac{c_s}{(4\pi)^2} \tr\bigg\{\frac{1}{2}  \bigg[- \ln\left[\frac{M^2}{\mu^2}\right] \, U^2 -\frac{1}{6} \ln\left[\frac{M^2}{\mu^2}\right] \, (G_{\mu\nu})^2\bigg] \nn \\ & + \frac{1}{M^2} \frac{1}{6}  \,\bigg[ -U^3 - \frac{1}{2} (\P_\mu U)^2 \bigg] + \frac{1}{M^4} \frac{U^4 }{24} \bigg\}.
\end{align}
Terms of $\mathcal{O} (M^0)$ which contribute to the dimension four operators are given by,
\begin{align}
    \L_\eff=&\frac{c_s}{(4\pi)^2} \tr\bigg\{\frac{1}{2}  \bigg[- \ln\left[\frac{M^2}{\mu^2}\right] \, Y^2 -\frac{1}{6} \ln\left[\frac{M^2}{\mu^2}\right] \, (F_{\mu\nu}+\Gamma_{\mu\nu})^2\bigg] \nn \\ & + \frac{1}{3}  \,\bigg[ -6 Y\Sigma^2 -  (\P_\mu \Sigma)^2 \bigg] + \frac{2}{3} \Sigma^4  \bigg\}.
\end{align}
Simplifying the dimension four operators by performing the spinor trace and writing it in terms of the interaction functionals gives,
\begin{align}
    \L_\eff^{\Psi(D4)}=&\frac{c_s}{(4\pi)^2} \tr^i\bigg\{ \frac{2}{3}\ln\left[\frac{M^2}{\mu^2}\right] (F_{\mu\nu})^2 - 8 \ln\left[\frac{M^2}{\mu^2}\right] S^2R^2 - \bigg(\frac{16}{3}+2\ln\left[\frac{M^2}{\mu^2}\right]\bigg)S^4 \nn\\
   & + \bigg(\frac{16}{3}-2\ln\left[\frac{M^2}{\mu^2}\right]\bigg)R^4 
     + 4\ln\left[\frac{M^2}{\mu^2}\right] SRSR- \bigg(\frac{4}{3}+2\ln\left[\frac{M^2}{\mu^2}\right]\bigg) (P_\mu S)^2 \nn \\
     & + \bigg(\frac{4}{3}-2\ln\left[\frac{M^2}{\mu^2}\right]\bigg) (\P_\mu R)^2 \bigg\}.
\end{align}

The results for operators of dimensions one to six derived in Secs.~\ref{sec:D5}, \ref{sec:D6} and \ref{sec:renorm} have been verified with the results provided in the Refs.~\cite{Angelescu:2020yzf,Ellis:2020ivx} \footnote{In these references, authors have employed the BMHV regularisation scheme \cite{tHooft:1972tcz,Breitenlohner:1977hr,Chanowitz:1979zu,Jegerlehner:2000dz} to compute the finite contribution to the renormalisable IR Lagrangian, i.e.,  D1 to D4 operators. Though, we have used dimensional regularisation which is equivalent to zeta function regularisation \cite{Hawking:1976ja}, our finite parts are in good agreement with the results given in Refs.~~\cite{Angelescu:2020yzf,Ellis:2020ivx}.}.


\section{Flavors, CP conservation, and violation in FUOLEA} \label{sec:discussion}
In this section, we discuss some of the salient features of the results stated in the previous section. The one-loop effective action up to dimension eight computed in our previous paper \cite{Banerjee:2023iiv} is true for any strong elliptic operator of the form $(D^2+M^2+U)$. In the case of scalars, this is automatically satisfied. But for fermions, the Dirac operator which is a weakly elliptic operator needs to be brought into that form through bosonization.  From a mathematical sense, our effective action is universal. While computing that for fermions in the presence of scalar and pseudo-scalar Yukawa interactions, we find some specific characteristics of the effective action that are absent when heavy scalar is integrated out.

In case of heavy scalar integration out, the term $U_s$ in the elliptic operator is computed from 
$$ \cfrac{\delta^2\L^s}{\delta \Phi^\dagger \delta \Phi} \supset U_s\; $$ having mass dimension +2 and is a functional of light fields (IR DOFs).
At this point, we must recall that the operator dimension is not directly related to the mass dimension of $U_s$ as it may contain mass-dimensionful couplings. While in the case of fermion integration out, the term $U_f = Y + 2M_f \Sigma$, see Eq.~\eqref{eq:U_f_def} is an artefact of the bosonization.  This has also mass dimension +2 and it contains (pseudo-)scalar fields along with mass-dimensionless couplings in a very convoluted manner, see Eq.~\eqref{eq:U_f_def}. Thus, the dimension of effective operators can not be directly computed by looking into  $U_f$, see Eq.~\eqref{eq:U_to_muti_op}. 
\vskip 0.2cm
\noindent
In the case of fermion which possesses generation (flavor) indices, we have 
\begin{equation*}
 [U_f ]_{ij}= [Y+M_f(S + i \gamma^5 R)]_{ij}; \quad \quad \text{where, $i,j$ are the flavor indices}.
\end{equation*}
The interaction terms $S$ and $R$ stem from Yukawa-like interactions. This is further reflected in the Wilson coefficients (WCs) that possess now the flavor indices as well. From a phenomenological perspective, this suggests that the role of low-energy experiments, such as flavor physics observations, must leave a significant impact on these WCs.

\vskip 0.2cm
\noindent
Our computed effective action contains information about both scalar $(S)$ and pseudo-scalar $(R)$ interactions. We have noted that if we switch off the pseudo-scalar part, i.e., set $R = 0$, we can only generate the CP-conserving (CPC) effective operators. It is as per our expectation as the CP-violation (CPV) arises through the non-universal Yukawa couplings between different chiralities of fermions. This serves as a simple consistency test of our result. 

\vskip 0.2cm
\noindent
Here, we emphasize, in detail, some of the interesting features of CPV operators:
\begin{itemize} 

    \item Only for operator classes that do not involve the covariant derivatives $D$, the CPV nature can be directly identified, if any. For such definite cases, with the $\cpc$ and $\cpv$ labels, we have delineated the operator classes consisting of only $\Phi$ and $F$. Regarding the remaining classes, we have simply classified them based on the presence of the Levi-Civita tensor. For example, the operators in Eq.~\eqref{eq:phi4d4-2} are not always CP-violating despite the fact that each operator is associated with a Levi-Civita tensor. A careful inspection reveals that by the IBP, these operators can be combined with the $F \Phi^4 D^2$ classes. Let us consider the operator ($\varepsilon _{\alpha \beta \mu \nu } \,R_{\mu }R_{\nu }S_{\alpha }R_{\beta }$) in the $\Phi^4 D^4$ class. By IBP, this operator can be rewritten as,
    \begin{align*}
        \varepsilon _{\alpha \beta \mu \nu } \,R_{\mu }R_{\nu }S_{\alpha }R_{\beta }=-\varepsilon _{\alpha \beta \mu \nu } \,R_{\mu }R_{\nu }S_{\alpha \beta }R-\varepsilon _{\alpha \beta \mu \nu } \,R_{\mu }R_{\nu\beta }S_{\alpha }R - \varepsilon _{\alpha \beta \mu \nu } \,R_{\mu\beta }R_{\nu }S_{\alpha }R.
    \end{align*}
    Using the anti-symmetric property of the Levi-Civita tensor, operators of the form $\varepsilon _{\alpha \beta \mu \nu } \Phi_{\alpha\beta}$ can be written as,
    \begin{align*}
        \varepsilon _{\alpha \beta \mu \nu } \Phi_{\alpha\beta} = \frac{1}{2}\varepsilon _{\alpha \beta \mu \nu } F_{\beta\alpha} \Phi - \frac{1}{2}\varepsilon _{\alpha \beta \mu \nu } \Phi F_{\beta\alpha}.
    \end{align*}
    Here, $\Phi$ is a scalar, and a tensor of any order can be constructed through repeated action of covariant derivative on it, e.g., a rank two tensor is constructed as $P_\beta P_\alpha \Phi = \Phi_{\alpha\beta}$. Using the above identity, we can transform $\Phi^4 D^4$ class to $F \Phi^4 D^2$ class:
    \begin{align*}
        \varepsilon _{\alpha \beta \mu \nu } \,R_{\mu }R_{\nu }S_{\alpha }R_{\beta } & =-\varepsilon _{\alpha \beta \mu \nu } \,R_{\mu }R_{\nu }S_{\alpha \beta }R-\varepsilon _{\alpha \beta \mu \nu } \,R_{\mu }R_{\nu\beta }S_{\alpha }R - \varepsilon _{\alpha \beta \mu \nu } \,R_{\mu\beta }R_{\nu }S_{\alpha }R\\
        & =\frac{1}{2}\big\{\varepsilon _{\alpha \beta \mu \nu } \,R_{\mu }R_{\nu }SR F_{\beta\alpha} - \varepsilon _{\alpha \beta \mu \nu } \,R_{\mu }R_{\nu }F_{\beta\alpha}SR - \varepsilon _{\alpha \beta \mu \nu } \,R_{\mu } F_{\beta\nu} R S_{\alpha }R \\
        &\quad +\varepsilon _{\alpha \beta \mu \nu } \,R_{\mu } R F_{\beta\nu} S_{\alpha }R - \varepsilon _{\alpha \beta \mu \nu } \,F_{\beta\mu} R R_{\nu }S_{\alpha }R + \varepsilon _{\alpha \beta \mu \nu } \, R F_{\beta\mu} R_{\nu }S_{\alpha }R\big\}\\
        & = R_{\mu }R_{\nu }\tilde F_{\mu\nu}SR - R_{\mu }R_{\nu }SR \tilde F_{\mu\nu} + R_{\mu } \tilde F_{\alpha\mu} R S_{\alpha }R -R_{\mu } R \tilde F_{\alpha\mu} S_{\alpha }R \\
        &\quad - \tilde F_{\alpha\nu} R R_{\nu }S_{\alpha }R + R \tilde F_{\alpha\nu} R_{\nu }S_{\alpha }R.
    \end{align*}

    \item We note that in the process of integrating out the heavy fermions at the one-loop level, only the CP-conserving operator class $F^4$ emerges. This was expected as even at the dimension six level, only the CP-conserving operator class emerges up to one-loop. We expect CPV $F^4$ operators to be generated, for the first time, at the two-loop level, as per the lesson noted in Refs.~\cite{Bakshi:2021prd,Naskar:2022rpg,Guedes:2023azv} for dimension six CPV class $ F^3 $.

    \item The CPV operators in the $F^3$ class first appear at two-loop \cite{Bakshi:2021prd,Naskar:2022rpg,Guedes:2023azv}, whereas the CPV operators in the $\Phi^2 F^3$ operator class at dimension eight appear at one-loop level. In the case of SMEFT, the $F^3$ operator class may receive a contribution from $\Phi^2 F^3$ class after electroweak symmetry breaking and thus the respective WC will be suppressed by a factor of $v_{ew}^2/\Lambda^2$, where $v_{ew}$ is the vacuum expectation value (vev) of SM Higgs and $\Lambda$ is the scale of new physics. However, since the $F^3$ operator class will have an additional $(4\pi)^{-2}$ loop factor suppression, it is possible that $\Phi^2 F^3$ offers a dominant contribution to $F^3$ class compared to the two-loop generated contributions to the same. 
\end{itemize}
\section{Conclusion and Outlook} \label{sec:conclusion}
In the hunch of achieving more precision and extracting more information from a theory, we are in an attempt to compute the one-loop effective action up to dimension eight for the first time. As a follow-up of our previous paper \cite{Banerjee:2023iiv}, we have now computed the universal one-loop effective action (UOLEA) up to dimension eight, achieved after integrating out heavy fermions having any allowed SM gauge quantum numbers. We have emphasized though the Dirac operator is a weak elliptic operator, still, it is possible to use the HK method after the successful bosonization of the fermionic operator. This enables us to use the UOLEA computed in our previous work  \cite{Banerjee:2023iiv} and highlights its true universal features. This also displays the robustness of the HK method for a model-independent computation of one-loop effective action. We agree with the results, for lower dimensional effective action, computed in the existing literature.

We have explicitly computed the fermionic UOLEA using the HK method that captures the footprint of both CP-conserving and violating interactions in UV theories. Our result is equally applicable to any UV as well as low energy theory. We have discussed many features of CP violations that can be captured in our generic effective action. For example, this will be very useful to compute the CPV effects very precisely at low energy after integrating our top quark. At this point, our result captures the effect of heavy fermion loops only and the CPV through Yukawa interactions. As our future endeavour, we are in the process of adding the contributions from loops containing information about the mixed spin propagators along with the light-heavy ones as well. But those are beyond the scope of this article.

\section*{Acknowledgements}
We acknowledge the useful discussions with Shamik Banerjee, Diptarka Das, and Nilay Kundu. SR would like to thank the Institute of Physics Bhubaneswar for the hospitality where part of this work was done.
\appendix
\section{D7 operators in terms of $\Sigma$ and $Y$}\label{App:D7}

Expanding the generalised functional $U$ in UOLEA and collecting the $\mathcal O(1/M^3)$ terms, we get,
\begin{align}
    \L_\eff^{\Psi{(D7)}}=&\frac{c_s}{(4\pi)^{2}}\,\tr\bigg[\frac{1}{M^3}\,\bigg\{-\frac{64}{105} \Sigma ^7 + \frac{8}{5} Y\,\Sigma ^5 - \frac{2}{3} Y^2\,\Sigma ^3 - \frac{1}{12} Y^2\,(\P^2\Sigma )+\frac{1}{3} Y^3\,\Sigma  - \frac{2}{15} \Sigma ^3\,F_{\mu \nu }^2  \nn\\
    & - \frac{2}{15} \Sigma ^3\,\Gamma _{\mu \nu }^2 + \frac{4}{15} \Sigma ^3\,(\P^2Y)-\frac{8}{15} \Sigma ^3\,(\P_{\nu }\Sigma )^2 - \frac{4}{5} \Sigma ^4\,(\P^2\Sigma )-\frac{1}{90} F_{\rho \sigma }^2\,(\P^2\Sigma )\nn\\
    & -\frac{1}{90} \Gamma _{\rho \sigma }^2\,(\P^2\Sigma ) + \frac{1}{60} (\P_{\mu}\,\P_{\mu }Y)(\P^2\Sigma )-\frac{4}{45} (\P_{\mu}\,\P_{\mu }\Sigma )(\P_{\nu}\Sigma )^2 + \frac{1}{60} (\P_{\mu}\,\P_{\mu }\Sigma )(\P^2Y) \nn\\
    & +\frac{1}{15} Y\,\Sigma \,F_{\mu \nu }^2 + \frac{1}{15} Y\,\Sigma \,\Gamma _{\mu \nu }^2 + \frac{2}{15} Y\,\Sigma \,(\P_{\mu }\Sigma )^2 - \frac{1}{12} Y\,\Sigma \,(\P^2Y)+\frac{4}{15} Y\,\Sigma ^2\,(\P^2\Sigma ) \nn\\
    & +\frac{1}{15} Y\,F_{\mu \nu }^2\,\Sigma  + \frac{1}{15} Y\,\Gamma _{\mu \nu }^2\,\Sigma - \frac{1}{30} Y\,(\P_{\mu }\Sigma )\,(\P_{\nu }\,F_{\nu \mu })-\frac{1}{30} Y\,(\P_{\mu }\Sigma )\,(\P_{\nu }\,\Gamma_{\nu \mu })\nn\\
    & +\frac{2}{15} Y\,(\P_{\mu }\Sigma )^2\,\Sigma  - \frac{1}{12} Y\,(\P^2Y)\Sigma + \frac{4}{15} Y\,(\P^2\Sigma )\Sigma ^2 - \frac{1}{30} \Sigma \,(\P_{\mu }Y)\,(\P_{\nu }\,F_{\nu \mu }) \nn\\
    & -\frac{1}{30} \Sigma \,(\P_{\mu }Y)\,(\P_{\nu }\,\Gamma_{\nu \mu })-\frac{2}{15} \Sigma \,(\P^2\Sigma )(\P^2\Sigma ) +\frac{1}{30} \Sigma \,(\P_{\nu }\,F_{\nu \mu })(\P_{\rho }\,F_{\rho \mu }) \nn \\
    & +\frac{1}{30} \Sigma \,(\P_{\nu }\,F_{\nu \mu })\P_{\rho }\,\Gamma _{\rho \mu }+\frac{1}{30} \Sigma \,(\P_{\nu }\,\Gamma_{\nu \mu })(\P_{\rho }\,F_{\rho \mu})+\frac{1}{30} \Sigma \,(\P_{\nu }\,\Gamma_{\nu \mu })(\P_{\rho }\,\Gamma _{\rho \mu }) \nn\\
    & +\frac{2}{15} \Sigma ^2\,(\P_{\mu }Y)\,(\P_{\mu }\Sigma ) + \frac{2}{15} \Sigma ^2\,(\P_{\mu }\Sigma )\,(\P_{\mu }Y) + \frac{2}{45} \Sigma ^2\,(\P_{\mu }\Sigma )\,(\P_{\nu }\,F_{\nu \mu })\nn\\
    & +\frac{2}{45} \Sigma ^2\,(\P_{\mu }\Sigma )\,(\P_{\nu }\,\Gamma_{\nu \mu }) -\frac{2}{45} \Sigma ^2\,\P_{\mu }\,F_{\mu \nu }(\P_{\nu}\Sigma ) - \frac{2}{45} \Sigma ^2\,\P_{\mu }\,\Gamma _{\mu \nu }(\P_{\nu}\Sigma ) - \frac{2}{15} \Sigma ^3\,F_{\mu \nu }\,\Gamma _{\mu \nu }  \nn\\
    & - \frac{2}{15} \Sigma ^3\,\Gamma _{\mu \nu }\,F_{\mu \nu } - \frac{2}{45} F_{\rho \mu }\,F_{\rho \nu }\,(\P_{\mu}\,\P_{\nu }\Sigma )-\frac{2}{45} F_{\rho \mu }\,\Gamma _{\rho \nu }\,(\P_{\mu}\,\P_{\nu }\Sigma )-\frac{2}{45} F_{\rho \nu }\,(\P_{\mu}\,\P_{\nu }\Sigma )\Gamma _{\rho \mu } \nn\\
    & - \frac{1}{90} F_{\rho \sigma }\,\Gamma _{\rho \sigma }\,(\P^2\Sigma ) -\frac{1}{90} F_{\rho \sigma }\,(\P^2\Sigma )\Gamma _{\rho \sigma } - \frac{2}{45} \Gamma _{\rho \mu }\,\Gamma _{\rho \nu }\,(\P_{\mu}\,\P_{\nu }\Sigma )-\frac{2}{3} Y\,\Sigma \,Y\,\Sigma ^2  \nn\\
    & + \frac{1}{15} Y\,\Sigma \,F_{\mu \nu }\,\Gamma _{\mu \nu } + \frac{1}{15} Y\,\Sigma \,\Gamma _{\mu \nu }\,F_{\mu \nu } + \frac{4}{15} Y\,\Sigma \,(\P^2\Sigma )\Sigma  + \frac{1}{30} Y\,F_{\mu \nu }\,\Sigma \,F_{\mu \nu } \nn\\
    & ++ \frac{1}{30} Y\,F_{\mu \nu }\,\Sigma \,\Gamma _{\mu \nu }  \frac{1}{15} Y\,F_{\mu \nu }\,\Gamma _{\mu \nu }\,\Sigma  + \frac{1}{30} Y\,\Gamma _{\mu \nu }\,\Sigma \,F_{\mu \nu } + \frac{1}{30} Y\,\Gamma _{\mu \nu }\,\Sigma \,\Gamma _{\mu \nu }  \nn\\
    & + \frac{1}{15} Y\,\Gamma _{\mu \nu }\,F_{\mu \nu }\,\Sigma  - \frac{1}{45} \Sigma \,F_{\mu \nu }\,F_{\nu \rho }\,F_{\rho \mu } - \frac{1}{45} \Sigma \,F_{\mu \nu }\,F_{\nu \rho }\,\Gamma _{\rho \mu } - \frac{1}{45} \Sigma \,F_{\mu \nu }\,\Gamma _{\nu \rho }\,F_{\rho \mu }  \nn\\
    &  - \frac{1}{45} \Sigma \,F_{\mu \nu }\,\Gamma _{\nu \rho }\,\Gamma _{\rho \mu } - \frac{2}{45} \Sigma \,F_{\mu \nu }\,(\P_{\mu }\Sigma )\,(\P_{\nu }\Sigma ) - \frac{1}{45} \Sigma \,\Gamma _{\mu \nu }\,F_{\nu \rho }\,F_{\rho \mu } - \frac{1}{45} \Sigma \,\Gamma _{\mu \nu }\,F_{\nu \rho }\,\Gamma _{\rho \mu }   \nn\\
    & - \frac{1}{45} \Sigma \,\Gamma _{\mu \nu }\,\Gamma _{\nu \rho }\,F_{\rho \mu } - \frac{1}{45} \Sigma \,\Gamma _{\mu \nu }\,\Gamma _{\nu \rho }\,\Gamma _{\rho \mu } - \frac{2}{45} \Sigma \,\Gamma _{\mu \nu }\,(\P_{\mu }\Sigma )\,(\P_{\nu }\Sigma ) - \frac{4}{45} \Sigma ^2\,F_{\mu \nu }\,\Sigma \,F_{\mu \nu } \nn \\
    & - \frac{2}{45} \Sigma \,(\P_{\mu }\Sigma )\,(\P_{\nu }\Sigma )\,F_{\mu \nu }  - \frac{4}{45} \Sigma ^2\,F_{\mu \nu }\,\Sigma \,\Gamma _{\mu \nu } - \frac{4}{45} \Sigma ^2\,\Gamma _{\mu \nu }\,\Sigma \,F_{\mu \nu } - \frac{4}{45} \Sigma ^2\,\Gamma _{\mu \nu }\,\Sigma \,\Gamma _{\mu \nu }\nn \\
    & - \frac{2}{45} \Sigma \,(\P_{\mu }\Sigma )\,(\P_{\nu }\Sigma )\,\Gamma _{\mu \nu }\bigg\}\bigg].
\end{align}
\section{D8 operators in terms of $\Sigma$ and $Y$}\label{App:D8}

Expanding the generalised functional $U$ in UOLEA and collecting the $\mathcal O(1/M^4)$ terms, we get,
\begin{align}
    \L_\eff^{\Psi{(D8)}} = &\frac{c_s}{(4\pi)^{2}}\,\tr\bigg[\frac{1}{M^4}\bigg\{\frac{1}{24}Y^4+\frac{16}{21} \Sigma ^8 -\frac{32}{15} Y\,\Sigma ^6 + \frac{4}{5} Y^2\,\Sigma ^4 + \frac{1}{30} Y^2\,F_{\mu \nu }^2 + \frac{1}{30} Y^2\,\Gamma _{\mu \nu }^2 \nn \\
    & + \frac{1}{15} Y^2\,(\P_{\mu}\Sigma )^2 - \frac{1}{24} Y^2\,(\P^2 Y)-\frac{1}{3} Y^3\,\Sigma ^2 + \frac{1}{15} \Sigma ^2\,(\P_{\mu }Y)^2 + \frac{2}{21} \Sigma ^4\,F_{\mu \nu }^2 + \frac{2}{21} \Sigma ^4\,\Gamma _{\mu \nu }^2  \nn \\
    & - \frac{32}{21} \Sigma ^4\,(\P_{\mu }\Sigma )^2 - \frac{2}{5} \Sigma ^4\,(\P^2 Y)+\frac{17}{5040} F_{\mu \nu }^2\,F_{\rho \sigma }^2+\frac{17}{5040} F_{\mu \nu }^2\,\Gamma _{\rho \sigma }^2 +\frac{11}{315} F_{\mu \nu }^2\,(\P_{\rho }\Sigma )^2 \nn \\
    & + \frac{17}{5040} F_{\rho \sigma }^2\,\Gamma _{\mu \nu }^2-\frac{1}{180} F_{\rho \sigma }^2\,(\P^2Y) +\frac{17}{5040} \Gamma _{\mu \nu }^2\,\Gamma _{\rho \sigma }^2+\frac{11}{315} \Gamma _{\mu \nu }^2\,(\P_{\rho }\Sigma )^2 - \frac{1}{180} \Gamma _{\rho \sigma }^2\,(\P^2Y) \nn \\
    & +\frac{6}{35} (\P_{\mu }\Sigma )^2\,(\P_{\nu }\Sigma )^2 - \frac{2}{45} (\P^2Y)(\P_{\nu }\Sigma )^2 + \frac{1}{120} (\P^2Y)(\P^2Y)-\frac{1}{210} (\P_{\mu}\,\P^2\Sigma )\,(\P_{\mu }\,\P^2\Sigma )  \nn \\
    & + \frac{1}{840} (\P_{\rho }\,\P_{\alpha }\,F_{\alpha \nu })^2 + \frac{1}{840} (\P_{\rho }\,\P_{\alpha }\,F_{\alpha \nu })\,(\P_{\rho }\,(\P_{\mu }\,\Gamma _{\mu \nu })) + \frac{1}{840} (\P_{\rho }\,\P_{\alpha }\,\Gamma _{\alpha \nu })\,(\P_{\rho }\,(\P_{\mu }\,F_{\mu \nu })) \nn \\
    & + \frac{1}{840} (\P_{\rho }\,\P_{\alpha }\,\Gamma _{\alpha \nu })^2 + \frac{4}{315} \Sigma \,F_{\mu \nu } (\P^2\Sigma )F_{\mu \nu } + \frac{4}{315} \Sigma \,\Gamma _{\mu \nu } (\P^2\Sigma )F_{\mu \nu } + \frac{4}{315} \Sigma \,F_{\mu \nu } (\P^2\Sigma )\Gamma _{\mu \nu }  \nn \\
    & + \frac{4}{315} \Sigma \,\Gamma _{\mu \nu } (\P^2\Sigma )\Gamma _{\mu \nu } - \frac{1}{15} Y\,\Sigma ^2\,F_{\mu \nu }^2 - \frac{1}{15} Y\,\Sigma ^2\,\Gamma _{\mu \nu }^2 + \frac{2}{15} Y\,\Sigma ^2\,(\P^2Y) -\frac{4}{15} Y\,\Sigma ^2\,(\P_{\nu}\Sigma )^2 \nn \\
    & - \frac{2}{5} Y\,\Sigma ^3\,(\P^2\Sigma )-\frac{1}{15} Y\,F_{\mu \nu }^2\,\Sigma ^2 - \frac{1}{15} Y\,\Gamma _{\mu \nu }^2\,\Sigma ^2 - \frac{1}{60} Y\,(\P_{\mu }Y)\,(\P_{\nu }\,F_{\nu \mu }) - \frac{2}{5} Y\,(\P^2\Sigma )\Sigma ^3 \nn \\
    & -\frac{1}{60} Y\,(\P_{\mu }Y)\,(\P_{\nu }\,\Gamma_{\nu \mu })+\frac{2}{15} Y\,(\P^2Y)\Sigma ^2 - \frac{1}{15} Y\,(\P^2\Sigma )(\P^2\Sigma )-\frac{4}{15} Y\,(\P_{\nu}\Sigma )^2\,\Sigma ^2  \nn \\
    & + \frac{1}{60} Y\,(\P_{\nu }\,F_{\nu \mu })((\P_{\rho }\,F_{\rho \mu }))+\frac{1}{60} Y\,(\P_{\nu }\,F_{\nu \mu })(\P_{\rho }\,\Gamma_{\rho \mu })+\frac{1}{60} Y\,(\P_{\nu }\,\Gamma_{\nu \mu })((\P_{\rho }\,F_{\rho \mu }))\nn \\
    & +\frac{1}{60} Y\,(\P_{\nu }\,\Gamma_{\nu \mu })(\P_{\rho }\,\Gamma_{\rho \mu }) +\frac{2}{15} Y^2\,\Sigma \,(\P^2\Sigma )+\frac{1}{30} Y^2\,F_{\mu \nu }\,\Gamma _{\mu \nu } + \frac{1}{30} Y^2\,\Gamma _{\mu \nu }\,F_{\mu \nu } \nn \\
    & + \frac{2}{15} Y^2\,(\P^2\Sigma )\Sigma  + \frac{4}{105} \Sigma \,F_{\nu \mu }^2\,(\P^2\Sigma ) +\frac{4}{105} \Sigma \,\Gamma _{\nu \mu }^2\,(\P^2\Sigma )-\frac{1}{15} \Sigma \,(\P^2Y)(\P^2\Sigma ) \nn \\
    & +\frac{4}{105} \Sigma \,(\P^2\Sigma )F_{\mu \nu }^2 + \frac{4}{105} \Sigma \,(\P^2\Sigma )\Gamma _{\mu \nu }^2 - \frac{1}{15} \Sigma \,(\P^2\Sigma )(\P^2Y)+\frac{1}{45} \Sigma ^2\,(\P_{\mu }Y)\,(\P_{\nu }\,F_{\nu \mu })\nn \\
    & +\frac{1}{45} \Sigma ^2\,(\P_{\mu }Y)\,(\P_{\nu }\,\Gamma_{\nu \mu })-\frac{1}{45} \Sigma ^2\,(\P_{\mu }\,F_{\mu \nu })(\P_{\nu}Y) + \frac{2}{21} \Sigma ^4\,\Gamma _{\mu \nu }\,F_{\mu \nu } + \frac{17}{5040} F_{\mu \nu }\,F_{\rho \sigma }^2\,\Gamma _{\mu \nu }  \nn \\
    & - \frac{1}{45} \Sigma ^2\,(\P_{\mu }\,\Gamma _{\mu \nu })(\P_{\nu}Y) + \frac{8}{35} \Sigma ^2\,(\P_{\mu }\,\P_{\nu}\Sigma )(\P_{\nu }\,\P_{\mu }\Sigma )-\frac{4}{105} \Sigma ^2\,(\P_{\nu }\,F_{\nu \mu })(\P_{\rho }\,F_{\rho \mu }) \nn \\
    & -\frac{4}{105} \Sigma ^2\,(\P_{\nu }\,F_{\nu \mu })(\P_{\rho }\,\Gamma_{\rho \mu })-\frac{4}{105} \Sigma ^2\,(\P_{\nu }\,\Gamma_{\nu \mu })(\P_{\rho }\,F_{\rho \mu })-\frac{4}{105} \Sigma ^2\,(\P_{\nu }\,\Gamma_{\nu \mu })(\P_{\rho }\,\Gamma_{\rho \mu })  \nn \\
    & -\frac{4}{15} \Sigma ^3\,(\P_{\nu}Y)\,(\P_{\nu}\Sigma ) - \frac{4}{15} \Sigma ^3\,(\P_{\nu}\Sigma )\,(\P_{\nu}Y) + \frac{2}{21} \Sigma ^4\,F_{\mu \nu }\,\Gamma _{\mu \nu } +\frac{17}{5040} F_{\mu \nu }\,\Gamma _{\mu \nu }\,F_{\rho \sigma }^2 \nn \\
    & +\frac{17}{5040} F_{\mu \nu }\,\Gamma _{\mu \nu }\,\Gamma _{\rho \sigma }^2+\frac{11}{315} F_{\mu \nu }\,\Gamma _{\mu \nu }\,(\P_{\rho}\Sigma )^2 + \frac{17}{5040} F_{\mu \nu }\,\Gamma _{\rho \sigma }^2\,\Gamma _{\mu \nu } + \frac{17}{5040} F_{\mu \nu }^2\,F_{\rho \sigma }\,\Gamma _{\rho \sigma } \nn \\
    & +\frac{2}{315} F_{\mu \nu }\,(\P_{\alpha }\,F_{\alpha \mu })(\P_{\rho }\,F_{\rho \nu })+\frac{2}{315} F_{\mu \nu }\,(\P_{\alpha }\,F_{\alpha \mu })(\P_{\rho }\,\Gamma _{\rho \nu })+\frac{17}{5040} F_{\mu \nu }^2\,\Gamma _{\rho \sigma }\,F_{\rho \sigma } \nn \\
    & +\frac{2}{315} F_{\mu \nu }\,(\P_{\alpha }\,\Gamma _{\alpha \mu })(\P_{\rho }\,F_{\rho \nu }) +\frac{2}{315} F_{\mu \nu }\,(\P_{\alpha }\,\Gamma _{\alpha \mu })(\P_{\rho }\,\Gamma _{\rho \nu })+\frac{11}{315} F_{\mu \nu }\,(\P_{\rho}\Sigma )^2\,\Gamma _{\mu \nu } \nn \\
    & +\frac{2}{315} F_{\nu \sigma }\,F_{\sigma \rho }\,(\P_{\rho }\,\P_{\mu }\,F_{\mu \nu }) + \frac{2}{315} F_{\nu \sigma }\,F_{\sigma \rho }\,(\P_{\rho }\,\P_{\mu }\,\Gamma _{\mu \nu }) + \frac{2}{315} F_{\nu \sigma }\,\Gamma _{\sigma \rho }\,(\P_{\rho }\,\P_{\mu }\,F_{\mu \nu })  \nn \\
    & + \frac{2}{315} F_{\nu \sigma }\,\Gamma _{\sigma \rho }\,(\P_{\rho }\,\P_{\mu }\,\Gamma _{\mu \nu }) - \frac{1}{45} F_{\rho \mu }\,F_{\rho \nu }\,(\P_{\mu }\,\P_{\nu}Y)-\frac{1}{45} F_{\rho \mu }\,\Gamma _{\rho \nu }\,(\P_{\mu }\,\P_{\nu}Y)\nn \\
    & + \frac{17}{5040} F_{\rho \sigma }\,\Gamma _{\mu \nu }^2\,\Gamma _{\rho \sigma }+\frac{17}{5040} F_{\rho \sigma }\,\Gamma _{\rho \sigma }\,\Gamma _{\mu \nu }^2-\frac{1}{180} F_{\rho \sigma }\,\Gamma _{\rho \sigma }\,(\P^2Y)-\frac{1}{180} F_{\rho \sigma }\,(\P^2Y)\Gamma _{\rho \sigma }  \nn \\
    & + \frac{2}{315} F_{\sigma \rho }\,(\P_{\rho }\,\P_{\mu }\,F_{\mu \nu })\,\Gamma _{\nu \sigma } + \frac{2}{315} F_{\sigma \rho }\,(\P_{\rho }\,\P_{\mu }\,\Gamma _{\mu \nu })\,\Gamma _{\nu \sigma } + \frac{2}{315} \Gamma _{\mu \nu }\,(\P_{\alpha }\,F_{\alpha \mu })(\P_{\rho }\,F_{\rho \nu }) \nn \\
    & +\frac{2}{315} \Gamma _{\mu \nu }\,(\P_{\alpha }\,F_{\alpha \mu })(\P_{\rho }\,\Gamma _{\rho \nu })+\frac{2}{315} \Gamma _{\mu \nu }\,(\P_{\alpha }\,\Gamma _{\alpha \mu })(\P_{\rho }\,F_{\rho \nu })+\frac{2}{315} \Gamma _{\mu \nu }\,(\P_{\alpha }\,\Gamma _{\alpha \mu })(\P_{\rho }\,\Gamma _{\rho \nu }) \nn \\
    & +\frac{2}{315} \Gamma _{\nu \sigma }\,\Gamma _{\sigma \rho }\,(\P_{\rho }\,\P_{\mu }\,F_{\mu \nu }) + \frac{2}{315} \Gamma _{\nu \sigma }\,\Gamma _{\sigma \rho }\,(\P_{\rho }\,\P_{\mu }\,\Gamma _{\mu \nu }) - \frac{1}{45} \Gamma _{\rho \mu }\,\Gamma _{\rho \nu }\,(\P_{\mu }\,\P_{\nu}Y) \nn \\
    & -\frac{1}{45} F_{\rho \nu }\,(\P_{\mu }\,\P_{\nu}Y)\Gamma _{\rho \mu } -\frac{2}{315} (\P_{\mu }\Sigma )\,(\P_{\mu }\,\P_{\rho }\,F_{\rho \nu })\,(\P_{\nu}\Sigma ) + \frac{1}{15} Y\,\Sigma \,(\P_{\mu }\Sigma )\,(\P_{\mu }Y)  \nn \\
    & + \frac{2}{315} (\P_{\mu }\Sigma )\,(\P_{\nu}\Sigma )\,(\P_{\mu }\,\P_{\rho }\,\Gamma _{\rho \nu }) - \frac{4}{315} (\P_{\mu }\Sigma )\,(\P_{\nu }\,F_{\nu \mu })(\P^2\Sigma ) - \frac{1}{15} Y\,\Sigma \,\Gamma _{\mu \nu }^2\,\Sigma \nn \\
    & +\frac{4}{315} (\P_{\mu }\Sigma )\,(\P^2\Sigma )(\P_{\nu }\,F_{\nu \mu })+\frac{4}{315} (\P_{\mu }\Sigma )\,(\P^2\Sigma )(\P_{\nu }\,\Gamma_{\nu \mu }) + \frac{2}{15} Y\,\Sigma \,(\P^2Y)\Sigma \nn \\
    & - \frac{2}{45} (\P^2\Sigma )(\P_{\nu}\Sigma )\,(\P_{\nu}Y) + \frac{4}{5} Y\,\Sigma \,Y\,\Sigma ^3 + \frac{2}{15} Y\,\Sigma \,Y\,(\P^2\Sigma )-\frac{1}{15} Y\,\Sigma \,F_{\mu \nu }^2\,\Sigma   \nn \\
    & - \frac{2}{315} (\P_{\mu }\Sigma )\,(\P_{\mu }\,\P_{\rho }\,\Gamma _{\rho \nu })\,(\P_{\nu}\Sigma ) -\frac{2}{45} (\P^2\Sigma )(\P_{\nu}Y)\,(\P_{\nu}\Sigma )  + \frac{1}{15} Y\,\Sigma \,(\P_{\mu }Y)\,(\P_{\mu }\Sigma ) \nn \\
    & + \frac{1}{45} Y\,\Sigma \,(\P_{\mu }\Sigma )\,(\P_{\nu }\,F_{\nu \mu })+\frac{1}{45} Y\,\Sigma \,(\P_{\mu }\Sigma )\,(\P_{\nu }\,\Gamma_{\nu \mu }) -\frac{1}{45} Y\,\Sigma \,(\P_{\mu }\,F_{\mu \nu })(\P_{\nu}\Sigma ) \nn \\
    & -\frac{4}{315} (\P_{\mu }\Sigma )\,(\P_{\nu }\,\Gamma_{\nu \mu })(\P^2\Sigma ) + \frac{2}{315} (\P_{\mu }\Sigma )\,(\P_{\nu}\Sigma )\,(\P_{\mu }\,\P_{\rho }\,F_{\rho \nu }) - \frac{2}{5} Y\,\Sigma ^2\,(\P^2\Sigma )\Sigma  \nn \\
    &- \frac{2}{5} Y\,\Sigma \,(\P^2\Sigma )\Sigma ^2 - \frac{4}{15} Y\,\Sigma \,(\P_{\nu}\Sigma )^2\,\Sigma + \frac{1}{120} Y\,F_{\mu \nu }\,Y\,F_{\mu \nu } + \frac{1}{15} Y\,(\P_{\mu }\Sigma )\,(\P_{\mu }Y)\,\Sigma \nn \\
    & - \frac{1}{45} Y\,\Sigma \,(\P_{\mu }\,\Gamma _{\mu \nu })(\P_{\nu}\Sigma ) + \frac{2}{5} Y\,\Sigma ^2\,Y\,\Sigma ^2 - \frac{1}{15} Y\,\Sigma ^2\,F_{\mu \nu }\,\Gamma _{\mu \nu } - \frac{1}{15} Y\,\Sigma ^2\,\Gamma _{\mu \nu }\,F_{\mu \nu }  \nn \\
    & + \frac{1}{60} Y\,F_{\mu \nu }\,Y\,\Gamma _{\mu \nu } - \frac{2}{45} Y\,F_{\mu \nu }\,\Sigma ^2\,F_{\mu \nu } - \frac{2}{45} Y\,F_{\mu \nu }\,\Sigma ^2\,\Gamma _{\mu \nu } - \frac{1}{90} Y\,F_{\mu \nu }\,F_{\nu \rho }\,F_{\rho \mu }  \nn \\
    & - \frac{1}{90} Y\,F_{\mu \nu }\,F_{\nu \rho }\,\Gamma _{\rho \mu } - \frac{1}{15} Y\,F_{\mu \nu }\,\Gamma _{\mu \nu }\,\Sigma ^2 - \frac{1}{90} Y\,F_{\mu \nu }\,\Gamma _{\nu \rho }\,F_{\rho \mu } - \frac{1}{90} Y\,F_{\mu \nu }\,\Gamma _{\nu \rho }\,\Gamma _{\rho \mu }  \nn \\
    & - \frac{1}{45} Y\,F_{\mu \nu }\,(\P_{\mu }\Sigma )\,(\P_{\nu}\Sigma ) + \frac{1}{120} Y\,\Gamma _{\mu \nu }\,Y\,\Gamma _{\mu \nu } - \frac{2}{45} Y\,\Gamma _{\mu \nu }\,\Sigma ^2\,F_{\mu \nu } - \frac{2}{45} Y\,\Gamma _{\mu \nu }\,\Sigma ^2\,\Gamma _{\mu \nu }  \nn \\
    & - \frac{1}{15} Y\,\Gamma _{\mu \nu }\,F_{\mu \nu }\,\Sigma ^2 - \frac{1}{90} Y\,\Gamma _{\mu \nu }\,F_{\nu \rho }\,F_{\rho \mu } - \frac{1}{90} Y\,\Gamma _{\mu \nu }\,F_{\nu \rho }\,\Gamma _{\rho \mu } - \frac{1}{90} Y\,\Gamma _{\mu \nu }\,\Gamma _{\nu \rho }\,F_{\rho \mu }  \nn \\
    & - \frac{1}{90} Y\,\Gamma _{\mu \nu }\,\Gamma _{\nu \rho }\,\Gamma _{\rho \mu } - \frac{1}{45} Y\,\Gamma _{\mu \nu }\,(\P_{\mu }\Sigma )\,(\P_{\nu}\Sigma ) + \frac{1}{15} Y\,(\P_{\mu }Y)\,(\P_{\mu }\Sigma )\,\Sigma \nn \\
    & - \frac{1}{45} Y\,(\P_{\mu }\Sigma )\,(\P_{\nu}\Sigma )\,F_{\mu \nu } - \frac{1}{45} Y\,(\P_{\mu }\Sigma )\,(\P_{\nu}\Sigma )\,\Gamma _{\mu \nu } + \frac{1}{45} Y\,(\P_{\mu }\Sigma )\,(\P_{\nu }\,F_{\nu \mu })\Sigma   \nn \\
    & + \frac{1}{45} Y\,(\P_{\mu }\Sigma )\,(\P_{\nu }\,\Gamma_{\nu \mu })\Sigma  - \frac{1}{45} Y\,(\P_{\mu }\,F_{\mu \nu })(\P_{\nu}\Sigma )\,\Sigma  - \frac{1}{45} Y\,(\P_{\mu }\,\Gamma _{\mu \nu })(\P_{\nu}\Sigma )\,\Sigma   \nn \\
    & - \frac{1}{45} \Sigma \,F_{\mu \nu }\,(\P_{\mu }Y)\,(\P_{\nu}\Sigma ) - \frac{1}{45} \Sigma \,F_{\mu \nu }\,(\P_{\mu }\Sigma )\,(\P_{\nu}Y) - \frac{2}{63} \Sigma \,F_{\mu \nu }\,(\P_{\mu }\Sigma )\,(\P_{\rho }\,F_{\rho \nu }) \nn \\
    & -\frac{2}{63} \Sigma \,F_{\mu \nu }\,(\P_{\mu }\Sigma )\,(\P_{\rho }\,\Gamma _{\rho \nu }) -\frac{2}{105} \Sigma \,F_{\mu \nu }\,(\P_{\rho }\,F_{\rho \nu })(\P_{\mu }\Sigma ) - \frac{2}{105} \Sigma \,F_{\mu \nu }\,(\P_{\rho }\,\Gamma _{\rho \nu })(\P_{\mu }\Sigma )  \nn \\
    & + \frac{4}{105} \Sigma \,F_{\nu \mu }\,\Gamma _{\nu \mu }\,(\P^2\Sigma )-\frac{1}{45} \Sigma \,\Gamma _{\mu \nu }\,(\P_{\mu }Y)\,(\P_{\nu}\Sigma ) - \frac{1}{45} \Sigma \,\Gamma _{\mu \nu }\,(\P_{\mu }\Sigma )\,(\P_{\nu}Y) \nn \\
    & - \frac{2}{63} \Sigma \,\Gamma _{\mu \nu }\,(\P_{\mu }\Sigma )\,(\P_{\rho }\,F_{\rho \nu })- \frac{1}{3} Y^2\,\Sigma \,Y\,\Sigma -\frac{2}{63} \Sigma \,\Gamma _{\mu \nu }\,(\P_{\mu }\Sigma )\,(\P_{\rho }\,\Gamma _{\rho \nu }) \nn \\
    & -\frac{2}{105} \Sigma \,\Gamma _{\mu \nu }\,(\P_{\rho }\,F_{\rho \nu })(\P_{\mu }\Sigma ) - \frac{2}{105} \Sigma \,\Gamma _{\mu \nu }\,(\P_{\rho }\,\Gamma _{\rho \nu })(\P_{\mu }\Sigma ) -\frac{1}{45} \Sigma \,(\P_{\mu }Y)\,(\P_{\nu}\Sigma )\,F_{\mu \nu }  \nn \\
    & + \frac{4}{105} \Sigma \,\Gamma _{\nu \mu }\,F_{\nu \mu }\,(\P^2\Sigma ) - \frac{1}{45} \Sigma \,(\P_{\mu }Y)\,(\P_{\nu}\Sigma )\,\Gamma _{\mu \nu } - \frac{4}{105} \Sigma \,(\P_{\mu }\Sigma )\,F_{\mu \nu }\,(\P_{\rho }\,F_{\rho \nu }) \nn \\
    & -\frac{4}{105} \Sigma \,(\P_{\mu }\Sigma )\,F_{\mu \nu }\,(\P_{\rho }\,\Gamma _{\rho \nu })-\frac{4}{105} \Sigma \,(\P_{\mu }\Sigma )\,\Gamma _{\mu \nu }\,(\P_{\rho }\,F_{\rho \nu }) - \frac{1}{45} \Sigma \,(\P_{\mu }\Sigma )\,(\P_{\nu}Y)\,\Gamma _{\mu \nu } \nn \\
    & -\frac{4}{105} \Sigma \,(\P_{\mu }\Sigma )\,\Gamma _{\mu \nu }\,(\P_{\rho }\,\Gamma _{\rho \nu })-\frac{1}{45} \Sigma \,(\P_{\mu }\Sigma )\,(\P_{\nu}Y)\,F_{\mu \nu } - \frac{2}{105} \Sigma \,(\P_{\mu }\Sigma )\,(\P_{\rho }\,F_{\rho \nu })\Gamma _{\mu \nu }  \nn \\
    & + \frac{52}{105} \Sigma \,(\P_{\mu }\Sigma )\,(\P_{\nu}\Sigma )\,(\P_{\mu }\,\P_{\nu}\Sigma )-\frac{2}{105} \Sigma \,(\P_{\mu }\Sigma )\,(\P_{\rho }\,F_{\rho \nu })F_{\mu \nu } + \frac{2}{105} \Sigma ^2\,F_{\mu \nu }\,\Gamma _{\nu \rho }\,F_{\rho \mu }  \nn \\
    & - \frac{2}{105} \Sigma \,(\P_{\mu }\Sigma )\,(\P_{\rho }\,\Gamma _{\rho \nu })F_{\mu \nu } - \frac{2}{105} \Sigma \,(\P_{\mu }\Sigma )\,(\P_{\rho }\,\Gamma _{\rho \nu })\Gamma _{\mu \nu } + \frac{4}{105} \Sigma \,(\P^2\Sigma )F_{\mu \nu }\,\Gamma _{\mu \nu }  \nn \\
    & + \frac{4}{105} \Sigma \,(\P^2\Sigma )\Gamma _{\mu \nu }\,F_{\mu \nu } + \frac{6}{35} \Sigma \,(\P_{\mu }\,\P_{\nu}\Sigma )\Sigma \,(\P_{\nu }\,\P_{\mu}\Sigma )+\frac{52}{105} \Sigma \,(\P_{\mu }\,\P_{\nu}\Sigma )(\P_{\mu }\Sigma )\,(\P_{\nu}\Sigma )  \nn \\
    & + \frac{12}{35} \Sigma \,(\P_{\nu}\Sigma )\,(\P_{\mu }\,\P_{\nu}\Sigma )(\P_{\mu }\Sigma ) - \frac{1}{35} \Sigma \,(\P_{\nu }\,F_{\nu \mu })\Sigma \,(\P_{\rho }\,F_{\rho \mu })-\frac{1}{35} \Sigma \,(\P_{\nu }\,F_{\nu \mu })\Sigma \,(\P_{\rho }\,\Gamma_{\rho \mu }) \nn \\
    & -\frac{1}{35} \Sigma \,(\P_{\nu }\,\Gamma_{\nu \mu })\Sigma \,(\P_{\rho }\,F_{\rho \mu })-\frac{1}{35} \Sigma \,(\P_{\nu }\,\Gamma_{\nu \mu })\Sigma \,(\P_{\rho }\,\Gamma_{\rho \mu })-\frac{4}{105} \Sigma \,(\P_{\rho }\,F_{\rho \nu })F_{\mu \nu }\,(\P_{\mu }\Sigma )  \nn \\
    & - \frac{4}{105} \Sigma \,(\P_{\rho }\,F_{\rho \nu })\Gamma _{\mu \nu }\,(\P_{\mu }\Sigma ) - \frac{2}{63} \Sigma \,(\P_{\rho }\,F_{\rho \nu })(\P_{\mu }\Sigma )\,F_{\mu \nu } - \frac{2}{63} \Sigma \,(\P_{\rho }\,F_{\rho \nu })(\P_{\mu }\Sigma )\,\Gamma _{\mu \nu }  \nn \\
    & - \frac{4}{105} \Sigma \,(\P_{\rho }\,\Gamma _{\rho \nu })F_{\mu \nu }\,(\P_{\mu }\Sigma ) - \frac{4}{105} \Sigma \,(\P_{\rho }\,\Gamma _{\rho \nu })\Gamma _{\mu \nu }\,(\P_{\mu }\Sigma ) - \frac{2}{63} \Sigma \,(\P_{\rho }\,\Gamma _{\rho \nu })(\P_{\mu }\Sigma )\,F_{\mu \nu }  \nn \\
    & - \frac{2}{63} \Sigma \,(\P_{\rho }\,\Gamma _{\rho \nu })(\P_{\mu }\Sigma )\,\Gamma _{\mu \nu } + \frac{2}{105} \Sigma ^2\,F_{\mu \nu }\,F_{\nu \rho }\,F_{\rho \mu } + \frac{2}{105} \Sigma ^2\,F_{\mu \nu }\,F_{\nu \rho }\,\Gamma _{\rho \mu }  \nn \\
    & + \frac{2}{105} \Sigma ^2\,F_{\mu \nu }\,\Gamma _{\nu \rho }\,\Gamma _{\rho \mu } + \frac{16}{35} \Sigma ^2\,F_{\mu \nu }\,(\P_{\mu }\Sigma )\,(\P_{\nu}\Sigma ) + \frac{2}{105} \Sigma ^2\,\Gamma _{\mu \nu }\,F_{\nu \rho }\,F_{\rho \mu } \nn \\
    & + \frac{2}{105} \Sigma ^2\,\Gamma _{\mu \nu }\,F_{\nu \rho }\,\Gamma _{\rho \mu } + \frac{2}{105} \Sigma ^2\,\Gamma _{\mu \nu }\,\Gamma _{\nu \rho }\,F_{\rho \mu }   + \frac{2}{105} \Sigma ^2\,\Gamma _{\mu \nu }\,\Gamma _{\nu \rho }\,\Gamma _{\rho \mu } +\frac{6}{35} \Sigma ^2 \,\Gamma _{\mu \nu }\, \Sigma ^2\, F_{\mu \nu } \nn \\
    & + \frac{16}{35} \Sigma ^2\,\Gamma _{\mu \nu }\,(\P_{\mu }\Sigma )\,(\P_{\nu}\Sigma ) - \frac{48}{35} \Sigma ^2\,(\P_{\mu }\Sigma )\,\Sigma ^2\,(\P_{\mu }\Sigma ) +\frac{3}{35} \Sigma ^2\, \Gamma _{\mu \nu }\, \Sigma ^2 \,\Gamma _{\mu \nu }\nn \\
    & +\frac{3}{35} \Sigma ^2 \,F_{\mu \nu }\, \Sigma ^2 \,F_{\mu \nu }+\frac{12}{35} \Sigma ^2 (\P_{\nu }\Sigma)\,F_{\mu \nu }\,(\P_{\mu } \Sigma) - \frac{256}{105} \Sigma ^3\,(\P_{\mu }\Sigma )\,\Sigma \,(\P_{\mu }\Sigma ) \nn\\
    &  - \frac{4}{105} \Sigma ^2\,(\P_{\mu }\Sigma )\,(\P_{\nu}\Sigma )\,F_{\mu \nu } - \frac{4}{105} \Sigma ^2\,(\P_{\mu }\Sigma )\,(\P_{\nu}\Sigma )\,\Gamma _{\mu \nu } + \frac{16}{105} \Sigma ^3\,F_{\mu \nu }\,\Sigma \,F_{\mu \nu } \nn \\
    & +\frac{12}{35} \Sigma ^2 (\P_{\nu }\Sigma) \Gamma _{\mu \nu } (\P_{\mu }\Sigma) + \frac{16}{105} \Sigma ^3\,F_{\mu \nu }\,\Sigma \,\Gamma _{\mu \nu } + \frac{16}{105} \Sigma ^3\,\Gamma _{\mu \nu }\,\Sigma \,F_{\mu \nu } + \frac{16}{105} \Sigma ^3\,\Gamma _{\mu \nu }\,\Sigma \,\Gamma _{\mu \nu } \nn \\
    & + \frac{1}{2520}F_{\alpha \mu }\,F_{\mu \nu }\,F_{\nu \rho }\,F_{\rho \alpha } +\frac{1}{2520}F_{\alpha \mu }\,F_{\mu \nu }\,F_{\nu \rho }\,\Gamma _{\rho \alpha }+\frac{1}{2520}F_{\alpha \mu }\,F_{\mu \nu }\,\Gamma _{\nu \rho }\,F_{\rho \alpha }\nn \\
    & +\frac{1}{2520}F_{\alpha \mu }\,F_{\mu \nu }\,\Gamma _{\nu \rho }\,\Gamma _{\rho \alpha } +\frac{1}{2520}F_{\alpha \mu }\,\Gamma _{\mu \nu }\,F_{\nu \rho }\,F_{\rho \alpha } +\frac{1}{2520}F_{\alpha \mu }\,\Gamma _{\mu \nu }\,F_{\nu \rho }\,\Gamma _{\rho \alpha }\nn \\
    & +\frac{1}{2520}F_{\alpha \mu }\,\Gamma _{\mu \nu }\,\Gamma _{\nu \rho }\,F_{\rho \alpha }+\frac{1}{2520}F_{\alpha \mu }\,\Gamma _{\mu \nu }\,\Gamma _{\nu \rho }\,\Gamma _{\rho \alpha } -\frac{1}{105} F_{\alpha \nu }\,(\P_{\mu }\Sigma )\,F_{\mu \nu }\,(\P_{\alpha}\Sigma )  \nn \\
    & - \frac{1}{105} F_{\alpha \nu }\,(\P_{\mu }\Sigma )\,\Gamma _{\mu \nu }\,(\P_{\alpha}\Sigma ) + \frac{1}{420} F_{\mu \nu }\,F_{\nu \rho }\,F_{\mu \sigma }\,F_{\sigma \rho } + \frac{1}{420} F_{\mu \nu }\,F_{\nu \rho }\,F_{\mu \sigma }\,\Gamma _{\sigma \rho }  \nn \\
    & + \frac{1}{2520}F_{\mu \nu }\,F_{\nu \rho }\,F_{\rho \alpha }\,\Gamma _{\alpha \mu }+\frac{1}{420} F_{\mu \nu }\,F_{\nu \rho }\,\Gamma _{\mu \sigma }\,F_{\sigma \rho } + \frac{1}{420} F_{\mu \nu }\,F_{\nu \rho }\,\Gamma _{\mu \sigma }\,\Gamma _{\sigma \rho }  \nn \\
    & + \frac{1}{2520}F_{\mu \nu }\,F_{\nu \rho }\,\Gamma _{\rho \alpha }\,\Gamma _{\alpha \mu }+\frac{1}{10080}F_{\mu \nu }\,F_{\rho \sigma }\,F_{\mu \nu }\,F_{\rho \sigma }+\frac{1}{5040}F_{\mu \nu }\,F_{\rho \sigma }\,F_{\mu \nu }\,\Gamma _{\rho \sigma }\nn \\
    & +\frac{1}{5040}F_{\mu \nu }\,F_{\rho \sigma }\,\Gamma _{\mu \nu }\,F_{\rho \sigma } +\frac{1}{5040}F_{\mu \nu }\,F_{\rho \sigma }\,\Gamma _{\mu \nu }\,\Gamma _{\rho \sigma }+\frac{17}{5040} F_{\mu \nu }\,F_{\rho \sigma }\,\Gamma _{\rho \sigma }\,\Gamma _{\mu \nu }\nn \\
    & +\frac{17}{5040} F_{\mu \nu }\,\Gamma _{\mu \nu }\,F_{\rho \sigma }\,\Gamma _{\rho \sigma }+\frac{17}{5040} F_{\mu \nu }\,\Gamma _{\mu \nu }\,\Gamma _{\rho \sigma }\,F_{\rho \sigma } +\frac{1}{420} F_{\mu \nu }\,\Gamma _{\nu \rho }\,F_{\mu \sigma }\,F_{\sigma \rho } \nn \\
    & + \frac{1}{420} F_{\mu \nu }\,\Gamma _{\nu \rho }\,F_{\mu \sigma }\,\Gamma _{\sigma \rho } + \frac{1}{2520}F_{\mu \nu }\,\Gamma _{\nu \rho }\,F_{\rho \alpha }\,\Gamma _{\alpha \mu }+\frac{1}{420} F_{\mu \nu }\,\Gamma _{\nu \rho }\,\Gamma _{\mu \sigma }\,F_{\sigma \rho }  \nn \\
    & + \frac{1}{420} F_{\mu \nu }\,\Gamma _{\nu \rho }\,\Gamma _{\mu \sigma }\,\Gamma _{\sigma \rho } + \frac{1}{2520}F_{\mu \nu }\,\Gamma _{\nu \rho }\,\Gamma _{\rho \alpha }\,\Gamma _{\alpha \mu }+\frac{1}{10080}F_{\mu \nu }\,\Gamma _{\rho \sigma }\,F_{\mu \nu }\,\Gamma _{\rho \sigma }\nn \\
    & +\frac{17}{5040} F_{\mu \nu }\,\Gamma _{\rho \sigma }\,F_{\rho \sigma }\,\Gamma _{\mu \nu } +\frac{1}{5040}F_{\mu \nu }\,\Gamma _{\rho \sigma }\,\Gamma _{\mu \nu }\,F_{\rho \sigma }+\frac{1}{5040}F_{\mu \nu }\,\Gamma _{\rho \sigma }\,\Gamma _{\mu \nu }\,\Gamma _{\rho \sigma } \nn \\
    & -\frac{1}{105} F_{\mu \nu }\,(\P_{\alpha}\Sigma )\,\Gamma _{\alpha \nu }\,(\P_{\mu }\Sigma ) - \frac{1}{630} F_{\mu \nu }\,(\P_{\rho}\Sigma )\,F_{\mu \nu }\,(\P_{\rho}\Sigma ) - \frac{1}{315} F_{\mu \nu }\,(\P_{\rho}\Sigma )\,\Gamma _{\mu \nu }\,(\P_{\rho}\Sigma )  \nn \\
    & - \frac{2}{315} F_{\mu \rho }\,F_{\rho \nu }\,(\P_{\mu }\Sigma )\,(\P_{\nu}\Sigma ) + \frac{2}{63} F_{\mu \rho }\,F_{\rho \nu }\,(\P_{\nu}\Sigma )\,(\P_{\mu }\Sigma ) - \frac{2}{315} F_{\mu \rho }\,\Gamma _{\rho \nu }\,(\P_{\mu }\Sigma )\,(\P_{\nu}\Sigma )  \nn \\
    & + \frac{2}{63} F_{\mu \rho }\,\Gamma _{\rho \nu }\,(\P_{\nu}\Sigma )\,(\P_{\mu }\Sigma ) + \frac{1}{420} F_{\mu \sigma }\,F_{\sigma \rho }\,\Gamma _{\mu \nu }\,F_{\nu \rho } + \frac{1}{420} F_{\mu \sigma }\,F_{\sigma \rho }\,\Gamma _{\mu \nu }\,\Gamma _{\nu \rho } \nn \\
    & + \frac{1}{420} F_{\mu \sigma }\,\Gamma _{\sigma \rho }\,\Gamma _{\mu \nu }\,F_{\nu \rho } + \frac{1}{420} F_{\mu \sigma }\,\Gamma _{\sigma \rho }\,\Gamma _{\mu \nu }\,\Gamma _{\nu \rho } + \frac{1}{2520}F_{\nu \rho }\,F_{\rho \alpha }\,\Gamma _{\alpha \mu }\,\Gamma _{\mu \nu } \nn \\
    & +\frac{1}{420} F_{\nu \rho }\,\Gamma _{\mu \sigma }\,F_{\sigma \rho }\,\Gamma _{\mu \nu } + \frac{1}{420} F_{\nu \rho }\,\Gamma _{\mu \sigma }\,\Gamma _{\sigma \rho }\,\Gamma _{\mu \nu } + \frac{1}{2520}F_{\nu \rho }\,\Gamma _{\rho \alpha }\,\Gamma _{\alpha \mu }\,\Gamma _{\mu \nu } \nn \\
    & +\frac{1}{2520}F_{\rho \alpha }\,\Gamma _{\alpha \mu }\,\Gamma _{\mu \nu }\,\Gamma _{\nu \rho }-\frac{1}{105} F_{\rho \mu }\,(\P_{\mu }\Sigma )\,F_{\rho \nu }\,(\P_{\nu}\Sigma ) - \frac{1}{105} F_{\rho \mu }\,(\P_{\mu }\Sigma )\,\Gamma _{\rho \nu }\,(\P_{\nu}\Sigma )  \nn \\
    & - \frac{2}{315} F_{\rho \nu }\,(\P_{\mu }\Sigma )\,(\P_{\nu}\Sigma )\,\Gamma _{\mu \rho } - \frac{1}{105} F_{\rho \nu }\,(\P_{\nu}\Sigma )\,\Gamma _{\rho \mu }\,(\P_{\mu }\Sigma ) + \frac{2}{63} F_{\rho \nu }\,(\P_{\nu}\Sigma )\,(\P_{\mu }\Sigma )\,\Gamma _{\mu \rho }  \nn \\
    & + \frac{1}{10080}F_{\rho \sigma }\,\Gamma _{\mu \nu }\,F_{\rho \sigma }\,\Gamma _{\mu \nu }+\frac{1}{5040}F_{\rho \sigma }\,\Gamma _{\mu \nu }\,\Gamma _{\rho \sigma }\,\Gamma _{\mu \nu }+\frac{1}{420} F_{\sigma \rho }\,\Gamma _{\mu \nu }\,\Gamma _{\nu \rho }\,\Gamma _{\mu \sigma } \nn \\
    & + \frac{1}{2520}\Gamma _{\alpha \mu }\,\Gamma _{\mu \nu }\,\Gamma _{\nu \rho }\,\Gamma _{\rho \alpha } -\frac{1}{105} \Gamma _{\alpha \nu }\,(\P_{\mu }\Sigma )\,\Gamma _{\mu \nu }\,(\P_{\alpha}\Sigma ) + \frac{1}{420} \Gamma _{\mu \nu }\,\Gamma _{\nu \rho }\,\Gamma _{\mu \sigma }\,\Gamma _{\sigma \rho } \nn \\
    & + \frac{1}{10080}\Gamma _{\mu \nu }\,\Gamma _{\rho \sigma }\,\Gamma _{\mu \nu }\,\Gamma _{\rho \sigma }-\frac{1}{630} \Gamma _{\mu \nu }\,(\P_{\rho}\Sigma )\,\Gamma _{\mu \nu }\,(\P_{\rho}\Sigma ) - \frac{2}{315} \Gamma _{\mu \rho }\,\Gamma _{\rho \nu }\,(\P_{\mu }\Sigma )\,(\P_{\nu}\Sigma )  \nn \\
    & + \frac{2}{63} \Gamma _{\mu \rho }\,\Gamma _{\rho \nu }\,(\P_{\nu}\Sigma )\,(\P_{\mu }\Sigma ) - \frac{1}{105} \Gamma _{\rho \mu }\,(\P_{\mu }\Sigma )\,\Gamma _{\rho \nu }\,(\P_{\nu}\Sigma ) - \frac{2}{45} Y\,\Sigma \,F_{\mu \nu }\,\Sigma \,F_{\mu \nu }  \nn \\
    & + \frac{17}{105} (\P_{\mu }\Sigma )\,(\P_{\nu}\Sigma )\,(\P_{\mu }\Sigma )\,(\P_{\nu}\Sigma ) - \frac{2}{45} Y\,\Sigma \,F_{\mu \nu }\,\Sigma \,\Gamma _{\mu \nu } - \frac{1}{15} Y\,\Sigma \,F_{\mu \nu }\,\Gamma _{\mu \nu }\,\Sigma  \nn \\
    & - \frac{2}{45} Y\,\Sigma \,\Gamma _{\mu \nu }\,\Sigma \,F_{\mu \nu } - \frac{2}{45} Y\,\Sigma \,\Gamma _{\mu \nu }\,\Sigma \,\Gamma _{\mu \nu } - \frac{1}{15} Y\,\Sigma \,\Gamma _{\mu \nu }\,F_{\mu \nu }\,\Sigma  - \frac{2}{45} Y\,F_{\mu \nu }\,\Sigma \,F_{\mu \nu }\,\Sigma   \nn \\
    & - \frac{2}{45} Y\,F_{\mu \nu }\,\Sigma \,\Gamma _{\mu \nu }\,\Sigma  - \frac{2}{45} Y\,\Gamma _{\mu \nu }\,\Sigma \,F_{\mu \nu }\,\Sigma  - \frac{2}{45} Y\,\Gamma _{\mu \nu }\,\Sigma \,\Gamma _{\mu \nu }\,\Sigma  + \frac{8}{315} \Sigma \,F_{\mu \nu }\,\Sigma \,F_{\nu \rho }\,F_{\rho \mu }  \nn \\
    & + \frac{8}{315} \Sigma \,F_{\mu \nu }\,\Sigma \,F_{\nu \rho }\,\Gamma _{\rho \mu } + \frac{8}{315} \Sigma \,F_{\mu \nu }\,\Sigma \,\Gamma _{\nu \rho }\,F_{\rho \mu } + \frac{8}{315} \Sigma \,F_{\mu \nu }\,\Sigma \,\Gamma _{\nu \rho }\,\Gamma _{\rho \mu }  \nn \\
    & + \frac{52}{105} \Sigma \,F_{\mu \nu }\,(\P_{\mu }\Sigma )\,\Sigma \,(\P_{\nu}\Sigma ) + \frac{8}{315} \Sigma \,F_{\nu \rho }\,F_{\rho \mu }\,\Sigma \,\Gamma _{\mu \nu } + \frac{8}{315} \Sigma \,F_{\nu \rho }\,\Gamma _{\rho \mu }\,\Sigma \,\Gamma _{\mu \nu }  \nn \\
    & + \frac{8}{315} \Sigma \,\Gamma _{\mu \nu }\,\Sigma \,\Gamma _{\nu \rho }\,\Gamma _{\rho \mu } + \frac{12}{35} \Sigma \,\Gamma _{\mu \nu }\,\Sigma \,(\P_{\mu }\Sigma )\,(\P_{\nu}\Sigma ) + \frac{52}{105} \Sigma \,\Gamma _{\mu \nu }\,(\P_{\mu }\Sigma )\,\Sigma \,(\P_{\nu}\Sigma )  \nn \\
    & + \frac{12}{35} \Sigma \,F_{\mu \nu }\,\Sigma \,(\P_{\mu }\Sigma )\,(\P_{\nu}\Sigma ) + \frac{8}{315} \Sigma \,\Gamma _{\mu \nu }\,\Sigma \,\Gamma _{\nu \rho }\,F_{\rho \mu } + \frac{16}{105} \Sigma \,(\P_{\mu }\Sigma )\,\Sigma \,(\P_{\nu}\Sigma )\,F_{\mu \nu } \nn \\
    & + \frac{16}{105} \Sigma \,(\P_{\mu }\Sigma )\,\Sigma \,(\P_{\nu}\Sigma )\,\Gamma _{\mu \nu }\bigg\}\bigg].
\end{align}
\section{Universal One-Loop Effective Lagrangian up to D8}
\label{App:UOLEA}
UOLEA derived in Ref.~\cite{Banerjee:2023iiv} in terms of the generalised covariant derivative and interaction functional defines in Eq.~\eqref{eq:unified_lag} is given below.
\begin{align}\label{eq:finite}
    \mathcal{L}_{\eff}^{d \leq 8} = & \cfrac{c_s}{(4\pi)^{2}}\   M^4\left[-\frac{1}{2}\,\left(\ln\left[\frac{M^2}{\mu^2}\right]-\frac{3}{2}\right)\right ]  + \cfrac{c_s}{(4\pi)^{2}} \, \tr\bigg\{ M^2\ \bigg[-\left(\ln\left[\frac{M^2}{\mu^2}\right]-1\right)\, U\bigg]  \nn \\
    & + M^0\ \frac{1}{2}  \bigg[- \ln\left[\frac{M^2}{\mu^2}\right] \, U^2 -\frac{1}{6} \ln\left[\frac{M^2}{\mu^2}\right] \, (G_{\mu\nu})^2\bigg] \nn\\ 
    & + \frac{1}{M^2} \frac{1}{6}  \,\bigg[ -U^3 - \frac{1}{2} (\P_\mu U)^2-\frac{1}{2}U\,(G_{\mu\nu})^2  - \frac{1}{10}(J_\nu)^2   + \frac{1}{15}\,G_{\mu\nu}\,G_{\nu\rho}\,G_{\rho\mu} \bigg]  \nn\\
    &+ \frac{1}{M^4} \frac{1}{24} \bigg[U^4 - U^2 (\P^2 U) + \frac{4}{5}U^2 (G_{\mu\nu})^2 + \frac{1}{5} (U\,G_{\mu\nu})^2 +  \frac{1}{5} (\P^2 U)^2  \nn\\ 
    & \hspace{2cm} -\frac{2}{5} U\, (\P_\mu U)\,J_{\mu} + \frac{2}{5} U(J_\mu)^2 - \frac{2}{15} (\P^2 U) (G_{\rho\sigma})^2 +\frac{1}{35}(\P_\nu J_{\mu})^2  \nn\\ 
    & \hspace{2cm} - \frac{4}{15} U\,G_{\mu\nu}G_{\nu\rho} G_{\rho\mu} - \frac{8}{15} (\P_\mu \P_\nu U)\, G_{\rho\mu} G_{\rho\nu} + \frac{16}{105}G_{\mu\nu}J_{\mu}J_{\nu}   \nn\\
    & \hspace{2cm} + \frac{1}{420} (G_{\mu\nu}G_{\rho\sigma})^2 +\frac{17}{210}(G_{\mu\nu})^2(G_{\rho\sigma})^2 +\frac{2}{35}(G_{\mu\nu}G_{\nu\rho})^2 \nn\\
    & \hspace{2cm} + \frac{1}{105} G_{\mu\nu}G_{\nu\rho}G_{\rho\sigma}G_{\sigma\mu} +\frac{16}{105} (\P_\mu J_{\nu}) G_{\nu\sigma}G_{\sigma\mu} \bigg]  \nn\\
    & + \frac{1}{M^6} \frac{1}{60}  \,\bigg[ -U^5 + 2\,U^3 (\P^2 U) + U^2(\P_\mu U)^2 - \frac{2}{3} U^2 G_{\mu\nu} U\,G_{\mu\nu}  - U^3 (G_{\mu\nu})^2  \nn \\
    & \hspace{2cm} + \frac{1}{3} U^2 (\P_\mu U)J_\mu - \frac{1}{3} U\,(\P_\mu U)(\P_\nu U)\,G_{\mu\nu}   - \frac{1}{3} U^2 J_\mu (\P_\mu U)  \nn\\
    & \hspace{2cm} -  \frac{1}{3} U\,G_{\mu\nu}(\P_\mu U)(\P_\nu U) - U\,(\P^2 U)^2  -  \frac{2}{3} (\P^2 U) (\P_\nu U)^2  - \frac{1}{7} ((\P_\mu U)G_{\mu\alpha})^2 \nn\\
    & \hspace{2cm} +\frac{2}{7} U^2 G_{\mu\nu}G_{\nu\alpha}G_{\alpha\mu}+\frac{8}{21}U\,G_{\mu\nu}U\,G_{\nu\alpha}G_{\alpha\mu}-\frac{4}{7}U^2(J_\mu)^2 -\frac{3}{7} (U\,J_\mu)^2 \nn \\
    & \hspace{2cm} +\frac{4}{7}U\,(\P^2U)(G_{\mu\nu})^2 +\frac{4}{7}(\P^2U)U(G_{\mu\nu})^2 -\frac{2}{7}U\,(\P_\mu U)J_\nu G_{\mu\nu} \nn \\
    & \hspace{2cm} -\frac{2}{7}(\P_\mu U)U\,G_{\mu\nu} J_\nu -\frac{4}{7}U\,(\P_\mu U)G_{\mu\nu} J_\nu -\frac{4}{7}(\P_\mu U)U\, J_\nu G_{\mu\nu} \nn \\
    & \hspace{2cm} +\frac{4}{21}U\,G_{\mu\nu}(\P^2U)G_{\mu\nu}  +\frac{11}{21}(\P_\alpha U)^2(G_{\mu\nu})^2 - \frac{10}{21}(\P_\mu U)J_\nu U\, G_{\mu\nu}  \nn \\
    & \hspace{2cm} - \frac{10}{21}(\P_\mu U) G_{\mu\nu} U \,J_\nu - \frac{2}{21} (\P_\mu U)(\P_\nu U)G_{\mu\alpha}G_{\alpha\nu} + \frac{10}{21} (\P_\nu U)(\P_\mu U)G_{\mu\alpha}G_{\alpha\nu}  \nn \\
    & \hspace{2cm}-\frac{1}{7} (G_{\alpha\mu}(\P_\mu U))^2 - \frac{1}{42} ((\P_\alpha U)G_{\mu\nu})^2 -\frac{1}{14} (\P_\mu \P^2 U)^2 -\frac{4}{21} (\P^2U) (\P_\mu U)J_\mu \nn \\
    & \hspace{2cm} +\frac{4}{21} (\P_\mu U)(\P^2U)J_\mu +\frac{2}{21} (\P_\mu U) (\P_\nu U)(\P_\mu J_{\nu}) - \frac{2}{21} (\P_\nu U) (\P_\mu U)(\P_\mu J_{\nu}) \bigg]  \nn\\
    & + \frac{1}{M^8} \frac{1}{120}  \,\bigg[U^6 - 3\,U^4 (\P^2 U) - 2\,U^3(\P_\nu U)^2 + \frac{12}{7}U^2 (\P_\mu \P_\nu U)(\P_\nu \P_\mu U)  \nn\\
    & \hspace{.5cm}  +\frac{26}{7} (\P_\mu \P_\nu U) U\,(\P_\mu U)(\P_\nu U) +\frac{26}{7} (\P_\mu \P_\nu U) (\P_\mu U)(\P_\nu U)U  + \frac{9}{7} (\P_\mu U)^2(\P_\nu U)^2  \nn\\
    & \hspace{2cm} + \frac{9}{7} U\,(\P_\mu \P_\nu U)U\,(\P_\nu \P_\mu U)  + \frac{17}{14} ((\P_\mu U)(\P_\nu U))^2 + \frac{8}{7} U^3G_{\mu\nu}U\,G_{\mu\nu}  \nn\\
    & \hspace{2cm} + \frac{5}{7} U^4(G_{\mu\nu})^2 + \frac{18}{7} G_{\mu\nu}(\P_\mu U)U^2(\P_\nu U) + \frac{9}{14} (U^2G_{\mu\nu})^2   \nn\\
    & \hspace{2cm} + \frac{18}{7} G_{\mu\nu}U\,(\P_\mu U)(\P_\nu U)U + \frac{18}{7} (\P_\mu \P_\nu U) (\P_\mu U)U\,(\P_\nu U)   \nn\\
    & \hspace{2cm} +  \bigg( \frac{8}{7} G_{\mu\nu}U\,(\P_\mu U)U\,(\P_\nu U) +  \frac{26}{7} G_{\mu\nu}(\P_\mu U)U\,(\P_\nu U)U \bigg) \nn\\
    & \hspace{2cm} +  \bigg( \frac{24}{7} G_{\mu\nu}(\P_\mu U)(\P_\nu U)U^2 - \frac{2}{7} G_{\mu\nu}U^2(\P_\mu U)(\P_\nu U)\bigg)\bigg] \nn\\
    & + \frac{1}{M^{10}} \frac{1}{210}  \,\bigg[-U^7 - 5\, U^4 (\P_\nu U)^2 - 8\,U^3(\P_\mu U)U(\P_\mu U)  -\frac{9}{2} (U^2 (\P_\mu U))^2 \bigg] \nn\\
    & + \frac{1}{M^{12}} \frac{1}{336} \,\bigg[U^8\bigg] \bigg\}\bigg].
\end{align}
Here, we consider that the tensors $G_{\mu\nu}$ and $J_\mu$ are functions of $\P$: $G_{\mu\nu} = [\P_\mu,\P_\nu]$, and $J_\mu = \P_\nu G_{\nu\mu} = [\P_\nu,[\P_\nu,\P_\mu]]$. Please note that the hermitian conjugates are already fed in the above expression such that the effective Lagrangian is self-hermitian.

\bibliographystyle{jhep}
\bibliography{ref.bib}

\end{document}